\documentclass[a4paper,oneside,submitp]{pioArxiv}

\definecolor{refcolor}{rgb}{0,0,0.4}    
\definecolor{anchorcolor}{rgb}{0,0,0}   
\definecolor{citecolor}{rgb}{0,0.4,0.4} 
\definecolor{urlcolor}{rgb}{0,0,0.690}   
\definecolor{filecolor}{rgb}{0,0,0.690}  
\definecolor{menucolor}{rgb}{0.75,0,0}  
\definecolor{pagecolor}{rgb}{0.75,0,0}  

\usepackage[square]{natbib}
\usepackage{afterpage}
\usepackage{here}
\usepackage{tabularx,pdflscape}
\usepackage{array,longtable}
\newcolumntype{Y}{>{\raggedright\arraybackslash}X}
\newcounter{Fig}

\newcommand\leqt[1]{\protect\label{eq:#1}}
\newcommand\reqtn[1]{\ref{eq:#1}}
\newcommand\reqt[1]{(\reqtn{#1})}

\newcommand\lsect[1]{\protect\label{sect:#1}}
\newcommand\rsect[1]{\ref{sect:#1}}

\newcommand\ltab[1]{\protect\label{tab:#1}}
\newcommand\rtab[1]{\ref{tab:#1}}

\newcommand\REMOVE[1]{}

\begin{document}
\sloppy

\title{Multistep Parametric Processes in Nonlinear Optics}
\shorttitle{Multistep Parametric Processes}

\author{Solomon M. Saltiel$^{1,2}$, Andrey A. Sukhorukov$^1$, and Yuri S. Kivshar$^1$\\
\begin{small}
$^1$ Nonlinear Physics Group and Center for Ultra-high bandwidth
Devices for Optical Systems (CUDOS), Research School of Physical
Sciences and Engineering, Australian National University,
Canberra ACT 0200, Australia\\
HomePage:~\href{http://wwwrsphysse.anu.edu.au/nonlinear}{wwwrsphysse.anu.edu.au/nonlinear}
\\
$^2$ Faculty of Physics, University of Sofia, 5 J. Bourchier Bld,
Sofia BG-1164, Bulgaria
\end{small}
}

\maketitle

\begin{abstract}
We present a comprehensive overview of different types of
parametric interactions in nonlinear optics which are associated
with simultaneous phase-matching of several optical processes in
quadratic nonlinear media, the so-called {\em multistep parametric
interactions}. We discuss a number of possibilities of double and
multiple phase-matching in engineered structures with the
sign-varying second-order nonlinear susceptibility, including (i)
uniform and non-uniform quasi-phase-matched (QPM) periodic optical
superlattices, (ii) phase-reversed and periodically chirped QPM
structures, and (iii) uniform QPM structures in non-collinear
geometry, including recently fabricated two-dimensional nonlinear
quadratic photonic crystals. We also summarize the most important
experimental results on the multi-frequency generation due to
multistep parametric processes, and overview the physics and basic
properties of multi-color optical parametric solitons generated by
these parametric interactions.
\end{abstract}

\tableofcontents

\section{Introduction} \lsect{intro}

Energy transfer between different modes and phase-matching
relations are the fundamental concepts in nonlinear optics. Unlike
nonparametric nonlinear processes such as self-action and
self-focusing of light in a nonlinear Kerr-like medium, parametric
processes involve several waves at different frequencies and they
require special relations between the wave numbers and wave group
velocities to be satisfied, the so-called {\em phase-matching
conditions}.

Parametric coupling between waves occurs naturally in nonlinear
materials without the inversion symmetry, when the lowest-order
nonlinear effects are presented by {\em quadratic nonlinearities},
often called $\chi^{(2)}$ nonlinearities because they are
associated with the second-order contribution ($\sim
\chi^{(2)}E^2$) to the nonlinear polarization of a medium.
Conventionally, the phase-matching conditions for most parametric
processes in optics are implemented either by using anisotropic
crystals (the so-called perfect phase-matching), or they occur in
fabricated structures with a periodically reversed sign of the
quadratic susceptibility (the so-called quasi-phase-matching or
QPM). The QPM technique is one of the leading technologies these
days, and it employs the spatial scales ($\sim 1 \div 30 \mu m$)
which are compatible with the operational wavelengths of optical
communication systems.

Nonlinear effects produced by quadratic intensity-dependent
response of a transparent dielectric medium are usually associated
with parametric frequency conversion such as the second harmonic
generation (SHG). The SHG process is one of the most well-studied
parametric interactions which may occur in a quadratic nonlinear
medium. Moreover, recent theoretical and experimental results
demonstrate that quadratic nonlinearities can also produce many of
the effects attributed to nonresonant Kerr nonlinearities via
cascading of several second-order parametric processes. Such
second-order cascading effects can simulate third-order processes and, in particular, those associated with the intensity-dependent change of the medium refractive index~\cite{Stegeman:1996-1691:OQE}. Importantly, the effective (or induced) cubic nonlinearity resulting from a cascaded SHG process
in a quadratic medium can be of the several orders of magnitude
higher than that usually measured in centrosymmetric Kerr-like
nonlinear media, and it is practically instantaneous.

The simplest type of the phase-matched parametric interaction is
based on the simultaneous action of two second-order parametric
sub-processes that belong to a single second-order interaction.
For example, the so-called two-step cascading associated with type
I SHG includes the generation of the second harmonic (SH), $\omega
+ \omega =2 \omega$, followed by the reconstruction of the
fundamental wave through the down-conversion frequency mixing
process, $2\omega -\omega = \omega$. These two sub-processes
depend only on a single phase-matching parameter $\Delta k$. In
particular, for the nonlinear $\chi^{(2)}$ media with a periodic
modulation of the quadratic nonlinearity, for the QPM periodic
structures, we have $\Delta k = k_2 -2k_1 + G_m$, where $k_1 =
k(\omega)$, $k_2 = k(2\omega)$ and $G_m$ is a reciprocal vector of
the periodic structure, $G_m = 2\pi m / \Lambda$, where $\Lambda$ is
the lattice spacing and $m$ is integer.  For a homogeneous bulk
$\chi^{(2)}$ medium, we have $G_m=0$.

Multistep parametric interactions and multistep cascading presents
a special type of the second-order parametric processes that
involve {\em several} different second-order nonlinear
interactions; they are characterized by at least two different
phase-matching parameters. For example, two parent processes of
the so-called third-harmonic cascading are: (i) second-harmonic
generation, $\omega + \omega =2 \omega$, and (ii) sum-frequency
mixing, $\omega + 2\omega =3 \omega$. Here, we may distinguish
five harmonic sub-processes, and the multistep interaction results
in their simultaneous action.

Different types of the multistep parametric processes include
third-harmonic cascaded generation, two-color parametric
interaction, fourth-harmonic cascading, difference-frequency
generation, etc. Various applications of the multistep parametric
processes have been mentioned in the literature. In particular,
the multistep parametric interaction can support multi-color
solitary waves, it usually leads to larger accumulated nonlinear
phase shifts, in comparison with the simple cascading, it can be
effectively employed for the simultaneous generation of
higher-order harmonics in a single quadratic crystal, and also for
the generation of a cross-polarized wave and frequency shifting in
fiber optics gratings. Generally, simultaneous phase-matching of
several parametric processes cannot be achieved by the traditional
methods such as those based on the optical birefringence effect.
However, the situation becomes different for the media with a
periodic change of the sign of the quadratic nonlinearity, as
occurs in the QPM structures or two-dimensional nonlinear photonic
crystals.

\pict[1]{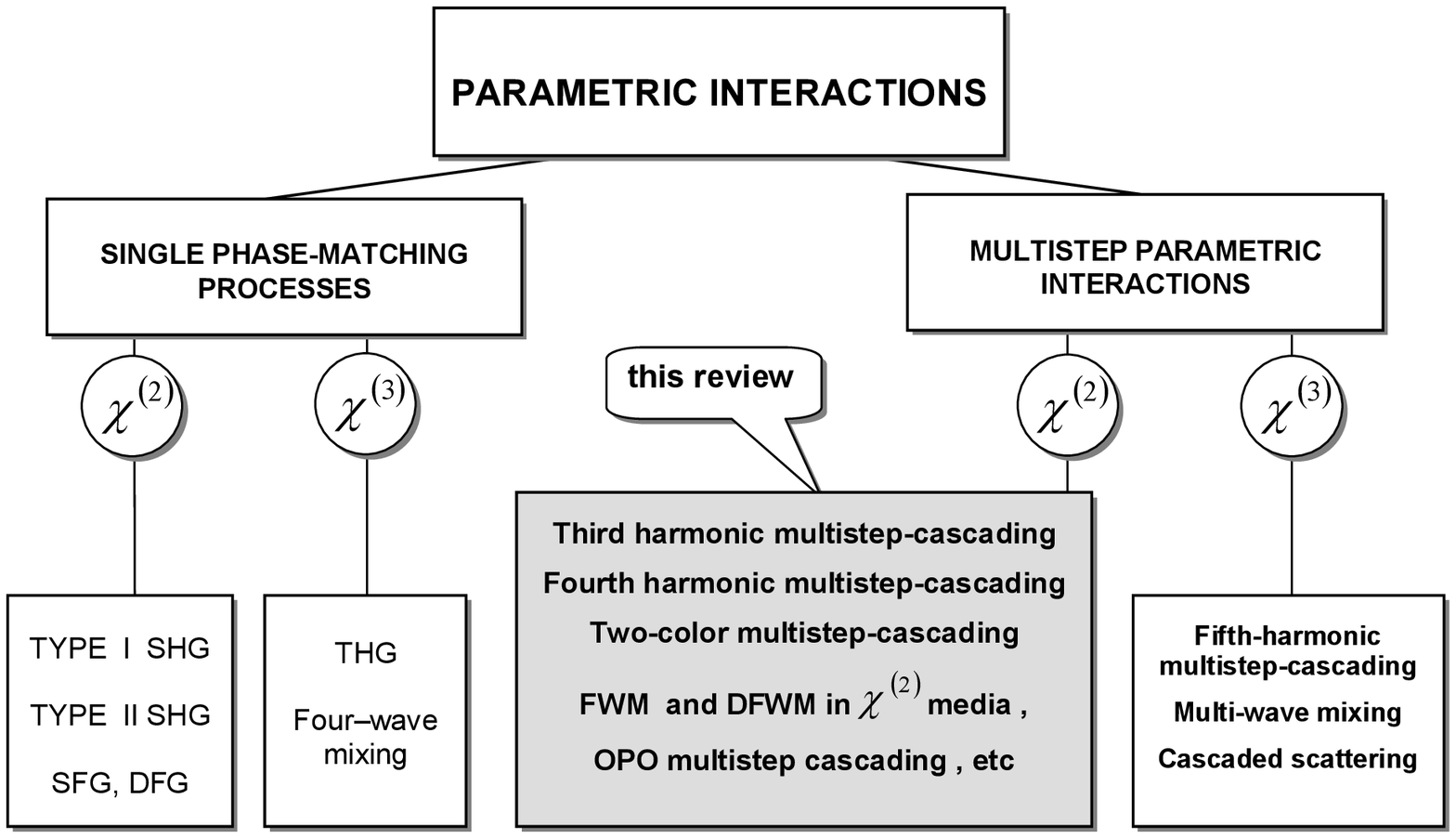}{cascading}{Different types of parametric
processes in nonlinear optics, and the specific topics covered by
this review paper. SHG: second-harmonic generation; SFG and DFG:
sum- and difference-frequency generation, THG: third-harmonic
generation; FWM: four-wave mixing; DFWM: degenerate FWM; OPO:
optical parametric oscillator.}

In this review paper, we describe the basic principles of
simultaneous phase-matching of two (or more) parametric processes
in different types of one- and two-dimensional nonlinear quadratic
optical lattices. We divide different types of phase-matched
parametric processes studied in nonlinear optics into {\em two
major classes}, as shown in Fig.~\rpict{cascading}, and discuss
different types of parametric interactions associated with
simultaneous phase-matching of several optical processes in
quadratic (or $\chi^{(2)}$) nonlinear media, the so-called
multistep parametric interactions. In particular, we overview the
basic principles of double and multiple phase-matching in
engineered structures with the sign-varying second-order nonlinear
susceptibility, including different types of QPM optical
superlattices, non-collinear geometry, and two-dimensional
nonlinear quadratic photonic crystals (which can be considered as
two-dimensional QPM lattices). We also summarize the most
important experimental results on the multi-frequency generation
due to multistep parametric processes, and overview the physics
and basic properties of multi-color optical solitons generated by
these parametric interactions.

\section[Single-step cascading]{Single-phase-matched processes}
          \lsect{single}

One of the simplest and first studied parametric process in
nonlinear optics is the second-harmonic generation (SHG). The SHG
process is a special case of a more general three-wave mixing
process which occurs in a dielectric medium with a quadratic
intensity-dependent response. The three-wave mixing and SHG
processes require only one phase-matching condition to be
satisfied and, therefore, they both can be classified as {\em
single phase-matched processes}.

In this section, we discuss briefly these single phase-matched
processes, and consider parametric interaction between three
continuous-wave (CW) waves with the electric fields $E_j =
\frac{1}{2} [A_j \exp(-i{\bf k}_j\cdot {\bf r} + i\omega_jt) +
{\rm c.c.}]$, where $j$ = 1,2,3, with the three frequencies
satisfying the energy-conservation condition, $\omega_1+\omega_2 =
\omega_3$. We assume that the phase-matching condition is nearly
satisfied, with a small mismatch $\Delta k$ among the three wave
vectors; i.e., $\Delta k = k_3(\omega_3) -k_1(\omega_1) -
k_2(\omega_2)$. In general, the three waves do not propagate along
the same direction, and the beams may walk off from each other as
they propagate inside the crystal. If all three wave vectors point
along the same direction (as, e.g., in the case of the QPM
materials), the waves have the same phase velocity and exhibit no
walk-off.

The theory of $\chi^{(2)}$-mediated three-wave mixing is available
in several books devoted to nonlinear optics
(\cite{Shen:1984:PrinciplesNonlinear,
Butcher:1992:ElementsNonlinear, Boyd:1992:NonlinearOptics}). The
starting point is the Maxwell wave equation written as
\begin{equation} \leqt{waveq}
     \nabla \times \nabla \times {\bf E} + \frac{1}{c^2}\frac {\partial^2\bf E}{\partial t^2}=
        - \frac{1}{\epsilon_0 c^2}\frac {\partial ^2 \bf P}{\partial t^2 },
\end{equation}
where $\epsilon_0$ is the vacuum permittivity and $c$ is the speed
of light in a vacuum. The induced polarization is written in in
the frequency domain as
\begin{equation} \leqt{waveq1}
     \tilde{\bf P}({\bf r}, \omega) = \epsilon_0\chi^{(1)}\tilde{\bf E} +
        \epsilon_0\chi^{(2)}\tilde{\bf E}\tilde{\bf E} + \cdots,
\end{equation}
where a tilde denotes the Fourier transform. Using the slowly varying envelope approximation and neglecting the walk-off, one can derive the following set of three coupled equations describing the parametric interaction of three waves under type II phase matching:
\begin{eqnarray}
   ik_1\frac{dA_1}{d z} - \omega_1^2 \Gamma A_3 A_2^* e^{-i\Delta  k
   z} &=&  0,
   \leqt{final_sys1}\\
   ik_2  \frac{d A_2}{d z} - \omega_2^2 \Gamma A_3 A_1^*
   e^{-i\Delta  k z} &=&  0, \\
  ik_3 \frac{d A_3}{d z} - \omega_3^2 \Gamma A_1 A_2 e^{i\Delta k z}
   &=&  0, \leqt{final_sys3}
\end{eqnarray}
where $\Gamma = d^{(2)}_{\rm eff}/c^2$, and $d^{(2)}_{\rm eff}$ is
a convolution of the second-order susceptibility tensor
$\hat{\chi}^{(2)}$ and the polarization unit vectors of the three
fields, $d^{(2)}_{\rm eff} = \frac{1}{2} <{\bf e}_3
\hat{\chi}^{(2)} {\bf e}_1 {\bf e}_2>$.

In the case of type I SHG, only a single beam at the pump
frequency $\omega_1$ is incident on the nonlinear crystal, and a
new optical field at the frequency $2 \omega_1$ is generated
during the SHG process. We can adapt
Eqs.~\reqt{final_sys1}--\reqt{final_sys3} to this case with
minor modifications. More specifically, we set $\omega_3 =
2\omega_1$ and $A_1 = A_2$. The first two equations then become
identical, and one of them can be dropped. The type I SHG process
is thus governed by the following set of of two coupled equations:
\begin{eqnarray}
    ik_1\frac{d A_1}{d z} - k_1 \sigma A_3 A_1^* e^{-i\Delta  k z} &=& 0, \\
    ik_3\ \frac{d A_3}{d z}  - 2k_1 \sigma A_1^2 e^{i\Delta  k z} &=& 0,
  \leqt{final_shg2}
\end{eqnarray}
where $\sigma = (\omega_1/n_1c)d^{(2)}_{\rm eff}$ is the nonlinearity parameter and $\Delta  k = k_3- 2 k_1$ is the phase-mismatch parameter.

Both the three-wave mixing and SHG processes present an example of
a single-phase-matched parametric process because it is controlled
by a single phase-matching parameter $\Delta k$. This kind of
parametric processes can be described as a two-step cascading
interaction which includes: (i) the generation of the
second-harmonic (SH) wave, $\omega + \omega =2 \omega$, followed
by (ii) the reconstruction of the fundamental wave through the
down-conversion frequency mixing process, i.e. $2\omega -\omega =
\omega$. Respectively, the first sub-process is responsible for
the generation of the SH field, with the most efficient conversion
observed at $\Delta k =0$, while the second sub-process, also
called cascading,  can be associated with an effective
intensity-dependent change of the phase of the fundamental
harmonic ($\sim d^2/\Delta k$), which is similar to that of the
cubic nonlinearity (\cite{Desalvo:1992-28:OL,
Stegeman:1996-1691:OQE, Assanto:1995-673:IQE}). This latter effect
is responsible for the generation of the so-called quadratic
solitons, two-wave parametric soliton composed of the mutually
coupled fundamental and second-harmonic components
(\cite{Sukhorukov:1988:NonlinearWave, Torner:1998-229:BeamShaping,
Kivshar:1997-451:AdvancedPhotonics,
Etrich:2000-483:ProgressOptics,
Torruellas:2001-127:SpatialOptical, Boardman:2001:SolitonDriven,
Buryak:2002-63:PRP}, and references therein).

Multistep parametric interactions and multistep cascading effects
discussed in this review paper are presented by different types of
phase-matched parametric interactions in quadratic (or
$\chi^{(2)}$) nonlinear media which involve several different
parametrically interacting waves, e.g. as in the case of the
frequency mixing and sum-frequency generation. However, all such
interactions can also be associated with the two major physical
mechanisms of the wave interaction discussed for the SH process
above: (i)~parametric energy transfer between waves determined by
the phase-mismatch between the wave vectors of the interacting
waves, and (ii)~phase changes due to this parametric interaction. Below, we discuss these interactions for a number
of physically important examples.

\section[Multi-step cascading]{Multistep phase-matched interactions}
          \lsect{multi}

\begin{small}
\begin{longtable}{cp{4cm}p{2.8cm}p{3cm}}

\caption{\ltab{multistep} Examples of multistep parametric
interactions involving SHG.}\\

\hline No.& Multistep parametric process &
Cascading $\chi^{(2)}$ steps &  Equivalent high-order parametric process\\
\hline 
\endfirsthead

\hline No.& Multistep parametric process &
Cascading $\chi^{(2)}$ steps &  Equivalent high-order parametric process\\
\hline 
\endhead

1 & Type I third-harmonic multistep process & {$\omega
+\omega =2\omega$}; {$\omega +2\omega =3\omega$}
& \ $\omega +\omega+\omega=3\omega$\\

2 & Type II third-harmonic multistep process &
 $\omega +\omega =2\omega$ ; $\omega_\perp +2\omega
=3\omega$ & \ $\omega+\omega+\omega_\perp=3\omega$ \\

3& 3:1 frequency conversion and division & $3\omega \rightarrow
2\omega +\omega$ ;
$\omega =2\omega -\omega$ &\\

4& Fourth-harmonic multistep process & $\omega +\omega =2\omega$ ;
$2\omega +2\omega
=4\omega$& $\omega+\omega+\omega + \omega=4\omega$ \\

5& Type I \& Type II two-color multistep process & $\omega +\omega
=2\omega$ ; $2\omega -\omega
=\omega_\perp$&$\omega+\omega-\omega =\omega_\perp$ \\

6& Type I \& Type I two-color multistep process & $\omega +\omega
=2\omega$ ; $\omega_\perp
+\omega_\perp=2\omega$& \\

7& wavelength conversion  & $\omega
+\omega =2\omega$ ; $2\omega - \omega_a
=\omega_b$&$\omega+\omega-\omega_a=\omega_b$ \\

8& Self-doubling OPO & $\omega_p \rightarrow \omega_i +\omega_s $
;
$\omega_s+\omega_s=\omega_{\rm s,SH}$& \\

9& Self-sum-frequency generation OPO & $\omega_p \rightarrow
\omega_i +\omega_s $ ;
$\omega_s+\omega_p=\omega_{\rm SFG}$& \\

10& Internally pumped OPO & $\omega_{p/2} +\omega_{p/2} =\omega_p$
; $\omega_p \rightarrow \omega_i +\omega_s $&
\end{longtable}
In the table above, the symbol $\omega_{\perp}$ stands for a wave
polarized in the plane perpendicular to that of the wave with the
main carrier frequency $\omega$.
~\\
~
\end{small}

In this part, we consider the nonlinear parametric interactions
that involve several processes, such that each of the processes is
described by an independent phase-matching parameter. In the early
days of nonlinear optics, the motivation to study this kind of
parametric interactions was to explore various possibilities for
the simultaneous generation of several harmonics in a single
nonlinear crystal (see, e.g.,
\cite{Akhmanov:1964:ProblemyNelineinoi,
Akhmanov:1972:ProblemsNonlinear}) as well as to use the cascading
of several parametric processes for measuring higher-order
susceptibilities in nonlinear optical crystals (see, e.g.,
\cite{Yablonovitch:1972-865:PRL, Akhmanov:1974-264:PZETF,
Akhmanov:1975-239:NonlinearSpectroscopy, Kildal:1979-5218:PRB,
Bloembergen:1982-685:RMP}). More recently, these processes were
proved to be efficient for the higher-order harmonic generation,
for building reliable standards for the third-order nonlinear
susceptibility measurements (see,  e.g.,
\cite{Bosshard:2000-10688:PRB} and references therein), and also
for generating multi-color optical solitons. Additionally, it is
expected that the multistep parametric processes and multistep
cascading will find their applications in optical communication
devices, for wavelength shifting and all-optical switching (see
discussions below). Another class of applications of the multistep
phase-matched parametric processes is the construction of optical
parametric oscillators (OPOs) and optical parametric amplifiers
(OPAs) with complimentary phase-matched processes in the same
nonlinear crystal where the main phase-matched parametric process
occurs. In this way,  the OPOs and OPAs may possess additional
coherent tunable outputs (see Sec.~\rsect{OPO} below). On the
basis of the third-harmonic multistep cascading process \{$\omega
+ \omega = 2\omega$; $\omega + 2\omega = 3\omega$ \}, real
advances have been made in the development of the nonlinear
optical systems for division by three. The multistep parametric
interactions governed by several phase-mismatching parameters can
also occur in centrosymmetric nonlinear media (see, e.g.,
\cite{Akhmanov:1975-143:PZETF, Reintjes:1984:NonlinearOptical,
Astinov:2000-853:OL, Crespo:2000-829:OL,
Misoguti:2001-13601:PRL}).

The first proposal for simultaneous phase-matching of two
parametric processes can be found in the pioneering book of
Akhmanov and Khokhlov (\cite{Akhmanov:1964:ProblemyNelineinoi,
Akhmanov:1972:ProblemsNonlinear}) who derived and investigated the
condition for the THG process in a single crystal with the
quadratic nonlinearity through the combined action of the SHG and
SFG parametric processes. For the efficient frequency conversion,
both the parametric processes, i.e. SHG and SFG, should be phase
matched simultaneously. Some earlier experimental attempts to
achieve simultaneously two phase-matching conditions for the SHG
and SFG processes were not very successful
(\cite{Sukhorukov:1970-266:IVR}, \cite{Orlov:1972-283:RAR}).
Recent development of novel techniques for efficient phase
matching, including the QPM technique, make many of such multistep
processes readily possible.

The multistep parametric processes investigated so far can be
divided into several groups, as shown in Fig.~\rpict{cascading}
and Table~\rtab{multistep}. These processes include:
third-harmonic multistep cascading; fourth-harmonic multistep
cascading; two-color multistep cascading; FWM and DFWM in
$\chi^{(2)}$ media , OPO and OPA multistep cascading. We will
review these groups separately.

Table~\rtab{multistep} does not provide a complete list of all
possible types of the multistep parametric processes. Here, we
pick up only several examples for each type of the multistep
process. As follows from Table~\rtab{multistep}, some of the
parametric processes simulate some known higher-order processes
but occur through several steps. However, some other multistep
parametric processes have no such simpler analogues. Below in the
review paper, we discuss some of those processes in more details.
For example, the THG multistep cascading is considered in
Sec.~\rsect{THG}, while the two-color multistep cascading is
discussed in Sec.~\rsect{2color}. The last three lines in
Table~\rtab{multistep} are the examples of the multistep cascading
processes in optical parametric oscillators and amplifiers, and
they will be discussed in Sec.~\rsect{OPO}.

To illustrate the physics responsible for the use of the
terminology "multistep interaction" or "multistep cascading", we
just point out the simultaneous action of SHG and SFG (line 1 in
Table~\rtab{multistep}); this three-wave interaction involves five
simpler parametric sub-processes in a quadratic medium. In order
to describe, for example, the nonlinear phase shift of the
fundamental wave accumulated in this interaction, we should
consider the following chain of parametric interactions: SHG
($\omega + \omega = 2\omega$), SFG ($\omega + 2\omega = 3\omega$),
DFM ($3\omega - \omega = 2 \omega$), and, finally, another DFM
($2\omega - \omega = \omega$).

\subsection{Third-harmonic multistep processes} \lsect{THG}

The multistep parametric interaction involving the THG process is
one of the most extensively studied multistep cascading schemes
(see Fig.~\rpict{THGscheme}). The two simpler parametric
interactions are the SHG process, $\omega + \omega = 2\omega$, and
the SFG process, $\omega + 2\omega = 3\omega$. Each of these
sub-processes is characterized by an independent phase matching
parameter, namely $\Delta k_{\rm SHG} = k_{2\omega} -
2k_{1\omega}$ and $\Delta k_{\rm SFG} = k_{3\omega} -
k_{2\omega}-k_{\omega}$.

\pict{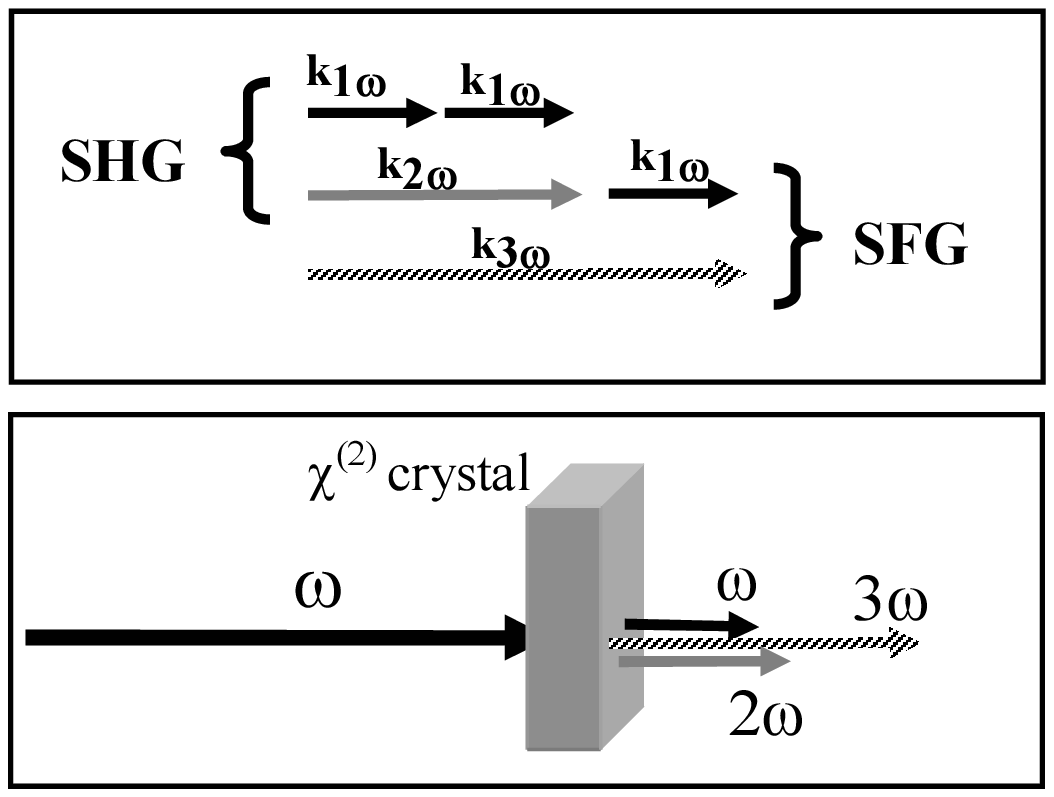}{THGscheme}{Schematic of the THG multistep
cascading. }

This kind of the multistep cascading appears in many  schemes of
parametric interactions in nonlinear optics, including: (i)
efficient generation of the third harmonic in a single quadratic
crystal; (ii) measurement of unknown $\chi^{(3)}$ tensor
components using known $\chi^{(2)}$ components of the crystal as a
reference; (iii) accumulation of a large nonlinear phase shift by
the fundamental wave; (iv) propagation of multi-color solitons;
(v) frequency division; (vi) generation of entangled and squeezed
photon states, etc.

\subsubsection{Efficient generation of a third-harmonic wave} \lsect{THGeffic}

First we consider the process of efficient generation of a
third-harmonic wave. In the approximation of plane waves, this
process is described by the following system of coupled equations
for the slowly varying amplitudes $A_{1}$, $A_{2}$, and $A_{3}$ of
the fundamental, second-, and third-harmonic waves, respectively:
\begin{equation} \leqt{THGeqns}
   \begin{array}{l} {\displaystyle
      \frac{d A_{1}}{d z}
      = - i \sigma_{1} A_{2} A_{1}^{\ast} e^{-i\Delta k_{\rm SHG}z}
      -   i \sigma_{3} A_{3} A_{2}^{\ast} e^{-i\Delta k_{\rm SFG}z} ,
   } \\*[9pt] {\displaystyle
      \frac{d A_{2}}{d z}
      = - i \sigma_{2} A_{1}^{2} e^{i\Delta k_{\rm SHG}z}
      -   i \sigma_{4} A_{3} A_{1}^{\ast} e^{-i\Delta k_{\rm SFG}z} ,
   } \\*[9pt] {\displaystyle
      \frac{d A_{3}}{d z}
      = - i \sigma_{5} A_{2} A_{1} e^{i\Delta k_{\rm SFG}z}
      -   i \gamma A_{1}^{3} e^{i\Delta k_{\rm THG}z} ,
   } \end{array}
\end{equation}
where $\Delta k_{\rm SHG} = k_{2} - 2k_{1}+ G_{p}$ and $\Delta
k_{\rm SFG} = k_{3} - k_{2}-k_{1}+ G_{q}$, $\sigma_{1,2}=
(2\pi/\lambda_{1}n_{1,2})d_{\rm eff,I}$ and
$\sigma_{j}=(\omega_{j-2}/\omega_{1})
(2\pi/\lambda_{1}n_{j-2})d_{\rm eff,II}$ (where $j=3,4,5$). Here,
the parameters $d_{\rm eff,I}$ and $d_{\rm eff,II}$ are the
effective quadratic nonlinearities corresponding to the two steps
of the multistep parametric process; its values depend on the
crystal orientation (see, e.g.,
\cite{Dmitriev:1999:HandbookNonlinear}) and the method of phase
matching (PM). The parameter $\gamma$ is found as $\gamma=
(3\pi/4\lambda_{1}n_{3})\chi^{(3)}_{\rm eff}$ (for calculation of
$\chi^{(3)}_{\rm eff}$ in crystals see \cite{Yang:1995-6130:AOP}).
The complementary wave vectors $G_p$ and $G_q$ are two vectors of
the QPM structure that can be used for achieving double phase
matching. A solution of this system (see, e.g.,
\cite{Kim:2002-33831:PRA, Qin:2003-73:JOSB}) gives maximum for THG
in several different situations, when (i) $\Delta k_{\rm SHG}
\longrightarrow 0$; (ii) $\Delta k_{\rm SFG} \longrightarrow 0$;
(iii) $\Delta k_{\rm THG} = \Delta k_{\rm SHG} + \Delta k_{\rm
SFG} = k_{3} - 3 k_{1} \longrightarrow 0$; and (iv) simultaneously
$\Delta k_{\rm SHG} \longrightarrow 0$
      and $\Delta k_{\rm SFG} \longrightarrow 0$.
The latter condition, for which both SHG and SFG steps should be
simultaneously phase matched, corresponds to the highest
efficiency for THG. The intensity of the third-harmonic (TH) wave
is proportional to the fourth power of the crystal length,
\begin{equation} \leqt{THGcasc}
   |A_{3}(L)|^2 = \frac{1}{4} \sigma_{2}^2\sigma_{5}^2 |A_{1}|^6 L^4 .
\end{equation}
This expression should be compared with the analogous result for
the centrosymmetric media:
\begin{equation} \leqt{THGdirect}
   |A_{3}(L)|^2 = \gamma^2|A_{1}|^6 L^2 .
\end{equation}
The advantage of the single-crystal phase-matched cascaded THG
becomes clear if we note that, in average, $(\sigma_{2}
\sigma_{5}L)^2$ is in $ 10^4-10^6$ times larger than $\gamma^2$,
even for the sample length as small as 1 mm.

\TableShiftLeft
\begin{small}
\begin{longtable}{>{\raggedright}p{1.8cm}c>{\raggedright}p{3.5cm}>{\raggedright}p{3cm}c>{\raggedright}p{1.3cm}p{1.4cm}p{5mm}}
\caption{ \ltab{THG}
Experimental results on cascaded THG processes} \\
\hline
  \centering Nonlinear & $\lambda_\omega$ & Phase-matched steps &\centering
  Phase matching  & L & \centering Regime& $\eta$,\%& \hspace {1.2 cm} Refs. \\
  \centering crystal & [$\mu m$] &  & \centering method & [cm] & & & \\
\hline
\endfirsthead

\hline
  \centering Nonlinear & $\lambda_\omega$ &
  Phase-matched steps &\centering Phase matching  & L & \centering Regime& $\eta$,\%& \hspace {1.2 cm} Refs. \\
  \centering crystal & [$\mu m$] &  & \centering method & [cm] & & & \\
\hline
\endhead

 LiTaO${_3}$ & 1.44 & $\Delta k_{\rm SHG}\approx0$; $\Delta
 k_{\rm SFG}\approx0$&
 QPOS &  1.5 & Pulsed (8~ns) &  27\% & a
\\

LiTaO${_3}$ & 1.570  & $\Delta k_{\rm SHG}\approx0$; $\Delta
 k_{\rm SFG}\approx0$&  QPOS  & 0.8 & Pulsed (8~ns) &  23\% &
 b, c
\\

LiTaO${_3}$ & 1.342 & $\Delta k_{\rm SHG}\approx0$; $\Delta
 k_{\rm SFG}\approx0$ &  SHG-QPM 1st-ord. SFG-QPM 3rd-ord. &  1.8 & quasi-cw (30~ns)  &  19.3 \% & d
\\

 LiTaO${_3}$ & 1.342 & $\Delta k_{\rm SHG}\approx0$; $\Delta
 k_{\rm SFG}\approx0$ &  SHG-QPM 1st-ord. SFG-QPM 3rd-ord.  & 1.2 & Pulsed (90~ns) &  10.2\% &  e\\

$\beta$-BBO & 1.055 &${\Delta k_{\rm THG}\approx0}$; ${\Delta
 k_{\rm SFG}=-\Delta k_{\rm SHG}\neq0}$ &  BPM   & 0.3 & Pulsed (350 fs) & 6\% & f, g \\

 LiTaO${_3}$ & 1.442 &$\Delta k_{SHG}\approx0$; $\Delta
 k_{SFG}\approx0$   &   QPOS & 0.6 & Pulsed (10~ns) &  5.8\%&  h\\

KTP& 1.8 &$\Delta k_{\rm SHG}\neq0$; $\Delta
 k_{\rm SFG}\approx0$   &   BPM &  & Pulsed (35~ps) &  5\%&  i\\

 LiTaO${_3}$ & 1.064  &  $\Delta k_{\rm SHG}\approx0$; $\Delta
 k_{\rm SFG}\approx0$ & PR-QPM & 1.2 & quasi-cw
(150~ns)  &  2.8\% & j
\\

 KTP & 1.618 &  ${\Delta k_{\rm THG}\approx0}$; ${\Delta
 k_{\rm SFG}=-\Delta k_{\rm SHG}\neq0}$ &
BPM &0.11  &  Pulsed (22~ps) & 2.4\%  &  k, l
\\

 SBN & 1.728  & $\Delta k_{\rm SHG}\approx0$; $\Delta k_{\rm SFG}\approx0$&
QPOS  &  0.75  &  Pulsed (15~ps) & 1.6\%  &  m
\\

  d-LAP & 1.055 &${\Delta k_{\rm THG}\approx0}$; ${\Delta
 k_{SFG}=-\Delta k_{\rm SHG}\neq0}$ &  BPM    & 0.1 & Pulsed (350 fs) & 1.2\% & f, g\\

   $\beta$-BBO & 1.05 & ${\Delta k_{\rm THG}\approx0}$; ${\Delta
 k_{\rm SFG}=-\Delta k_{\rm SHG}\neq0}$&
BPM &0.72  &  Pulsed (5~ps) &  0.8\% &   n

\\

KTP waveguide & 1.234 &$\Delta k_{SHG}\approx0$; $\Delta
 k_{SFG}\approx0$    &   QPM &  0.26 &  Pulsed (9~ns) &  0.4\%\footnote{backward THG} &  o \\

 KTP & 1.32
 &  $\Delta k_{\rm SHG}\neq0$; $\Delta k_{\rm SFG}\approx0$&
BPM & 0.47  &  pulsed (200 fs) & 0.17\% &  p,q
\\

LiNbO${_3}$ waveguide & 1.619 &  $\Delta k_{\rm SHG}\approx0$;
$\Delta
 k_{\rm SFG}\approx0$  &
SHG- QPM 1st-ord. SFG- QPM 3rd-ord.   & &  pulsed (7~ps)  & 0.055\%  & r
\\

LiNbO${_3}$& 1.534 &$\Delta k_{\rm SHG}\neq0$; $\Delta
 k_{\rm SFG}\approx0$   &   QPM &  0.6 & Pulsed (9~ns) & 0.016\%\footnote{backward THG} & s\\

KTP waveguide & 1.65 &  $\Delta k_{\rm SHG}\approx0$; $\Delta
 k_{\rm SFG}\approx0$  &
SHG- QPM 1st-ord. SFG- QPM 3rd-ord.  & 0.35 &  pulsed (6~ps)  & 0.011\%  & t \\

 $\beta$-BBO & 1.053 & ${\Delta k_{\rm THG}\approx0}$; ${\Delta
 k_{\rm SFG}=-\Delta k_{\rm SHG}\neq0}$&
BPM &0.7  &  Pulsed (45~ps) &  0.007\% &   u\\

  LiNbO${_3}$ 2D-NPC &  1.536  & $\Delta k_{\rm SHG}\approx0$; $\Delta
 k_{\rm SFG}\approx0$ &
2D-QPM & 1  & pulsed (5~ns) &  ~0.01\%\footnote{Type II cascaded
THG} &v
\\

  LiNbO${_3}$ & 3.561  & $\Delta k_{\rm SHG}\approx0$; $\Delta
 k_{\rm SFG}\approx0$ &
QPM& 2 &  CW & $10^{-4}$\%
[W$^{-2}$]  & w
\\

 KTP &1.55 & $\Delta k_{\rm SHG}\approx0$; $\Delta
 k_{\rm SFG}\approx0$  &
QPOS  &  1  & CW & 3.$10^{-5}$\%
[W.cm$^{-2}$] &  x
\\

 Y:LiNbO${_3}$ & 1.064 &$\Delta k_{\rm SHG}\approx0$; $\Delta
 k_{\rm SFG}\approx0$   &SHG-QPM 9th-ord. SFG-QPM 33rd-ord.   &  0.5&  pulsed (100~ns) &  $10^{-5}$\% & y \\

\end{longtable}
~\\
References:\\
$^a$ \cite{Zhang:2001-899:OL}\\
$^b$ \cite{Qin:1998-6911:JAP}\\
$^c$ \cite{Zhu:1997-843:SCI}\\
$^d$ \cite{He:2002-944:CHIL}\\
$^e$ \cite{Luo:2001-3006:APL}\\
$^f$ \cite{Banks:1999-4:OL}\\
$^g$ \cite{Banks:2002-102:JOSB}\\
$^h$ \cite{Chen:2001-577:APL}\\
$^i$ \cite{Takagi:2000-865:JLU}\\
$^j$ \cite{Liu:2002-1676:JOSB}\\
$^k$ \cite{Feve:2000-1373:OL}\\
$^l$ \cite{Boulanger:1999-475:JPB}\\
$^m$ \cite{Zhu:1998-432:APL}\\
$^n$ \cite{Qiu:1988-225:APB}\\
$^o$ \cite{Gu:1999-127:OL}\\
$^p$ \cite{Mu:2000-117:OL}\\
$^q$ \cite{Ding:2000-21:JNOM}\\
$^r$ \cite{Baldi:1995-1350:ELL}\\
$^s$ \cite{Gu:1998-323:OC}\\
$^t$ \cite{Sundheimer:1994-975:ELL}\\
$^u$ \cite{Tomov:1992-4172:AOP}\\
$^v$ \cite{Broderick:2002-2263:JOSB}\\
$^w$ \cite{Pfister:1997-1211:OL}\\
$^x$ \cite{Fradkin-Kashi:2002-23903:PRL}\\
$^y$ \cite{Volkov:1998-1046:KE}\\
Abbreviations:\\
QPOS~-- quasi periodical optical superlattices; BPM~--  birefringence phase matching;
PR-QPM~-- phase reversed QPM structure (see \cite{Chou:1999-1157:OL}); 2D-NPC~-- two-dimensional
photonic crystals; 2D-QPM~-- two-dimensional QPM structure.\\
~\\
\end{small}

Introducing normalized efficiency (measured in units of
$W^{-1}cm^{-2}$) for the first and second steps in separate
crystals as $\eta_{0,1}$ and $\eta_{0,2}$, respectively, we obtain
\begin{equation} \leqt{EFFthg}
   \eta_{3\omega} = \frac{1}{4} \eta_{0,1} \eta_{0,2} P_{1}^2 L^4
\end{equation}
The results for THG in a single quadratic crystal under the
condition of double phase matching were reported for the
efficiency exceeding 20\%, as shown in Table~\rtab{THG}.
Figure~\rpict{THGexper} shows the phase matching curves in the
experiment (\cite{Zhang:2001-899:OL}) where 27\% THG efficiency
has been achieved. The two PM curves are not perfectly overlapped.
We may expect that with an improvement of the superlattice
structure the achieved efficiencies will be higher. Numerical
solution of the system~\reqt{THGeqns} shows that, in general, the
third-harmonic output has an oscillating behavior as a function of
the length or input power, and generally it does not reach the
efficiency of 100\%. However, as shown in
\cite{Egorov:1998-2345:IANF, Chirkin:2000-847:KE} and
\cite{Zhang:2000-436:OL},  the total conversion of the fundamental
wave into the third-harmonic wave is possible if the ratio
$\sigma_{2}/\sigma_{5}$ is optimized.

\pict{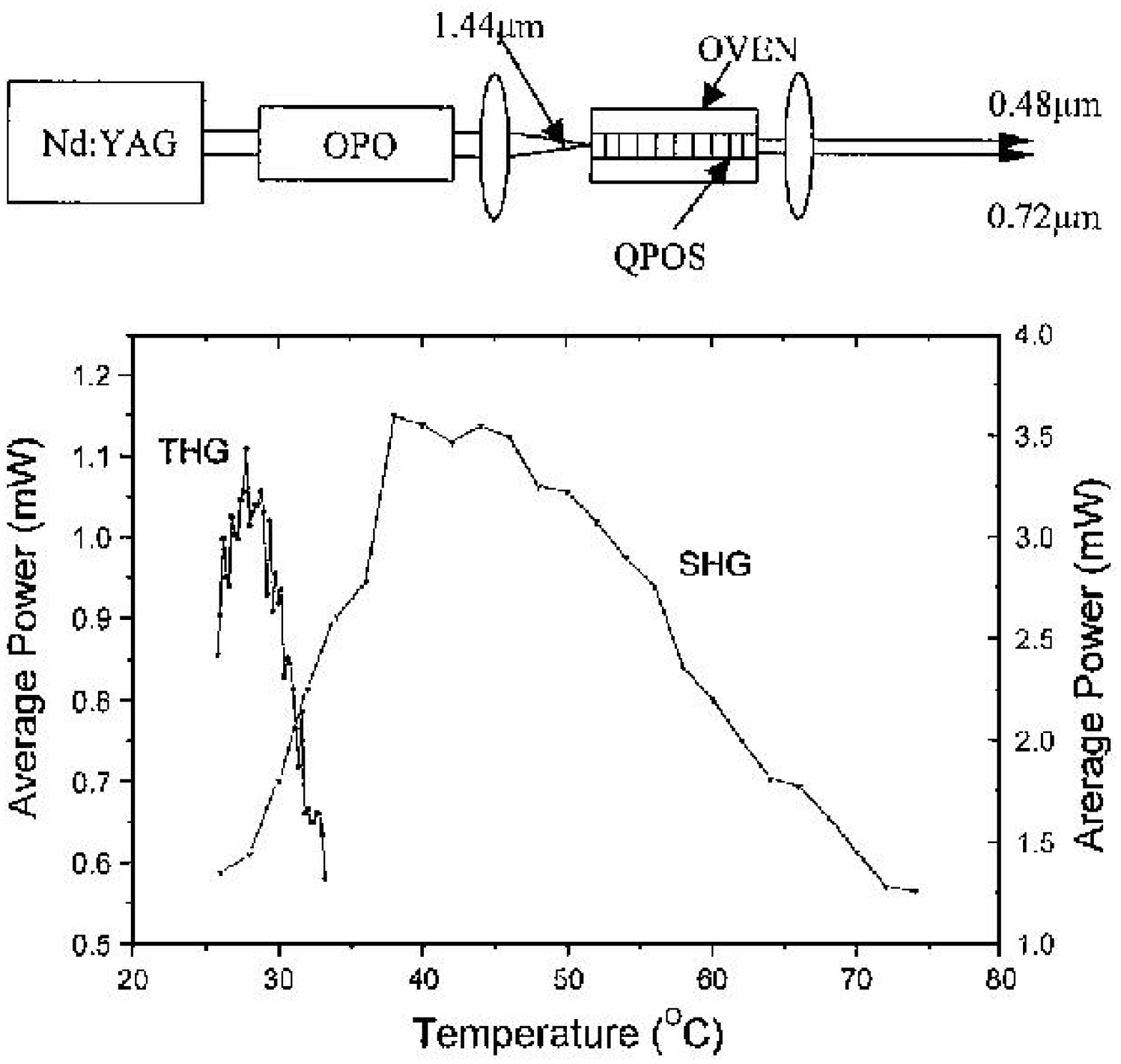}{THGexper}{ (above) Schematic of the experimental
setup. (below) Average powers of the second harmonic (right scale)
and third harmonic (left scale) vs. temperature. Average power of
the fundamental wave is 4.8 mW (\cite{Zhang:2001-899:OL}). }

In other cases, where only one phase-matched condition is
satisfied, the THG conversion efficiency is not so large, and it
should be compared to that of the direct THG process in a cubic
medium. The  efficiency of a cascaded THG process is inversely
proportional to the square of the wave vector mismatch of the
unmatched process. It is important to note that in such cases,
even when only one of the phase-matched parameters $\Delta k_{\rm
SHG}$ or $\Delta k_{\rm SFG}$ is close to zero (such conditions
can be easily realized in the birefringent phase matching in bulk
quadratic media or in an uniform QPM structure) the THG process
has the behavior of the phase matched process with the
characteristic dependence $[\sin(x)/x]^2$ for the TH intensity vs.
tuning and a cubic dependence on the input intensity. The
dependence of the sample thickness is quadratic but not periodic
function as for the totally non phase matched processes.

Situation (iii), that is, tuning where $\Delta k_{\rm SHG}+\Delta k_{\rm SFG} = 0$, also corresponds to the phase-matching condition for direct~THG where the fundamental wave is converted directly into a third-harmonic wave. In such a case, both the cascade process and direct process $k_3=3k_1$ contribute to the third harmonic wave. However, a relative contribution of the two processes could be different. In some cases, the direct THG process is stronger (see, e.g., \cite{Feve:2000-1373:OL}), in
other cases the cascading THG process is dominant (see, e.g.,
\cite{Banks:2002-102:JOSB, Bosshard:2000-10688:PRB}).

We wonder why the whole process is phase matched when both steps
are mismatched? The reason is that in the case (iii) we have the
situation similar to that of the quasi-phase-matching effect
(see, e.g., \cite{Reintjes:1984:NonlinearOptical, Banks:1999-4:OL,
Durfee:2002-822:JOSB}). Indeed, if the first step is mismatched,
inside a nonlinear medium it generates a periodically modulated
polarization at the frequency $2\omega$ ($P(2\omega)$) and the
modulation period is exactly a mismatch of the second step. 
Thus, in the regions where the phase of the polarization P(2omega) is
reversed, we have the minimum generation of second- and third-harmonic
waves, and the  generated TH components interfere constructively with the
propagating third-harmonic wave along the length of the crystal.
This leads to a quadratic dependence
of the THG efficiency on the crystal length.

For the cascaded THG processes, the optimal focusing  is an
important issue when the goal is the maximum conversion
efficiency. If only one of the steps is phase matched, the optimal
focusing is in the input face when SFG is phase matched, or at the
output face when SHG is phase matched
(\cite{Rostovtseva:1977-56:OC, Rostovtseva:1980-1081:KE}). If both
the steps are phase matched, the optimum focusing position is in
the center of the nonlinear medium (\cite{Ivanov:2002-397:OC}).

In Table~\rtab{THG}, we summarize the experimental results
obtained for the efficient single-crystal THG processes.
Conditions at which the third-harmonic wave included in double
phase-matched interaction can be transformed with 100\%
efficiency into the $2\omega$ wave or $\omega$ wave were found by
\cite{Komissarova:1993-1025:KE, Komissarova:1997-2298:IANF}; and
\cite{Egorov:1998-2345:IANF}. \cite{Volkov:1998-101:KE} show that
the same parametric interaction can be used for 100\% conversion
from $2\omega$ wave into $3\omega$ wave. Considering
the  situations with nonzero SH and TH boundary conditions it has to be
taken into account that, as shown by \cite{Alekseev:2002-206:PZETF}, the spatial evolution of three light waves participating simultaneously in SHG and SFM under the conditions of QPM double phase matching becomes chaotic at large propagation distances for many values of the complex input wave amplitudes. Thus, the possibility of transition to chaos exists in the application of an additional pump at frequency $2\omega$ in order to increase the efficiency of THG [\cite{Egorov:1998-2345:IANF}].
As shown in \cite{Longhi:2001-57:EPD} for the multistep cascading process
\{$\omega + \omega = 2\omega$; $\omega + 2\omega = 3\omega$ \} in
a cavity, the formation of spatial patterns is possible. Another
application of the multistep cascading interaction and, in
particular, the THG multistep cascading is the generation of the
entangled and squeezed quantum states (see Sec.~\rsect{OPO} below). The conditions for the simultaneous phase matching of both SHG and SFG processes were considered by \cite{Pfister:1997-1211:OL,
Grechin:2001-933:KE}, for uniform QPM structures; by
\cite{Zhu:1999-1093:OQE}, for quasi-periodic optical
superlattices; by \cite{Fradkin-Kashi:1999-1649:IQE} and
\cite{Fradkin-Kashi:2002-23903:PRL}, for the generalized Fibonacci
structures; and by \cite{Saltiel:2000-1204:OL}, for the
two-dimensional nonlinear photonic crystals. This topic will be
discussed below in Sec.~\rsect{matching}.

\subsubsection{Nonlinear phase shift in multistep cascading} \lsect{PhaseShift}

The multistep parametric process, that combines two phase-matched
interaction SHG $(\omega+\omega=2\omega)$ and SFG $(2\omega +
\omega = 3\omega)$, is described by the system of
equations~\reqt{THGeqns}, and it can be used for all-optical
processing and formation of parametric solitons. Indeed, the
nonlinear phase shift (NPS) accumulated by the fundamental wave in
the multistep cascading process is in several times larger than
that in the standard cascading interaction that involves only one
phase matched process (\cite{Koynov:1998-96:OC}).

\pict{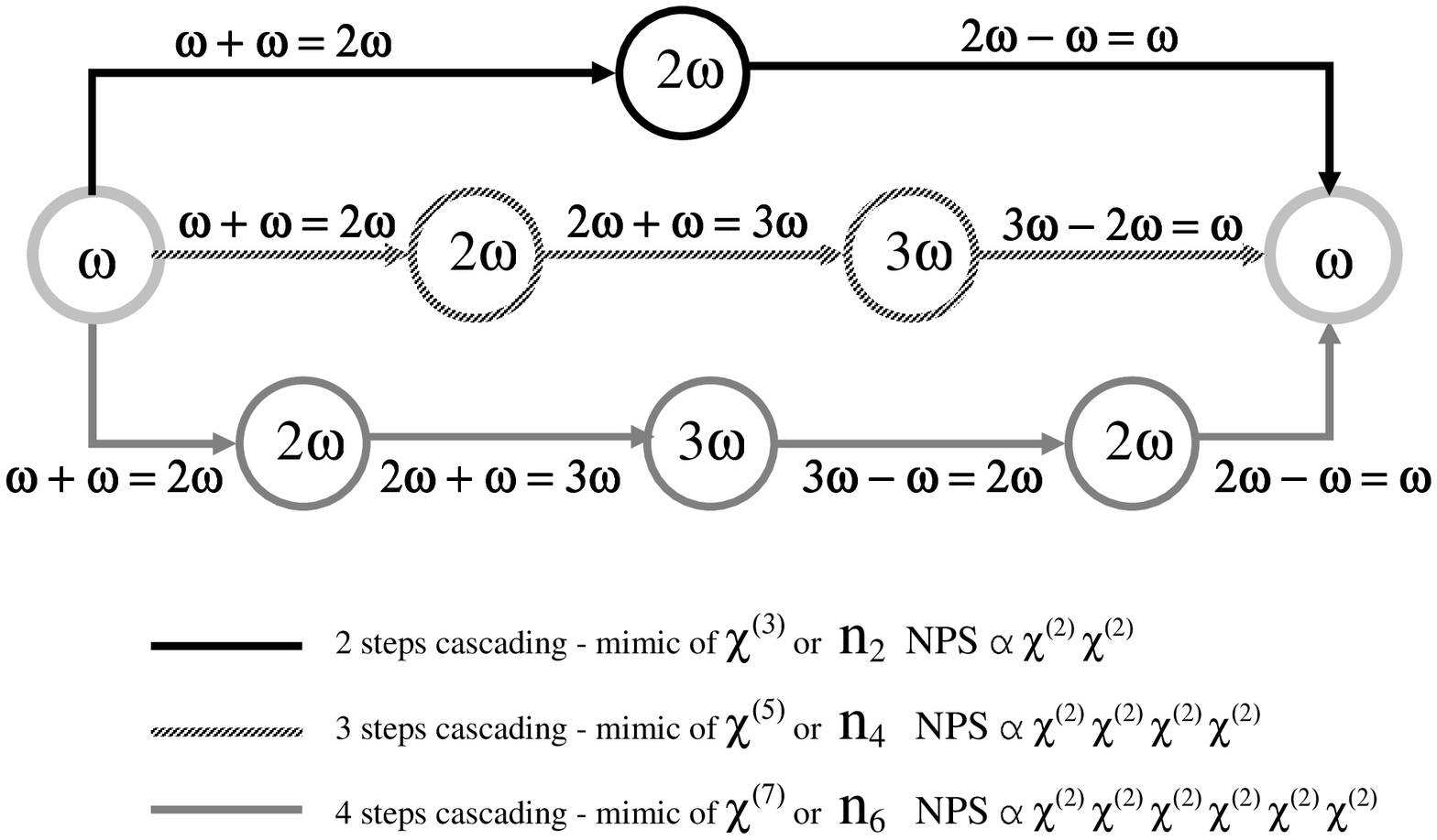}{THGphase}{Schematic of the possible channels for
the phase modulation of the fundamental wave
(\cite{Koynov:1998-96:OC}). }

\pict{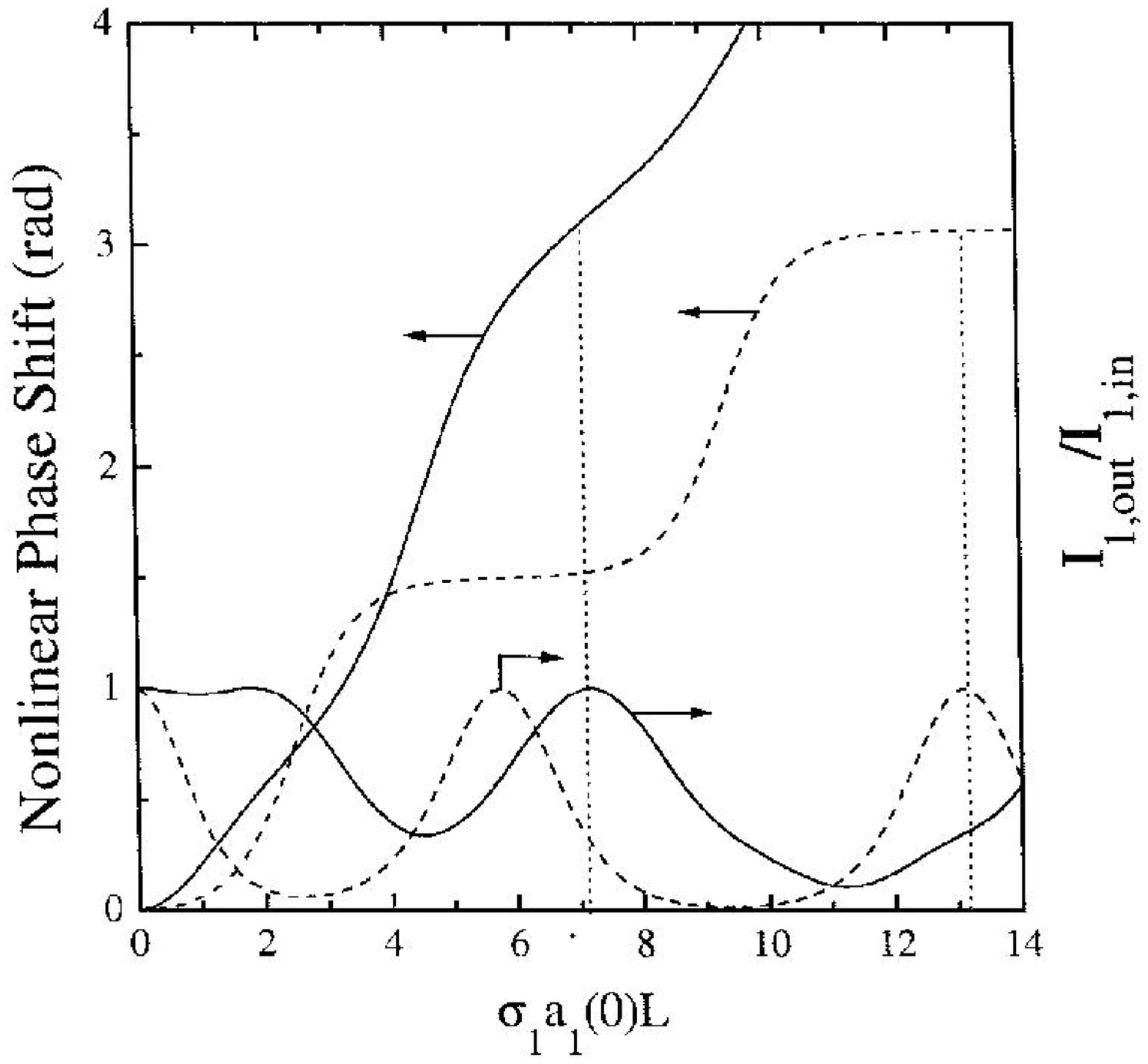}{THGphaseCompar}{Nonlinear phase shift and
depletion of the fundamental wave as a function of its normalized
input amplitude: solid line~-- multistep cascading, dash line~--
type I SHG case (\cite{Koynov:1998-96:OC}). }

To illustrate this result, we consider the fundamental wave with
the frequency $\omega$ entering a second-order nonlinear media
under appropriate phase-matching conditions. As the first step,
the wave with frequency $2\omega$ is generated via the type I SHG
process, and then, as the second step, the third-harmonic wave is
generated via the SFG process $(2\omega+\omega=3\omega)$. Both the
processes, SHG and SFG, are assumed to be nearly phase matched.
The generated second- and third-harmonic waves are down-converted
to the fundamental wave $\omega$ via the processes
$(2\omega-\omega)$; $(3\omega-2\omega)$, and
$(3\omega-\omega,2\omega-\omega)$,  contributing all to the
nonlinear phase shift that the fundamental wave collects. As
follows from numerical calculations, the total NPS is a result of
the simultaneous action of two-, three-, and four-step
$\chi^{(2)}$ cascading, and it can exceed the value of $\pi$ for
relatively low input intensities. The possible channels for the
phase modulation and NPS of the fundamental wave are shown in
Fig.~\rpict{THGphase}. The interpretation of the analytical result
obtained in the fixed-intensity approximation
(\cite{Koynov:1998-96:OC}) shows that the effective cascaded
fifth-order and higher-order nonlinearities are involved into the
accumulation of a total NPS, and the signs of the contributions
from different processes can be controlled by a small change of
the phase matching conditions. Schematically, the role of the
multistep cascading with three and four steps can be interpreted
as equivalent of a contribution from the higher-order nonlinear
corrections $n_4$ and $n_6$ to the refractive index, and they can
be linked to the cascaded $\chi^{(5)}$ and $\chi^{(7)}$ processes,
respectively. Indeed, for a relatively low intensity of the
fundamental wave, the refractive index can be written in the form
of expansion,
\begin{equation} \leqt{n-series}
   n(E) = n_0 + n_2 E^2 + n_4 E^4 + n_6 E^6 + \ldots.
\end{equation}

The advantage of the multistep cascading for accumulating large
nonlinear phase shift over the conventional two-step cascading is
illustrated in Fig.~\rpict{THGphaseCompar} obtained by numerical
integration of the system~\reqt{THGeqns}. This result is also
useful for studying the multi-color solitons supported by this
type of multistep parametric interaction
(\cite{Kivshar:1999-759:OL, Huang:2001-418:CHI}). We note again
the THG multistep cascading can be efficient only if the two
processes can be simultaneously phase-matched. Due to dispersion
in a bulk material, generally it is impossible to achieve double
phase matching. However, due to the recent progress in the design
of the QPM structures, the double phase matching can be achieved
in nonlinear photonic materials (\cite{Zhang:2001-899:OL,
Luo:2001-3006:APL, Fradkin-Kashi:2002-23903:PRL}), although there
exist a number of technical problems to be solved.

\subsection{Wavelength conversion}
         \lsect{conversion}

\pict{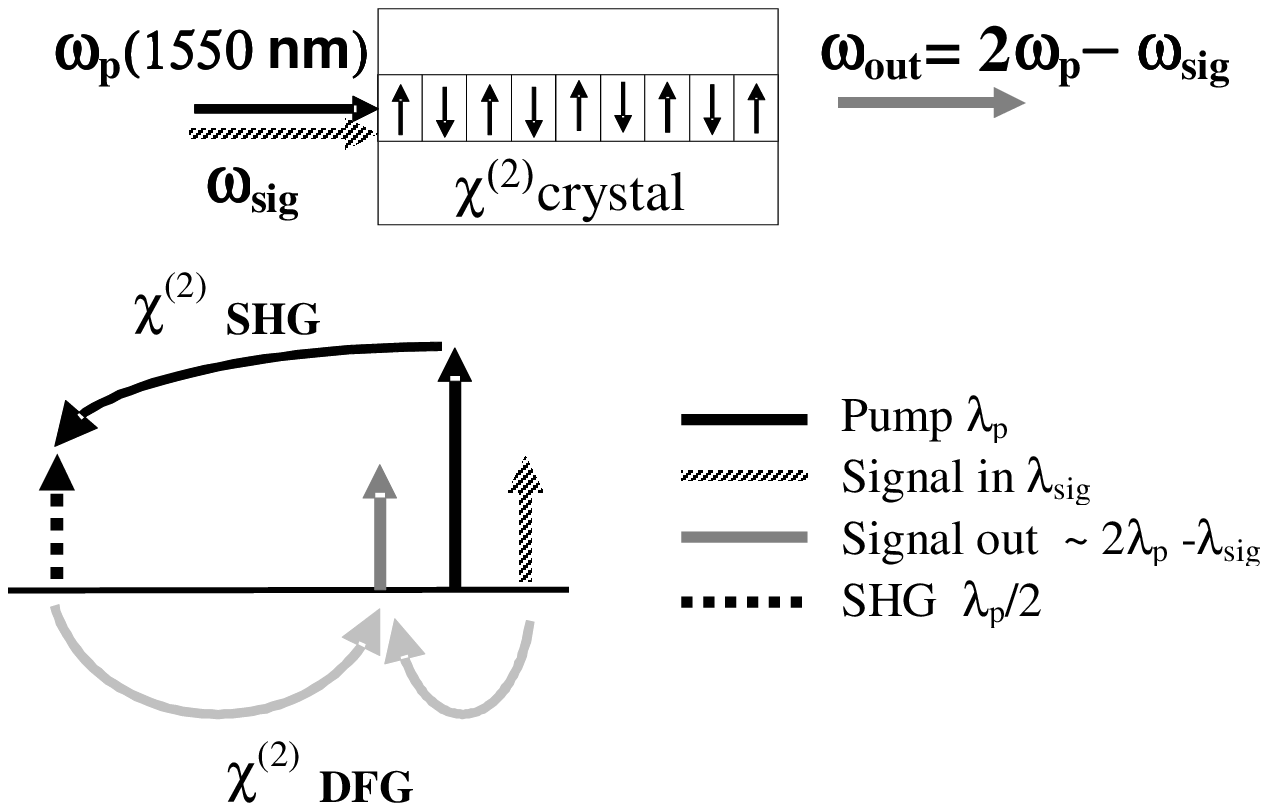}{conversion}{Wavelength conversion of optical
communication channels in a periodically-poled nonlinear crystal
using SHG/DFG multistep cascading. }

So far we discussed only the multistep cascading process that
allows to generate the third harmonic by combining the SHG and SFG
phase-matched parametric interactions in a single crystal. Let now
consider another single-crystal multistep process that combines
SHG and DFM and mimic in this way the four-wave mixing (FWM)
process with two input waves $\omega_p$ and $\omega_{\rm sig}$,
resulting in the generation of a signal at $\omega_{\rm out} =
2\omega_p - \omega_{\rm sig}$. The idea of this type multistep
cascading is illustrated in Fig.~\rpict{conversion}. Usually, the
difference $ \lambda_{\rm sig}- \lambda_p$ is smaller than 50 nm
(with $\lambda_p \sim 1550 nm$) that leads to the result that when
SHG process is phase matched the DFM process is very close to
exact phase matching. Here we have the situation of double phase
matched multistep cascading. This allows a very efficient
conversion from $\lambda_{\rm sig}$ to $\lambda_{\rm
sig}-\lambda_p/2$ which is in $10^4 \div 10^5$ times larger than
what can be obtained with the direct FWM process. To be more
specific, let us consider the parametric interaction shown in
Fig.~\rpict{conversion} which is described by the system of
parametrically coupled equations,
\begin{equation} \leqt{WCeqns}
   \begin{array}{l} {\displaystyle
      \frac{d A_{p}}{d z}
      = - i \sigma_{1} A_{2} A_{p}^{\ast} e^{-i\Delta k_{\rm SHG}z}  ,
   } \\*[9pt] {\displaystyle
      \frac{d A_{2}}{d z}
      = - i \sigma_{2} A_{p}^{2} e^{i\Delta k_{\rm SHG}z}
      -   i \sigma_{3} A_{\rm sig} A_{\rm out}e^{i\Delta k_{\rm DFG}z} ,
   } \\*[9pt] {\displaystyle
      \frac{d A_{\rm out}}{d z}
      = - i \sigma_{4} A_{2} A_{\rm sig}^{\ast} e^{-i\Delta k_{\rm DFG}z} ,
   } \\*[9pt] {\displaystyle
      \frac{d A_{\rm sig}}{d z}
      = - i \sigma_{5} A_{2} A_{\rm out}^{\ast} e^{-i\Delta k_{\rm DFG}z} ,
   } \end{array}
\end{equation}
where $\Delta k_{\rm SHG} = k_{2} - 2k_{1}+ G_{m}$ and $\Delta
k_{\rm DFG} = k_{2} - k_{\rm sig}-k_{\rm out}$, $\sigma_{1}$ to
$\sigma_{5}$ are the coupling coefficients proportional to the
second-order nonlinearity parameter $d_{eff}$. The vector $G_{m}$
is one of the QPM vectors used for achieving the phase matching
(\cite{Fejer:1992-2631:IQE}) in the QPM structure. If the
birefringence phase matching is used, then $G_{m}=0$. Two
phase-matching parameters, i.e. $\Delta k_{\rm SHG}$ and $\Delta
k_{\rm DFG}$,  are involved into this parametric cascaded
interaction. However, if the signal wavelength is sufficiently
close to that of the pump, i.e. $|\lambda_{\rm sig}-\lambda_p| \ll
\lambda_p$, the tuning curves for the two processes practically
overlap, and we have the situation in which if one of the
parametric processes is phase matched, the other one is also phase
matched. In other words, in this case the signal wavelength is
sufficiently close to that of the pump, and we work under the
conditions of double phase matching. If we neglect the depletion,
the amplitude of the phase-matched SH wave can be found as
$A_2(z)=-i\sigma_2A_p^2z$. Then, the output signal can be written
in the form,
\begin{equation} \leqt{WCampl}
   |A_{\rm out}(L)|^2 \simeq 4 \sigma_2^2 \sigma_4^2 
                    |A^2_p|^2 |A_{\rm sig}|^2
                    \frac{\sin^4 \left(\Delta k_{\rm DFG}L/2 \right)}{
                                                (\Delta k_{\rm DFG})^4} .
\end{equation}

In the limit $\Delta k_{\rm DFG} \longrightarrow 0$, the
efficiency becomes
\begin{equation} \leqt{WCampl2}
   |A_{\rm out}(L)|^2 = \frac{1}{4} \sigma_2^2 \sigma_4^2 
                        |A^2_p|^2 |A_{\rm sig}|^2 L^4 ,
\end{equation}
or
\begin{equation} \leqt{?}
   \eta_{\rm out} = \frac{1}{4} \eta_{0}^2 P_{p}^2 L^4 ,
\end{equation}
where $\eta_{0}$ is the normalized efficiency measured in
$W^{-1}cm^{-2}$; it depends on the overlap integral between the
interacting modes and the effective nonlinearity $d_{\rm eff}$,
and it is the same normalized value that describes the efficiency
of the first step, the SHG process (\cite{Chou:1999-653:IPTL}).
From Eq.~\reqt{WCampl} it follows that the output signal
is a linear function of the input signal---the property important
for communications.  Also, the efficiency is proportional to the
fourth power of the crystal length $L$. Detailed theoretical
description of this cascading process can be found in
\cite{Gallo:1997-1020:APL,Gallo:1999-741:JOSB,Chen:2002-675:ISQE}.

For the DFG process, the width of the phase-matching curve depends
on the type of the crystal, its length, and the phase-matching
method. Several proposal for increasing the phase-matching region have
been suggested, including the use of the phase-reversed QPM
structures (\cite{Chou:1999-1157:OL}); the pump deviation from the
exact phase matching (\cite{Chou:1999-978:ELL}); the periodically
chirped (phase modulated) QPM structures
(\cite{Asobe:2003-558:OL}); the phase shifting domain
(\cite{Liu:2003-239:OC}). In particular, the paper
\cite{Gao:2004-557:IPTL} reports that the use of
sinusoidally chirped QPM superlattices provides broader bandwidth and more
flat response compared to homogeneous and segmented QPM structures.

Novel cascaded $\chi^{(2)}$ wavelength conversion schemes are
based on the SFG and DFG processes and the use of two pump beams,
as proposed and demonstrated in \cite{Xu:2004-292:OL, Chen:2004-256:IQE}. The conversion efficiency is enhanced by 6 dB, as compared with the
conventional cascaded SHG+DFG wavelength conversion configuration.
The cascading steps in this SFG+DFG wavelength conversion method
are $\omega_{\rm SF} = \omega_{p1} + \omega_{p2}$, and
$\omega_{\rm out} = \omega_{\rm SF} - \omega_{\rm sig}$, so that
the effective third-order interaction is the totally nondegenerate
FWM process: $\omega_{\rm out} = \omega_{p1} + \omega_{p2} -
\omega_{\rm sig}$.

The second-order cascading wavelength conversion is one of the
best examples of the richness of the multistep cascading
phenomena. In the initial stage of development, this concept was
experimentally demonstrated in the media employing two types of
the phase matching techniques, i.e. the birefringence phase
matching in a bulk crystal (e.g. \cite{Tan:1993-2472:APL,
Banfi:1998-439:OL}) and periodically-poled nonlinear LiNbO$_3$
crystal (\cite{Banfi:1998-439:OL}). Since that, this parametric
process was extensively investigated not only as an interesting
multistep parametric effect that may occur in different nonlinear
media (see Table~\rtab{conversion}), but also as a proposal for
realistic all-optical communication devices
(\cite{Chou:1999-653:IPTL, Chou:2000-82:IPTL,
Kunimatsu:2000-1621:IPTL}). Nowadays, there is a real progress in
the suggesting this device for optical communication industry
(\cite{Cardakli:2002-200:IPTL, Cardakli:2002-1327:IPTL}), because
of clear advantages over all other devices used for the wavelength
shifting. Indeed, one such device can simultaneously shift several
channels. As is shown in Fig.~\rpict{conversion}, the spectra of
the shifted signal is a mirror image of the origin. This feature
can be used to invert the signal chirp for dispersion management
in transmission systems. A successful experimental demonstration
of this property has been reported recently in
\cite{Kunimatsu:2000-1621:IPTL}: a 600-fs pulse transmission over
144 km using midway frequency inversion with this type
second-order cascaded wavelength conversion resulted in a
negligible pulse distortion.

\pict{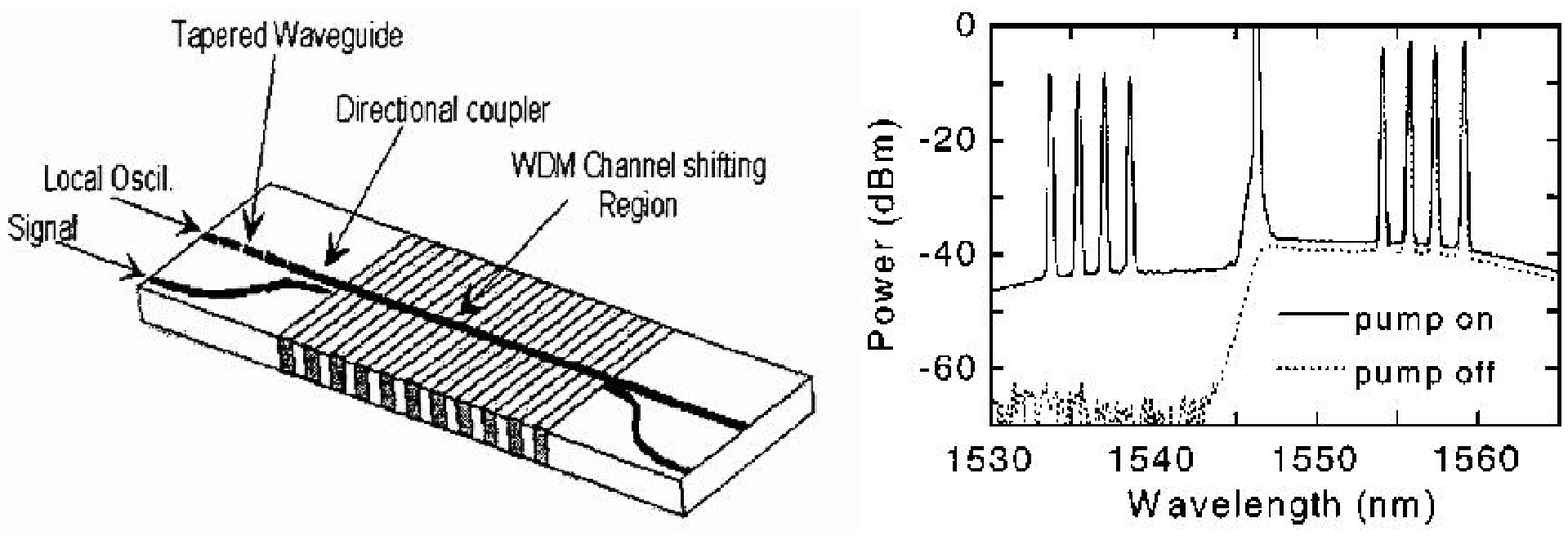}{PPLN}{Right: periodically poled LiNbO$_3$ crystal for
the cascaded wavelength conversion. Left: wavelength conversion
with 1545 nm pump and four inputs in the range 1555-1560 nm
(\cite{Chou:2000-82:IPTL}). }

The waveguide made in the LiNbO$_3$ crystals (see
Fig.~\rpict{PPLN}) are by now the most suitable nonlinear
structures for the cascaded simulation of the FWM wavelength
shifting. The important features of this device is an almost
perfect linear dependence between the input and output signals for
more than 30 dB of the dynamic range, instantaneous memoryless
transparent wavelength shifting that can be used at rates of several
tera-Hertz, and transparent crosstalk-free operation
(\cite{Cardakli:2002-1327:IPTL}). Additionally,
\cite{Couderc:2002-421:OC} demonstrated that the wavelength
conversion multistep cascading system can support parametric
solitons in the waveguiding regime, we discuss these results in more detail in Sec.~\rsect{solitonConversion} below.
Experimental works reporting the
second-order cascaded wavelength conversion are summarized in
Table~\rtab{conversion}.

\setlength{\LTleft}{0cm}
\begin{small}
\begin{longtable}{p{1.5cm}cp{1.5cm}p{2cm}p{1.5cm}l}

\caption{\ltab{conversion}
Experimental results on the multistep SHG and DFM cascading} \\
\hline
Nonlinear crystal & $\lambda_p (\lambda_{s_{max}})$ [$\mu m$]
& L [cm] & Regime & Phase matching method & Refs. \\\hline
\endfirsthead

\hline
Nonlinear crystal & $\lambda_p (\lambda_{s_{max}})$ [$\mu m$]
& L [cm] & Regime & Phase matching method & Refs. \\\hline
\endhead

BBO     & 1.064 (1.090) & 1  &   Pulsed (30ps) &
BPM & a \REMOVE{\cite{Tan:1993-2472:APL}} \\

MBA-NP & 1.064 (1.090)&   0.32  &   Pulsed (30ps) &
BPM & b \REMOVE{\cite{Nitti:1994-263:OC}} \\

LiNbO$_3$ waveguide &   1.533 (1.535) &  1 &  CW\&Pulsed (7ps) &
QPM & c \REMOVE{\cite{Trevino-Palacios:1998-2157:ELL}} \\

LiNbO$_3$ & 1.8 (1.863)& 1.9  &  Pulsed
(20 ps)&
QPM & d \REMOVE{\cite{Banfi:1998-136:APL}} \\

NPP & 1.148 (1.158) &   0.28   & Pulsed
(20 ps)&
NCPM & e \REMOVE{\cite{Banfi:1998-439:OL}} \\

LiNbO$_3$ waveguide  &  1.562 (1.600) &  4 &  CW &
QPM & f \REMOVE{\cite{Chou:1999-653:IPTL}} \\

Ti:LiNbO$_3$ waveguide &1.103 (1.107) &   5.8 &    Pulsed
(20 ps) &
NCPM & g, h \REMOVE{\cite{Cristiani:1999-1198:APL, Banfi:2000-260:JOA}} \\

LiNbO$_3$ waveguide &   1.545 (1.580) &  5  & CW &
QPM & i \REMOVE{\cite{Chou:2000-82:IPTL}} \\

Ti:LiNbO$_3$ waveguide &   1.556 (1.565)  &  7.8 (8.6)  &   CW \&
pulsed (6 ps)&
QPM & j \REMOVE{\cite{Schreiber:2001-501:APB}} \\

LiNbO$_3$ waveguide  & 1.565 (1.585)  & 2 &  CW &
QPM & k \REMOVE{\cite{Cristiani:2001-263:OC}} \\

LiNbO$_3$ waveguide  & 1.542 (1.562)  & 1  & CW \& pulsed &
QPM & l \REMOVE{\cite{Ishizuki:2001-953:OQE}} \\

  LiNbO$_3$
waveguide  &  1.550 (1.560)   & &   CW & QPM & m \\

  LiNbO$_3$ waveguide  & 1.553 (1.565) &  6 &  CW & QPM & n\\

  Ti:LiNbO$_3$
waveguide  &  1.557 (1.553) &  6  &  CW & QPM & o\\

 LiNbO$_3$
waveguide  &  1.532 (1.565) &  2 &  CW & QPM & p\\

  LiNbO$_3$
waveguide  & 1.537 (1570) &   4.5   &  CW & QPM & q\\

  LiNbO$_3$
waveguide  & 1.558..1.568 (1.600)  &  3.4  &   CW & 
PM QPM\footnote{phase modulated QPM} & r\\

  {\footnotesize MgO:LiNbO$_3$
waveguide}  & 1.543(1.573) &   5 &  CW & QPM & s\\

  Ti:LiNbO$_3$ waveguide  & 1.55(1.62) & 3 &  CW & 
  QPM\footnote{three types of QPM grating are compared: homogeneous, segmented gratings and sinusoidally chirped} & t\\

    LiNbO$_3$ waveguide  & 1.545(1.580) & 5 &  CW & QPM & u

\end{longtable}
~\\
References:\\
$^a$ \cite{Tan:1993-2472:APL}\\
$^b$ \cite{Nitti:1994-263:OC}\\
$^c$ \cite{Trevino-Palacios:1998-2157:ELL}\\
$^d$ \cite{Banfi:1998-136:APL}\\
$^e$ \cite{Banfi:1998-439:OL}\\
$^f$ \cite{Chou:1999-653:IPTL}\\
$^g$ \cite{Cristiani:1999-1198:APL}\\
$^h$ \cite{Banfi:2000-260:JOA}\\
$^i$ \cite{Chou:2000-82:IPTL}\\
$^j$ \cite{Schreiber:2001-501:APB}\\
$^k$ \cite{Cristiani:2001-263:OC}\\
$^l$ \cite{Ishizuki:2001-953:OQE}\\
$^m$ \cite{Cardakli:2002-1327:IPTL}\\
$^n$ \cite{Harel:2002-849:JOSB}\\
$^o$ \cite{Cristiani:2002-669:IPTL}\\
$^p$ \cite{Zeng:2003-187:OLT}\\
$^q$ \cite{Zhou:2003-846:JOSB}\\
$^r$ \cite{Asobe:2003-558:OL}\\
$^s$ \cite{Bracken:2003-954:IPTL}\\
$^t$ \cite{Gao:2004-557:IPTL}\\
$^u$ \cite{Sun:2003-125:OC}
~\\
~
\end{small}

\subsection{Two-color multistep cascading} \lsect{2color}

By two-color multistep cascading in a quadratic medium we
understand the multi phase-matched parametric interaction  between
several waves which possess two frequencies (or wavelengths)
only. One of the ways to introduce a parametric process involving
more than one phase-matched interaction with two wavelengths is to
consider a vectorial interaction between the waves with different
polarizations, or degenerate interaction between allowed modes in
a waveguide. We denote two waves at the fundamental frequency (FF)
(at $\lambda=\lambda_{\rm fund}$)  as A and B, and the two waves
of the second harmonic (SH) field (at
$\lambda_{sh}=\lambda_{fund}/2$ ), as S and T. Each pair of the
eigenmodes [(A,B) and (S,T)] can be, for example, two orthogonal
polarization states or two different waveguide modes at the
fundamental and second-harmonic wavelengths, respectively. There
exists a finite number of possible multistep parametric
interactions that can coupled these waves. For example, if we
consider the AA-S\&AB-S cascading, then the multistep cascading is
composed of the following sub-processes. First, the fundamental
wave A generates the SH wave S via the type I SHG process. Then,
by the down-conversion process SA-B, the other fundamental
eigenmode B is generated. At last, the initial FF wave A is
reconstructed by the processes SB-A or AB-S, SA-A. When we deal
with two orthogonal polarizations, the two principal second-order
processes AA-S and AB-S are governed by two different components
(or two different combinations of the components) of the
$\chi^{(2)}$ susceptibility tensor, thus introducing additional
degrees of freedom into the parametric interaction. The
classification of different types of the multistep parametric
interactions has been introduced by \cite{Kivshar:1999-5056:PRE}
and \cite{Saltiel:2000-959:JOSB}.

\begin{small}
\begin{longtable}{p{3.cm}p{8mm}p{2.5cm}p{2cm}p{10mm}p{7mm}}
\caption{\ltab{2colour} Two-color multistep cascading processes}\\

\hline
  Multistep-cascading schemes & No of waves
  & SHG processes & Equivalent cascading schemes
  & WC/SC & Refs. \\\hline
\endfirsthead
\hline
  Multistep-cascading schemes & No of waves
  & SHG processes & Equivalent cascading schemes
  & WC/SC & Refs. \\\hline
\endhead

 AA-S:AB-S &3 &  Type~I \& Type II  & \centering  BB-S:AB-S;  AA-T:AB-T; BB-T:AB-T & SC & a,b,c\\

  AA-S:AB-T &4  & Type~I \& Type II & \centering   BB-S:AB-T;  AA-T:AB-S; BB-T:AB-S & WC & d,e \\

  AA-S:BB-S & 3&   Type~I \& Type I   & \centering  AA-T:BB-T  &    WC & f,g,h,i
 \\

  AA-S:AA-T &3&   Type~I \& Type I &  \centering   BB-S:BB-T   &   SC & h,j
 \\

  AB-S:AB-T & 4  & Type~II \& Type II  &             &SC &
 \\

  { AA-S:BB-S:AB-S}  &  3 &  Type~I \& Type II  &  \centering     { AA-T:BB-T:AB-T } &  &        k,l,m\\

\end{longtable}
~\\
$^a$ \cite{Saltiel:1999-1296:OL}\\
$^b$ \cite{Saltiel:2000-959:JOSB}\\
$^c$ \cite{Petrov:2001-355:OL, Petrov:2002-268:JOSB}\\
$^d$ \cite{DeRossi:1997-53:OQE}\\
$^e$ \cite{Pasiskevicius:2002-1628:OL}\\
$^f$ \cite{Assanto:1994-1720:OL}\\
$^g$ \cite{Kivshar:1999-5056:PRE}\\
$^h$ \cite{Grechin:1999-155:KE, Grechin:2001-929:KE}\\
$^i$ \cite{Grechin:2001-929:KE}\\
$^j$ \cite{TrevinoPalacios:1995-170:APL}\\
$^k$ \cite{Trillo:1994-1825:OL}\\
$^l$ \cite{Towers:1999-1738:OL, Towers:2000-2018:JOSB}\\
$^m$ \cite{Boardman:1997-1899:PRE, Boardman:1998-891:OQE} \\
~\\
~
\end{small}

Different types of the multistep parametric interactions can be
divided into two major groups. The first group is composed by the
parametric interactions with two common waves in both cascading
processes, and the two processes are strongly coupled (SC). For
the other group, both the parametric processes share one common
wave, and these processes are weakly coupled (WC). The same
classification can be applied to other types of the multistep
parametric interactions. In Table~\rtab{2colour}, we present an
updated classification of two-color multistep parametric processes
and the publications where they are analyzed.

The first analysis of the multistep parametric interactions of
this type has been carried out by \cite{Assanto:1994-1720:OL} and
\cite{Trillo:1994-1825:OL}. These authors studied the simultaneous
phase-matching of two SHG processes of the type AA-S and BB-S,
where S denotes the SH wave whereas A and B stand for the the FF
waves polarized in the perpendicular planes. The two orthogonal FF
fields interact through the generated SH wave. All-optical
operations and the polarization switching can be performed on the
base of this scheme. \cite{Kivshar:1999-5056:PRE} demonstrated
that this two-color parametric interaction can support two-color
spatial solitary waves. \cite{TrevinoPalacios:1995-170:APL}
considered the interference between the parametric processes AA-S
and AA-T, where S and T are two different modes of the waveguide
at the frequency $2\omega$. In both the cases mentioned above, two
different parametric processes share one and the same fundamental
wave. However, it is possible that the multistep cascading
interaction involves the SHG process of the type AB-S. Such a
four-wave multistep cascading process was considered by
\cite{DeRossi:1997-53:OQE}. The two SHG processes were AB-S and
AA-T, where (A,B) and (S,T) are pairs of the modes at the
frequencies $\omega$ and $2\omega$, respectively.
\cite{DeRossi:1997-53:OQE} concluded that a devise based on this
type of multistep cascading can operate as an all-optical
modulator or as an all-optical switch, with a good switching
contrast at 1.55 $\mu$m.

Another interesting process of the multistep interaction was
considered by \cite{Boardman:1997-1899:PRE,Boardman:1998-891:OQE}
who studied the parametric mode coupling in a nonlinear waveguide
placed in a magnetic field. In this case, the simultaneous
coexistence of six SHG processes is possible, namely, ooo, ooe,
oee, eee, eeo, and eoo. The exact number of allowed parametric
interactions depends on the symmetry point group of the material.
It was shown that, by controlling a ratio of the input fundamental
components, one of the SH components can be controlled and
switched off. Importantly, repulsive and collapsing regimes for
the interacting parametric solitons can be produced by switching
the direction of the magnetic field.

\cite{Saltiel:1999-1296:OL} considered the possibility to realize
the efficient polarization switching in a quadratic crystal that
supports simultaneous phase matching for both Type I and Type II
SHG processes (e.g. ooo and oeo). At certain conditions, the
fundamental beam involved in such a process can accumulate a large
nonlinear phase shift at relatively low input power
(\cite{Saltiel:2000-959:JOSB}). The SHG process that can be
realized by two possible pairs of the simultaneously phase-matched
processes (ooo, ooe) and (ooe, eee) was studied theoretically in
\cite{Grechin:1999-155:KE}. The effect of SHG by simultaneous
phase-matching of three parametric processes (ooe, eee, and oee)
in a crystal of LiNbO$_3$ was estimated theoretically in
\cite{Grechin:2001-929:KE}. It was shown that, at certain
conditions, one can obtain polarization insensitive SHG.

Also, we would like to mention  several experimental studies of
the two-color multistep cascading interactions. In the paper of
\cite{TrevinoPalacios:1995-170:APL}, an interplay of two Type I
SHG processes with a common fundamental wave was observed.
\cite{Petrov:2001-355:OL} and \cite{Petrov:2002-268:JOSB}
performed the experiment with a BBO crystal in which, as a result
of the simultaneous action of the Type I SHG and Type II SHG
interactions, the generation of the wave orthogonally polarized to
the input fundamental wave was observed.
\cite{Couderc:2002-421:OC} demonstrated that the multistep
cascading interaction can support parametric solitons in the
waveguiding regime. In this latter case, the multistep interaction
simulates an effective nearly-degenerate FWM process and, in this
sense, it is almost "two-color". \cite{Pasiskevicius:2002-1628:OL}
realized experimentally the simultaneous SHG process that gives
two SH waves with the orthogonal polarizations in the blue
spectral region by use of the Type II and Type I QPM phase
matching in a periodically poled KTP crystal.

As a simple example of the two-color multistep parametric
interaction, we consider the cascading of the Type I and Type II
SHG processes according to the scheme: BB-S and AB-S. Here, the SH
wave is generated by two interactions. Thus, we can expect an
increase of the SHG efficiency when both A and B waves (i.e., the
ordinary and extraordinary waves) are involved. However, if only
one of the waves, say the wave A, is launched at the input, this
type of the double phase-matched interaction will lead to the
generation of a wave perpendicular to the input wave, through the
parametric process SA $\rightarrow$ B. Additionally, the
fundamental wave A accumulates a strong nonlinear phase shift.
This parametric interaction is described by the following
equations for plane waves,
\begin{equation} \leqt{TCMSCeqns}
   \begin{array}{l} {\displaystyle
      \frac{d A}{d z}
      = - i \sigma_{1} S A^{\ast} e^{-i \Delta k_{\rm SHG} z}
      -   i \sigma_{3} S B^{\ast} e^{-i \Delta k_{\rm DFG} z} ,
   } \\*[9pt] {\displaystyle
      \frac{d S}{d z}
      = - i \sigma_{2} A^{2} e^{i \Delta k_{\rm SHG} z}
      -   i \sigma_{4} A B e^{i \Delta k_{\rm DFG} z} ,
   } \\*[9pt] {\displaystyle
      \frac{d B}{d z}
      = - i \sigma_{5} S A^{\ast} e^{-i \Delta k_{\rm DFG} z} ,
   } \end{array}
\end{equation}
where $A$, $S$, and $B$ are the complex amplitudes of the input
fundamental wave, the second-harmonic wave, and the orthogonally
polarized wave at the fundamental frequency, $\sigma_1$ and
$\sigma_2$ are defined above, and
\[
\sigma_{3,4,5}= \frac{2\pi d_{\rm
eff,II}}{\lambda_{1}n_{A,2,B}}\left(\frac{\omega_{1,2,1}}{\omega_{1}}\right).
\]
If we neglect dispersion of the index of refraction, i.e.
$n_A\simeq n_B\simeq n_2$, then we can assume that
$\sigma_1\simeq\sigma_2$ and
$\sigma_3\simeq\sigma_5\simeq\sigma_4/2$. The phase-mismatch
parameters are defined as $\Delta k_{\rm SHG} = k_{2} - 2k_{A}+
G_{p}$ and $\Delta k_{\rm DFG} = k_{2} - k_{A}-k_{B}+ G_{q}$,
where $G_{p}$ and $G_{q}$ are two QPM vectors used for the phase
matching.

Solution of this system, with respect of the amplitude of the wave
B, has been obtained by \cite{Saltiel:1999-1296:OL}) in the
approximation of nondepleted pump. It gives the following result,
\begin{equation} \leqt{Yana}
   B(L) =  \frac{i \sigma_1 \sigma_3}{
                        D_- \Delta k_{\rm SHG}}
        |A|^2 A \sin(D_- L)e^{iD_+ L},
\end{equation}
where
$$D_{\pm}=\frac{1}{2}(\Delta k_{\rm SHG}-\Delta k_{\rm DFG}) \pm
\frac{\sigma_1^2}{\Delta k_{\rm SHG}}|A|^2.$$

Equation~\reqt{Yana} shows that the generation of the component B
by this multistep interaction mimic the FWM process of the type
$AAA^{\ast}-B$, governed by the cascaded cubic nonlinearity. This
effective cubic nonlinearity leads to the accumulation of a
nonlinear phase shift by the wave A that, similar to the case of
the cascaded THG process, includes a contribution of high-order
$(>3)$ nonlinearities (\cite{Saltiel:2000-959:JOSB}). Existence of
two-color multistep parametric solitons and waveguiding effects
are the result of the nonlinear phase shift collected by the
interacting waves (\cite{Kivshar:1999-5056:PRE}).

\subsection{Fourth-harmonic multistep cascading} \lsect{FHG}

Another fascinating example of the multistep cascading interaction
is  the fourth-harmonic generation (FHG) in a single crystal with
the second-order nonlinearity.  There are two possible parametric
processes of the second order that lead to the generation of a
fourth harmonic. In both the cases the fourth harmonic (FH)
amplitude is proportional to the factor $(\chi^{(2)})^3$:
\begin{enumerate}
\item
   SHG + SHG:  $  \omega +   \omega = 2 \omega$;
               $2 \omega + 2 \omega = 4 \omega$ ;
\item
   SHG + SFG +SFG:  $\omega  + \omega = 2 \omega$;
                    $2\omega + \omega = 3 \omega$;
                    $3\omega + \omega = 4 \omega$.
\end{enumerate}
The phase matching conditions define which of these processes will
be more effective. Obviously, the former process is easier to
realize technically because it requires only two phase-matching
conditions to be fulfilled simultaneously, and therefore many
papers deal with this case.

The first experiment on the generation of a forth-harmonic wave by
cascading was reported in the paper by
\cite{Akhmanov:1974-264:PZETF}, where the FHG cascaded process in
Lithium Formiate crystal was studied.  As pointed out by the
authors, the generation is due to the simultaneous action of two
and three processes with the involvement of the quadratic and
cubic nonlinearities. The FHG process in CdGeAs$_2$ was observed
by \cite{Kildal:1979-5218:PRB}. Both these pioneer papers were
motivated by the idea to estimate the magnitude of the direct
fourth-order nonlinearity in terms of $(\chi^{(2)})^3$ . One of
the first studies that reported the efficiency of the cascaded FHG
process is the paper by \cite{Hooper:1994-6980:AOP} where  the
efficiency of 3.3 x 10$^{-4}$\% was obtained in a single LiNbO$_3$
crystal. The other paper by \cite{Sundheimer:1994-975:ELL}
reported the efficiency of 0.012\%. However, in those experiments
only the first step $\omega+\omega = 2\omega$ was phase-matched
while the other one was not matched, leading to the overall low
conversion efficiency. Somewhat larger efficiency of 0.066\% was
reported in the work of \cite{Baldi:1995-1350:ELL} where a
periodically poled LiNbO$_3$ waveguide was used. Both steps of the
multistep parametric interaction, namely $\omega+\omega = 2\omega$
and $2\omega+2\omega = 4\omega$, were phase matched: the first
step was realized through the first-order QPM process, while the
second step was realized through the 7-th order QPM process. As is
shown by
\cite{Norton:2003-1008:OE} and \cite{Sukhorukov:2001-34:PLA}, if both
steps are phase-matched, the resulting efficiency should be close
to 100\%. In the second paper, the existence and stability of the
normal modes for such a multistep cascading system have been
studied. Some possibilities for the double-phase-matched FHG
process for certain input wavelengths in single-crystals of
LiNbO$_3$, LiTaO$_3$, KTP, and GaAs have been shown by
\cite{Pfister:1997-1211:OL} and \cite{Grechin:2001-933:KE}. The
possibility for FHG by double phase-matching in broader spectral
region by use of the phase-reversed QPM structure was discussed by
\cite{Sukhorukov:2001-34:PLA}.

In a very interesting experiment, \cite{Broderick:2002-2263:JOSB}
demonstrated cascaded FHG in a 2D nonlinear photonic crystal with
the efficiency 0.01\% in a (not optimized) 2D planar QPM
structure. Several useful efficient schemes for FHG in 2D
nonlinear photonic crystals have been proposed by
\cite{Saltiel:2000-1204:OL} and \cite{deSterke:2001-539:OL}. An
optimal design of 2D nonlinear photonic crystals for achieving the
maximum efficiency of FHG has been discussed by
\cite{Norton:2003-188:OL} and \cite{Norton:2003-1008:OE}. The use
of the FHG multistep interaction for the frequency division
schemes (4:1) and (4:2) have been discussed in
\cite{Dmitriev:1998-228:ICONO98} and
\cite{Sukhorukov:2001-34:PLA}.

The basic equations describing the FHG parametric process in a
double-phase-matched QPM structure for the interaction of plane
waves  can be written in the form (see, e.g.,
\cite{Hooper:1994-6980:AOP})
\begin{equation} \leqt{FoHGeqns}
   \begin{array}{l} {\displaystyle
      \frac{d A_{1}}{d z}
      = - i \sigma_{1} A_{2} A_{1}^{\ast} e^{-i \Delta k_{\rm SHG}z},
   } \\*[9pt] {\displaystyle
      \frac{d A_{2}}{d z}
      = - i \sigma_{2} A_{1}^{2} e^{i \Delta k_{\rm SHG} z}
      - i \sigma_{6} A_{4} A_{2}^{\ast} e^{- i \Delta k_{4} z},
   } \\*[9pt] {\displaystyle
      \frac{d A_{4}}{d z}
      = - i \sigma_{7} A_{2}^{2} e^{i \Delta k_{4} z},
   } \end{array}
\end{equation}
where $A_{1}$, $A_{2}$ and $A_{4}$ are the complex amplitudes of
the fundamental, second-harmonic, and fourth-harmonic waves,
respectively. The parameters $\sigma_{1}$ and $\sigma_{2}$ are
defined above, and
\[
\sigma_{6,7}= \frac{4\pi }{\lambda_{1}n_{2,4}}d_{\rm eff,II}.
\]
As before, if we neglect the index of refraction dispersion
($n_1\simeq n_2\simeq n_4 $) then we can accept that
$\sigma_1\simeq\sigma_2$ and $\sigma_6\simeq\sigma_7$. Phase
mismatch parameters are $\Delta k_{\rm SHG} = k_{2} - 2k_{A}+
G_{p}$ and $\Delta k_{4} = k_{4} - 2k_{2}+ G_{q}$ , where $G_{p}$
and $G_{q}$ are two QPM reciprocal vectors. The role of high order
nonlinearities are not included in~\reqt{FoHGeqns} since their
contribution is rather small when one works in conditions close to
double or triple phase-matching.

Solution of the system~\reqt{FoHGeqns} can be found neglecting the
depletion effects. It reveals that the phase-matched FHG wave is
generated when either one of the following phase-matching
conditions is satisfied,
\begin{enumerate}
\item $\Delta k_{\rm SHG}\longrightarrow0$;
\item $\Delta k_{4}\longrightarrow0$;
\item $\Delta k_{\rm SHG} + \Delta k_{4}
       = k_{4} - k_{2} - 2 k_{1} + G_{p} + G_{q} \longrightarrow 0$; and
\item $2 \Delta k_{\rm SHG} + \Delta k_{4}
       = k_{4} - 4 k_{1} + 2 G_{p} + G_{q} \longrightarrow 0$
\end{enumerate}
The case $\Delta k_{\rm SHG} \longrightarrow 0$ gives the
strongest phase-matching process among all those cases, and the
generated FH wave exceeds by several orders of magnitude FH generated by other
schemes. For the double-phase-matching process (when $\Delta
k_{\rm SHG}\longrightarrow 0$ and $\Delta k_{4}\longrightarrow 0$)
the squared amplitude of forth harmonic is given by the expression
(\cite{deSterke:2001-539:OL}):
\begin{equation} \leqt{F0HGampl}
   |A_{4}(L)|^2 = \frac{1}{9}\sigma_2^4 \sigma_6^2 |A_1|^8 L^6.
\end{equation}

Introducing again the normalized efficiency (measured in
W$^{-1}$cm$^{-2}$) for the first and second steps as $\eta_{0,1}$
and $\eta_{0,2}$, respectively, we can obtain the efficiency of
the cascaded FHG process,
\begin{equation} \leqt{EFFFoHG}
   \eta_{4\omega} = \frac{1}{9} \eta_{0,1}^2 \eta_{0,2} P_{1}^3 L^6 .
\end{equation}
Thus, the efficiency of the cascaded FHG process in a single
$\chi^{(2)}$ is proportional to the 6-th power of the length and
the 3-rd power of the pump,  and it can be estimated with the
known efficiencies for the separated steps. For the pump
intensities at which the depletion effect of the fundamental and
second-harmonic waves can not be neglected, the
system~\reqt{FoHGeqns} should be solved numerically. As shown in
\cite{Zhang:2001-317:JOA}, the 100\% conversion of the fundamental
wave into the fourth-harmonic wave is possible independently on
the ratio of the nonlinear coupling coefficients
$\sigma_2/\sigma_6$. This behavior is in contrast with the
$\chi^{(2)}$-based cascaded THG process where the 100\% conversion
is possible only for a specific ratio of the nonlinear coupling
coefficients (see Sec.~\rsect{THG}).

\subsection{OPO and OPA Multistep parametric processes} \lsect{OPO}

Many studies dealing with optical parametric oscillators (OPOs)
and optical parametric amplifiers (OPAs) have observed, in
addition to the expected signal and idler waves, other waves at
different wavelengths that are coherent and are generated with a
good efficiency. These waves were identified as being a result of
additional phase-matched (or nearly phase-matched) second-order
parametric processes in OPOs and OPAs. In some of the cases, up to
five additional waves coming out from the nonlinear crystal were
detected and measured. By tuning the wavelength of the signal and
idler waves simultaneously, the additional outputs from OPO can be
frequency tuned as well. An example is shown in Fig.~\rpict{OPO}
adapted from \cite{Zhang:2002-2479:JOSB}. The study of the
simultaneous phase matching in OPO was started more than 30 years
ago. As shown in \cite{Ammann:1971-5618:JAP}, in many nonlinear
crystals the phase-matching tuning curves for OPO and SHG of the
signal and the idler cross and the experimental observations of
the corresponding double phase-matching parametric processes are
possible.

\pict{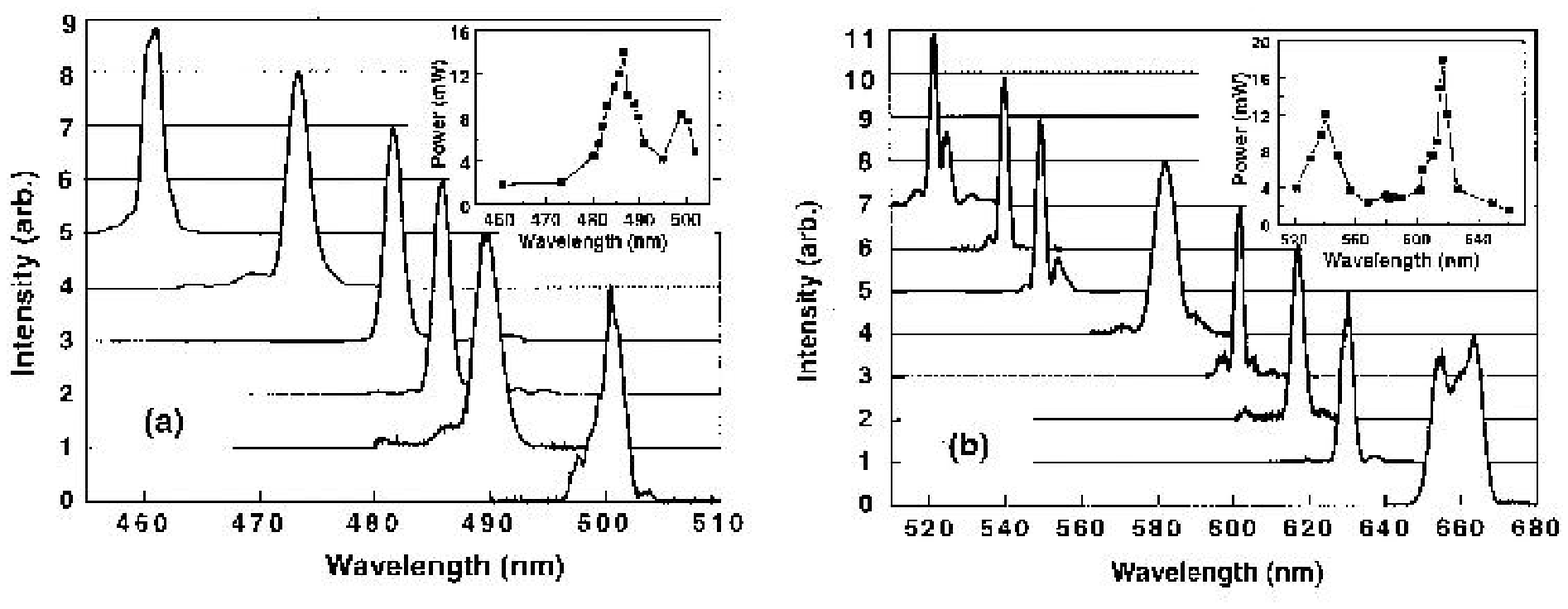}{OPO}{The tuning characteristics of OPO with the
simultaneously phase matched SHG and SFG processes. (a) SFG
between the signal and pump, and (b) SHG of the signal. Inset:
output power vs. wavelength (\cite{Zhang:2002-2479:JOSB}). }

Plane-wave single-pass theory of the new frequency generation with
OPO with the simultaneous frequency doubling of the signal wave in
the OPO crystal was presented in \cite{Aytur:1998-447:IQE}. The
case of OPO with the simultaneous frequency mixing between the
pump and signal waves~--- in \cite{Dikmelik:1999-897:IQE} and
\cite{Morozov:2003-233:JOA}. \cite{Moore:1998-803:IQE} considered
theoretically the simultaneous phased matched tandem of OPOs in a
single nonlinear crystal. The signal wave of the first OPO process
becomes a pump wave of the second process. The simultaneous action
of the parametric generation, $\omega_p \rightarrow \omega_i +
\omega_s$, and SFG between the pump and signal wave, $\omega_{\rm
SF} = \omega_p + \omega_s$, was also analyzed by
\cite{Huang:2002-13899:JPCM}, where the main goal was to show that
this type of the multistep cascading can be used for the
simultaneous generation of three fundamental colors.

\begin{small}
\begin{longtable}{c c p{2.4 cm}p{1.5cm}c p{1.5cm}p{5mm}}
\caption {\ltab{OPO} Experimental results on the OPO and OPA multistep cascading}\\
  \hline
   Nonlinear & $\lambda_{pump}$ & Phase matching & Phase & L &  Regime&   Refs \\
   crystal & [$\mu m$] & processes &  matching method & [cm] &  & \\
  \hline
\endfirsthead

  \hline
   Nonlinear & $\lambda_{pump}$ & Simultaneously & Phase & L &  Regime&   Refs \\
   crystal & [$\mu m$] & PM processes &  matching method & [cm] &  & \\
  \hline
\endhead

 ADP  & 1.06 & $\omega_{1}=\omega_{p}+\omega_{s}$ &
 BMP &  5& { Pulsed (Q~pulse)} &  a
\\

LiNbO$_3$ & 1.06 & $\omega_{1}=\omega_{s}+\omega_{s}$ or $\omega_{2}=\omega_{i}+\omega_{i}$&
 BMP & 0.38& { Pulsed (Q~pulse)}&  b
\\

LiNbO$_3$ & 1.064 &  \mbox{$\omega_{1}=\omega_{s}+\omega_{s}$}  \mbox{$\omega_{2}=\omega_{1}-\omega_{i}$}  &
 BPM & 3(5) &pulsed
 (35~ps)  &  c\\

KTP &0.79 &  $\omega_{1}=\omega_{s}+\omega_{s}$  $\omega_{2}=\omega_{p}+\omega_{s}$  $\omega_{3}=\omega_{p}+\omega_{i}$&    BPM & 0.115 & pulsed (115~fs) &
    d\\

LiNbO$_3$ & 0.532  & { $\omega_{p}=\omega_{p}/2+\omega_{p}/2$} &
NCPM& 3.2 & CW & e\\

BBO & 0.74-0.89& $\omega_{1}=\omega_{p}+\omega_{i}$  & BPM & 0.4 & pulsed (150~fs)& f \\

KTP& 0.82-0.92 & $\omega_{1}=\omega_{s}+\omega_{s}$  $\omega_{2}=\omega_{p}+\omega_{s}$  $\omega_{3}=\omega_{p}+\omega_{i}$ &
 NCPM  & 0.2 &  pulsed (120~fs) & g
\\

LiNbO$_3$ & 1.064 & $\omega_{1}=\omega_{p}+\omega_{p}$  $\omega_{2}=\omega_{p}+\omega_{s}$  $\omega_{3}=\omega_{p}+2\omega_{s}$ &
  QPM & 1.5 & Pulsed (7~ns) & h
\\

LiNbO$_3$ & 0.532 & { $\omega_{p}=\omega_{p}/2+\omega_{p}/2$} &
 NCPM  & 0.75 &  CW & i
\\

LiNbO$_3$ &  0.78-0.80& $\omega_{1}=\omega_{p}+\omega_{p}$  $\omega_{2}=\omega_{s}+\omega_{s}$  $\omega_{3}=\omega_{p}+\omega_{s}$  $\omega_{4}=\omega_{p}+\omega_{i}$  $\omega_{5}=\omega_{s}+\omega_{s}+\omega _s$&
 QPM & 0.6 & pulsed
(2~ps) & j\\

LiNbO3 & 0.793 & $\omega_{1}=\omega_{p}+\omega_{s}$&
QPM & 0.08 & pulsed
(85~fs) &   k
\\

$\beta$-BBO  &   0.53  &$\omega_1=\omega_{s}+\omega_{s}$
$\omega_{s1}=\omega_1-\omega_{i}$
$\omega_{i1}=\omega_{p}-\omega_{s1}$ 
$\omega_{2}=\omega_{s1}+\omega_{s1}$ 
{$\omega_{s2}=\omega_2-\omega_{i}$
$\omega_{i2}=\omega_{p}-\omega_{s2}$} ...... &  BPM & 0.8 & pulsed
(1~ps) & l\\

KTP & 0.74-0.76 & $\omega_{1}=\omega_{s}+\omega_{s}$&
 BPM & 0.5 & pulsed
 (150~fs)   &  m
\\

LiNbO$_3$ &  1.064  & $\omega_{s}\rightarrow\omega_{s2}+\omega_{i2}$  &
QPM & 2.5 & pulsed
 (43~ns)  & n\\

LiNbO$_3$ &   0.79-0.81  & $\omega_{1}=\omega_{s}+\omega_{s}$ &
QPM & 0.1 & pulsed
 (100~fs)  & o
\\

KTP & 0.827 &  $\omega_{1}=\omega_{p}+\omega_{s}$ &
  BPM  & 0.5 & pulsed (170~fs) & p
\\

KTP & 0.76-0.84 & $\omega_{1}=\omega_{s}+\omega_{s}$ $\omega_{2}=\omega_{p }+\omega_{s}$ $\omega_{3}=\omega_{1}+\omega_{i}$&
QPM & 0.05 & pulsed
 (30~fs)  & q, r
\\

LiTaO$_3$ & 0.532 & $\omega_{1}=\omega_{p }+\omega_{s}$&
QPOS & 2 & pulsed
(40~ps) &  s
\\

LiNbO3  & 0.8  & $\omega_{1}=\omega_{p}+\omega_{s}$ $\omega_{2}=\omega_{s }+\omega_{s}$&
QPM & 0.05 & pulsed
(40~fs) &     t
 \\

LiNbO$_3$ &  1.064  & $\omega_{1}=\omega_{s}+\omega_{s}$  $\omega_{2}=\omega_{s}+\omega_{p}$   { $\omega_{3}=\omega_{s}+\omega_{s}+\omega_{s}$} $\omega_{4}=\omega_{p}+\omega_{p}$ $\omega_{5}=\omega_{p}+\omega_{1}$&

 QPM & 2 & pulsed
 (17.5~ns)   &   u\\

KTA &       0.796  & $\omega_{1}=\omega_{s}+\omega_{s}$&
BPM & 2 & pulsed
 (140~fs)  & v
\\

LiNbO$_3$ &  0.79   & $\omega_{1}=\omega_{s}+\omega_{s}$&
APQPM & 1.8 &    pulsed
(5~ns)  & w

\\

$\beta$-BBO & 0.405 (0.81) & $\omega_{1}=\omega_{p}/2+\omega_{s}$ $\omega_{2}=\omega_{i}+\omega_{i}$ &

BPM & 0.2 & pulsed
 (90~fs)  & x

\end{longtable}
~\\
$^a$ \cite{Andrews:1970-605:PRL}\\
$^b$ \cite{Ammann:1971-5618:JAP}\\
$^c$ \cite{Bakker:1989-398:OC}\\
$^d$ \cite{Powers:1993-2162:JOSB}\\
$^e$ \cite{Schiller:1993-1696:JOSB}\\
$^f$ \cite{Petrov:1995-2171:OL}\\
$^g$ \cite{Hebling:1995-919:OL}\\
$^h$ \cite{Myers:1995-2102:JOSB}\\
$^i$ \cite{Schiller:1996-3374:APL}\\
$^j$ \cite{Butterworth:1997-618:OL}\\
$^k$ \cite{Burr:1997-3341:APL}\\
$^l$ \cite{Varanavicius:1997-1603:OL}\\
$^m$ \cite{Kartaloglu:1997-280:OL}\\
$^n$ \cite{Vaidyanathan:1997-49:OE}\\
$^o$ \cite{McGowan:1998-694:JOSB}\\
$^p$ \cite{Koprulu:1999-1546:JOSB}\\
$^q$ \cite{Zhang:2001-2005:OL}\\
$^r$ \cite{Zhang:2002-2479:JOSB}\\
$^s$ \cite{Du:2002-1573:APL}\\
$^t$ \cite{Zhang:2002-1873:APL}\\
$^u$ \cite{Xu:2002-801:CHIL}\\
$^v$ \cite{Kartaloglu:2003-65:IQE}\\
$^w$ \cite{Kartaloglu:2003-343:JOSB}\\
$^x$ \cite{Lee:2003-1702:OE}\\
~\\
Abbreviations: BPM~--  birefringence phase matching; NCPM~--
noncritical phase matched; QPM~-- uniformly poled
quasi-phase-matched structure; QPOS~-- quasi-periodical optical
superlattices; APQPM~-- aperiodically poled QPM structure.\\
~\\
~

\end{small}

In Table~\rtab{OPO}, we have summarized different experimental
results on the multistep cascading processes observed in OPOs and
OPAs. For each of this work, we also mention the additional
second-order parametric processes.

In the special case when all frequencies become phase related,
i.e. $\omega_p:\omega_s:\omega_i=3:2:1$, OPO displays the unique
properties (\cite{Kobayashi:2000-856:OL, Longhi:2001-57:EPD}). As
a matter of fact, this case corresponds to the third-harmonic
multistep cascading (see Sec.~\rsect{THG}) but realized in a
cavity. The second harmonic of the idler simultaneously generated
in the OPO crystal (or by externally frequency doubling)
$\omega_{2i}=\omega_i+\omega_i$ will interfere with the signal
wave of the frequency $\omega_i$. The resulting beat signal can be
used for locking OPO at this particular tuning point and for
realizing the frequency divisions (3:2) and (3:1), or vice versa.
More details about the frequency division with these types OPO can
be found in the papers by \cite{Lee:2003-13808:PRA,
Douillet:2001-548:ITIM, Slyusarev:1999-1856:OL,
Zondy:2001-23814:PRA, Zondy:2003-35801:PRA}. For the double
phase-matched parametric interactions for the frequencies
$3\omega;2\omega;\omega$, the total conversion from a $3\omega$
wave to a $2\omega$ wave was predicted when the pump is the
$3\omega$ wave (\cite{Komissarova:1993-1025:KE}),  and from a
$2\omega$ wave to a $3\omega$ wave,  when the pump is the
$2\omega$ wave (\cite{Volkov:1998-101:KE}) for the OPA scheme.
Additionally, the theoretical studies in \cite{Longhi:2001-57:EPD,
Longhi:2001-713:OL, Longhi:2001-55202:PRE} show that OPO, when the
signal and idler wave have the frequencies $\omega$ and $2\omega$,
respectively, can produce different types of interesting patterns
including the spiral and hexagonal patterns. The internally pumped
OPOs, due to the simultaneous action of the SHG process and the
parametric down-conversion, also shows the formation of spatial
patterns and the parametric instability dynamics
(\cite{Lodahl:1999-3251:PRA, Lodahl:2000-4506:PRL,
Lodahl:2001-23815:PRA}).

Additionally, the multistep cascaded OPOs and OPAs can be an
efficient tool for generating the entangled and squeezed photon
states. In particular, \cite{Smithers:1974-790:PRA} considered the
quantum properties of light generated during simultaneous action
of the parametric processes
$\omega_{p}\rightarrow\omega_i+\omega_s $ and
$\omega_{1}=\omega_i+\omega_p$. The theoretical predictions by
\cite{Marte:1995-4815:PRL, Marte:1995-2296:JOSB, Eschmann:1997-247:QSO} suggest that, due to the multistep cascading
in internally pumped OPO, this type of OPO can be an excellent
source for generating twin photons with the sub-Poissonian
statistics, and the generated SH wave exhibits a perfect noise
reduction. Several other phase-matched schemes for generating the
entangled and squeezed photon states have been proposed by
\cite{Chirkin:2002-S91:JOB, Nikandrov:2002-333:PZETF} and
\cite{Chirkin:2003-169:JOB} who utilizes OPA with two
simultaneously phase-matched processes, $2\omega=\omega+\omega$
and $3\omega=2\omega+\omega$. In the theoretical study
(\cite{Nikandrov:2002-81:JRLR}), a possibility of generating
squeezed light with the utilization of a single quadratic crystal
was compared with that of two different crystals for each step.
The conclusion is that a single-crystal cascaded method is more
efficient. New possibilities for the generation of squeezed
polarized light have been discussed by
\cite{Dmitriev:physics/0308025:ARXIV} who considered the five-wave generation via the four simultaneously phase-matched parametric processes.

\subsection{Other types of multistep interactions} \lsect{other}

The multistep parametric interactions allow building compact
frequency converters with several visible beams as the output.
\cite{He:2003-228:APL} reports on the simultaneous generation of
all three "traffic signal lights". The simultaneous generation of
a pair blue and green waves  have been achieved in
\cite{Capmany:2000-1225:APL} by exploring self-doubling and
self-frequency mixing active media.

\pict{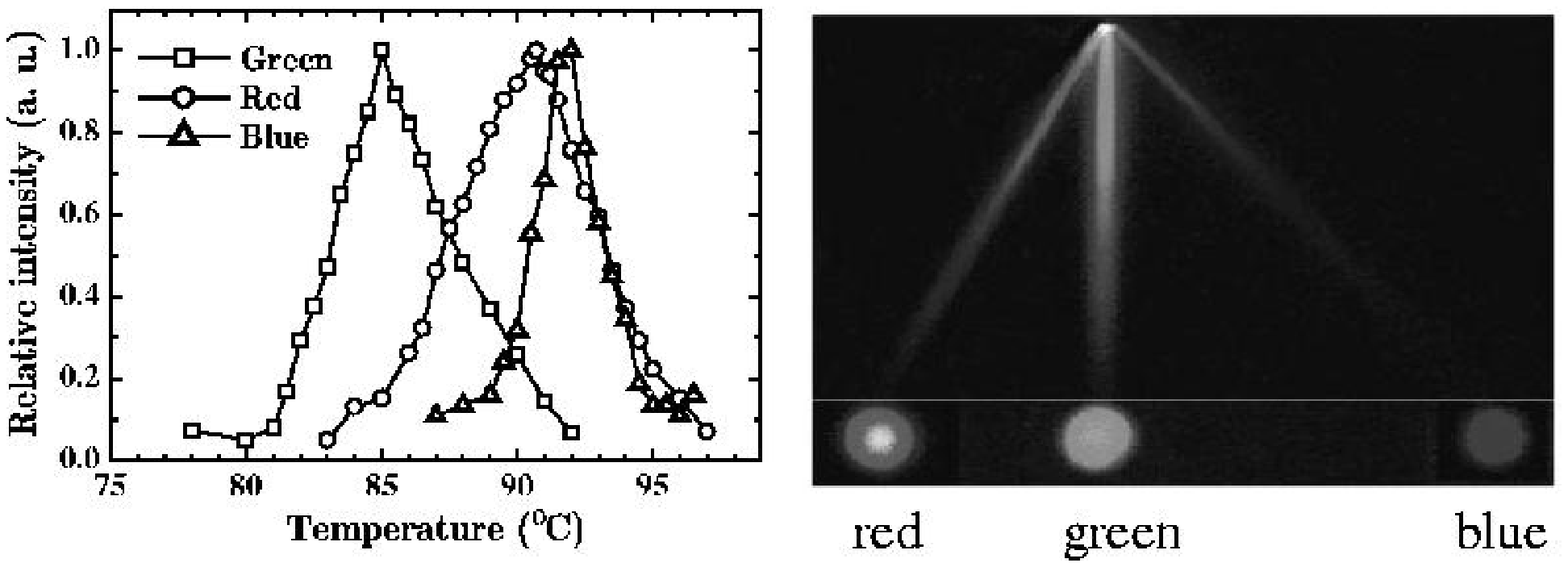}{RGB}{ Left: the phase matching curves for three
simultaneously phase-matched processes in a single
aperiodically-poled LiTaO$_3$ crystal. Right: visible red, green
and blue outputs from a nonlinear crystal diffracted by a prism
(\cite{Liao:2003-3159:APL}). }

\begin{small}
\begin{longtable}{p{1.6 cm}p{1.0 cm}p{1.8 cm}p{1.6 cm}cp{1.2 cm}p{0.6 cm}}

\caption{ \ltab{RGB} Experiments on the simultaneous generation of
several visible beams}\\
\hline
  Nonlinear crystal & $\lambda_p$ [$\mu m$] & COLORS & PM method
  & L [cm] & Regime& Ref.  \\
\hline
\endfirsthead
\hline
  Nonlinear crystal & $\lambda_p$ [$\mu m$] & COLORS & PM method
  & L [cm] & Regime& Ref.  \\
\hline
\endhead
 KTP waveguide &1.023
0.716  & RED, GREEN, BLUE  & { BPM/QPM} & 0.45 & CW  & a
\\

  KTP waveguide &1.650 & RED, GREEN, BLUE &   QPM & 0.35 & pulsed
(6 ps) & b
\\

  LiNbO$_3$ waveguide & 1.620 &  RED, GREEN, BLUE &   QPM & & pulsed
(7 ps)     & c

 \\

  NYAB&  1.338
0.807 0.755  & RED, GREEN, BLUE   & BPM & 0.5 &CW  &   d

 \\

  { Nd:LiNbO$_3$} &  1.084
0.744 &  GREEN, BLUE &  APQPM & 0.095 &  CW    & e

 \\

  { Nd:LiNbO$_3$}  & 1.372
1.084 0.744 &  RED, ORANGE, GREEN, BLUE(2) &   APQPM & 0.3  & CW &
f
\\

  SBN & 1.34
0.88  &  RED, GREEN, BLUE &   APQPM & 0.7 & CW & g
 \\

  LiTaO$_3$ & 1.342
1.064 &  RED, GREEN, BLUE  &  APQPM & 1 & CW &h
 \\

  LiTaO$_3$ & 1.342
1.064  & {\footnotesize RED, YELLOW, GREEN} & APQPM & 1 & CW  & i \\

\end{longtable}
~\\
$^a$ \cite{Laurell:1993-1872:APL}\\
$^b$ \cite{Sundheimer:1994-975:ELL}\\
$^c$ \cite{Baldi:1995-1350:ELL}\\
$^d$ \cite{Jaque:1999-325:APL}\\
$^e$ \cite{Capmany:2000-1225:APL}\\
$^f$ \cite{Capmany:2001-144:APL}\\
$^g$ \cite{Romero:2002-4106:APL}\\
$^h$ \cite{Liao:2003-3159:APL}\\
$^i$ \cite{He:2003-228:APL}\\
~\\
~
\end{small}

Recently, several studies presented successful attempts to obtain
the simultaneous generation of red, green and blue radiation(the
so-called RGB radiation) from a single nonlinear quadratic
crystal. This is an important target for building compact
laser-based projection displays. Theoretically, the parametric
process for achieving the generation of three primary colors as
OPO outputs was considered in \cite{Huang:2002-13899:JPCM}.
\cite{Liao:2003-3159:APL} used a single aperiodically-poled
LiTaO$_3$ crystal for generating 671, 532 and 447 nm (see
Fig.~\rpict{RGB}) with three simultaneously phase-matched
processes: SHG of 1342 and 1064 nm (the output of the dual-output
Nd:YVO$_4$ laser) and SFG of 671 and 1342 nm.
\cite{Jaque:1999-325:APL} realized a different method for
achieving the three-wavelength output. In their experiments, a
nonlinear medium is the Nd:YAl$_3$(BO$_3$)$_4$ crystal that was
pumped by a Ti:sapphire laser. The red signal at 669 nm was
obtained by self-frequency doubling of the fundamental laser line.
The green signal at 505 nm and a blue signal at 481 nm were
obtained by self-SFG of the fundamental laser radiation at 1338 nm
and the pump radiation (807 nm, for green, and 755 nm, for blue).
All three processes were simultaneously phase matched by
birefringence phase matching due to an exceptional situation that
the three phase-matchings appear extremely close to each other and
their tuning curves overlap. The generation of red, green and blue
signals by triple phase-matching in LiNbO$_3$ and KTP periodically
poled waveguides was reported also by
\cite{Sundheimer:1994-1400:ELL} and \cite{Baldi:1995-1350:ELL}.
The experimental efforts to build optical devices with the
simultaneous generation of several visible harmonics are
summarized in Table~\rtab{RGB}.

\subsection{Measurement of  the $\chi^{(3)}$-tensor components } \lsect{chi3}

As an important application of the multistep parametric processes
in nonlinear optics, we would like to mention the possibility
making a calibration link between the second- and third-order
nonlinearity in a nonlinear medium. Both THG and FWM processes in
non-centrosymmetric nonlinear media can be used for this purpose.
Measurements can be done in the phase-matched or non-phase-matched
regimes. The non-phase-matched regime allows achieving a higher
accuracy, however, it is more complicated since the signal is weak
and additional care should to be taken to avoid the influence of
the respective third-order effect in air. The basic idea of these
types of measurements is to compare the THG or FWM signal in
several configurations that include a proper choice of the
direction and polarization of the input and output waves. In some
of the configurations the output signal is generated in result of
the direct contribution of the inherent cubic nonlinearity of a
sample. In other cases, the signal is generated due to the cascade
contribution, while in the third group,  the contribution of both
direct and cascaded processes is comparable. In this way, we can
access the ratio
$$\frac{\chi^{(3)}(-3\omega,\omega,\omega,\omega)}{\chi^{(2)}(-3\omega,2\omega,\omega)\chi^{(2)}(-2\omega,\omega,\omega)}$$
for the case of the THG multistep interaction,  and the ratio
$$\frac{\chi^{(3)}(-\omega_3,\omega_1,\omega_1,-\omega_2)}{\chi^{(2)}(-\omega_3,2\omega_1,-\omega_2)\chi^{(2)}(-2\omega_1,\omega_1,\omega_1)}$$
for the case of the FWM multistep cascading. Because the
information about the $\chi^{(2)}$ components is more available,
we can determine quite accurate parameters of the cubic
nonlinearity of non-centrosymmetric materials by using this
internal calibration procedure which does not require the
knowledge of the parameters of the laser beam.

To illustrate that let us consider the phase-matched THG. In
condition of non-depletion of the fundamental and second harmonic
wave and neglecting the temporal and spatial walk off effect the
equations~\reqt{THGeqns} has following solution for phase matching
$\Delta k_{\rm THG} = k_{3} - 3 k_{1} \longrightarrow 0$:
\begin{equation} \leqt{MeasCHI3}
   A_{3}(L) = -i \left(\gamma + \frac{\sigma_{2}\sigma_{5}}{\Delta k_{\rm SFG}}\right)
           \frac{\sin(\Delta k_{\rm THG}L/2)}{\Delta k_{\rm THG}L/2}|A_{1}|^3 L.
\end{equation}

Apparent cubic nonlinearity consists two parts direct and cascading:
\begin{equation} \leqt{CHI3eff}
   {\chi^{(3)}_{\rm tot} = \chi^{(3)}_{\rm eff,dir} + \chi^{(3)}_{\rm eff,casc}} ,
\end{equation}
with
\[
\chi^{(3)}_{\rm eff,casc}= \frac{16 \pi d_{\rm eff,I} d_{\rm
eff,II}}{\lambda_1 n_1 \Delta k_{\rm SFG}}.
\]

One of the ways to separate the contribution of the two
nonlinearities and express of $\chi^{(3)}_{\rm eff,dir}$ in terms
of the product $(d_{\rm eff,I})(d_{\rm eff,II})$ is to use the
azimuthal (\cite{Banks:2002-102:JOSB}) or input polarization
(\cite{Kim:2002-33831:PRA}) dependence of $\chi^{(3)}_{\rm
eff,dir}$ and $\chi^{(3)}_{\rm eff,casc}$. The other way is to
compare the TH signal obtained under the condition $\Delta k_{\rm
THG}\longrightarrow0$ with the TH signal obtained under one of the
conditions $\Delta k_{\rm SHG} \longrightarrow 0$ or $\Delta k_{\rm
SFG}\longrightarrow0$, where the signal is only proportional to
the factor $|\chi^{(3)}_{\rm eff,casc}|^2$
(\cite{Akhmanov:1977-1710:ZETF, Chemla:1974-71:IQE}), and calculate
$\chi^{(3)}_{\rm eff,dir}$; however this procedure gives two possible values due to an indeterminate sign. Considering all symmetry classes
\cite{Feve:2002-63817:PRA} found the crystal directions for which
the second-order cascade processes give no contribution and,
therefore, they are suitable for the measurement of the value of
$\chi^{(3)}_{\rm eff,dir}$.

The parametric interaction that occurs when the degenerated FWM
(DFWM) process is in non-centrosymmetric media is special because
it consists of the steps of the optical rectification and linear
electro-optic effects (\cite{Bosshard:1995-2816:PRL,
Unsbo:1995-43:JOSB, Zgonik:1996-570:JOSB, Biaggio:1999-193:PRL,
Biaggio:2001-63813:PRA}). Then, the measured $\chi^{(3)}$
component can be expressed through the squared electro-optic
coefficient of the medium. The cascaded $\chi^{(3)}$ contribution
in the crystals with a large electro-optic effect leads to a very
strong cascaded DFWM effect which can exceed by many times the
contribution of the inherent direct $\chi^{(3)}$ nonlinearity
(\cite{Bosshard:1999-196:OL}).

Table~\rtab{chi3measure} presents a summary of the experimental
results on the measurement of the cubic nonlinearities by the use
of the cascaded THG or cascaded FWM processes.

\begin{small}
\begin{longtable}{p{2cm}p{0.9 cm}p{1.6cm}p{1.3cm}p{5cm}}
\caption{ \ltab{chi3measure} Experimental results for the
$\chi^{(3)}$-tensor components measured through the second-order
multistep cascading processes}\\

\hline
  Nonlinear  crystal
 & $\lambda_{fund}$ [$\mu m$] & PM/NPM/ NCPM$^\ast$
 & Cascading scheme & Reference \\\hline
\endfirsthead
\hline
  Nonlinear  crystal
 & $\lambda_{fund}$ [$\mu m$] & PM/NPM/ NCPM$^\ast$
 & Cascading scheme & Reference \\\hline
\endhead

 ADP & 1.06  &  PM &   THG  & \cite{Wang:1969-396:APL}
\\

  GaAs  &  10.6 &   NPM & FWM & \hspace{0mm}\cite{Yablonovitch:1972-865:PRL}
\\

 CdGeAs$_2$ &10.6  &   PM &  THG & \hspace{0mm}\cite{Chemla:1974-71:IQE}
\\

  KDP &1.064 &   PM &  THG & \hspace{0mm}\cite{Akhmanov:1977-1710:ZETF}
\\

   $\alpha$-quartz &   1.91 &   NPM & THG & \hspace{0mm}\cite{Meredith:1981-5522:PRB}
\\

      $\beta$-BBO & 1.054    & PM  & THG & \hspace{0mm}\cite{Qiu:1988-225:APB}
\\

$\beta$-BBO &1.053  &  PM  &  THG& \hspace{0mm}\cite{Tomov:1992-4172:AOP}
\\

KTP & 1.06  &  PM  &  DFWM& \hspace{0mm}\cite{Desalvo:1992-28:OL}
\\

$\beta$-BBO    KD*P   d-LAP  & 1.055
&    PM &  THG & \hspace{0mm}\cite{Banks:1999-4:OL, Banks:2002-102:JOSB}
\\

 KTP & 1.62 &     NCPM &   THG &
 \hspace{0mm}\cite{Feve:2000-1373:OL, Boulanger:1999-475:JPB}
\\

DAST$^\#$   & 1.064  &  PM &  DFWM  & \hspace{0mm}\cite{Bosshard:1999-196:OL}
\\

    $\alpha$-quartz  KNbO$_3$  KTaO$_3$  SF59 BK7 \mbox{fused silica}
  & 1.064
1.318
1.907
2.100 &
NPM

 &
 THG
  &
  \hspace{0mm}\cite{Bosshard:2000-10688:PRB}
\\
 KNbO$_3$ DAST BaTiO$_3$ &  1.06 &    PM &  DFWM  &
 \hspace{0mm}\cite{Biaggio:1999-193:PRL, Biaggio:2001-63813:PRA}
\\

AANP$^\#$  &  1.390
1.402&    NCPM &   FWM & \hspace{0mm}\cite{Taima:2003-83:OMT}
\\

KDP BBO LiNbO$_3$ &  1.064 0.532 &   NPM &  DFWM  &
 \hspace{0mm}\cite{Ganeev:2003-561:OPS}
\end{longtable}
~\\
$^\ast$ PM~-- phase matched; NPM~-- non phase-matched; NCPM~-- noncritically phase-matched;\\
$^\#$ organic crystals.
\end{small}

\section[Phase-matching of cascaded interactions]{Phase-matching for multistep cascading} \lsect{matching}

In general, the simultaneous phase-matching of several parametric
processes is hard to achieve by using the traditional
phase-matching methods, such as those based on the optical
birefringence effect, except some special cases discussed above.
However, the situation becomes quite different for the nonlinear
media with a periodic variation of the sign of the quadratic
nonlinearity, as it occurs in the fabricated one-dimensional (1D)
QPM structures (\cite{Fejer:1992-2631:IQE}) or two-dimensional
(2D) $\chi^{(2)}$ nonlinear photonic crystals
(\cite{Berger:1998-4136:PRL, Broderick:2000-4345:PRL,
Saltiel:2000-1204:OL}). In this part of the review, we describe
the basic principles of the simultaneous phase-matching of two (or
more) parametric processes in different types of 1D and 2D
nonlinear optical superlattices.

\subsection{Uniform QPM structures } \lsect{QPM}

\pict{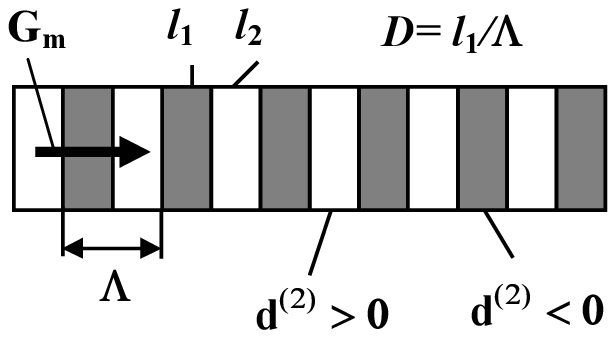}{QPM}{Schematic of the uniform QPM structure: $G_m$ is
the reciprocal vector of the $m$-th order; $D$ is the filling
factor, and $\Lambda$ is the period of the QPM structure. }

A bulk homogeneous nonlinear crystal possessing a quadratic
nonlinearity is usually homogeneous everywhere. Several methods
 have been suggested and employed (\cite{Fejer:1998-375:BeamShaping})
for creating a periodic change of the sign of the second-order
nonlinear susceptibility $d^{(2)}$, as occurs in the QPM structure
shown in Fig.~\rpict{QPM}. From the mathematical point of view,
such a periodic sequence of two domains can be described by a
simple periodic function,
\begin{equation} \leqt{SUMd(z)}
   d(z) = d_0 \sum_{m \neq 0} g_m e^{i G_m z},
\end{equation}
\begin{equation} \leqt{gm}
   g_m = \left(\frac{2}{m \pi}\right) \sin(\pi m D),
\end{equation}
where $G_m = (2\pi m/\Lambda)$ is the reciprocal QPM vector. The
uniform QPM structure is characterized by a set of the reciprocal
vectors: $\pm 2\pi/\Lambda, \pm 4\pi/\Lambda, \pm 6\pi/\Lambda,
\pm 8\pi/\Lambda, ...$, which can be used to achieve the
phase-matching conditions when $\Delta k\longrightarrow 0$. The
integer number $m$ (that can be both \textit{positive} and
\textit{negative}) is called \textit{the order of the wave vector
phase-matching}. According to Eq.~\reqt{SUMd(z)}, the smaller is
the order of the QPM reciprocal wave vector, the larger is the
effective nonlinearity. If the filling factor $D = 0.5$, the
effective quadratic nonlinearities (proportional to the parameter
$d_0 g_m$) that correspond to the even orders QPM vectors vanish.
Importantly, such uniform QPM structures can be used for
simultaneous phase-matching of two parametric processes when the
interacting waves are collinear or non-collinear to the reciprocal
wave vectors of the QPM structure.

\subsubsection{Collinear case with two commensurable periods}
        \lsect{collinear}

As an example of the multistep QPM interaction, we consider the
third-harmonic multistep-cascading process under the condition
that the interacting waves are {\em collinear} to the reciprocal
wave vectors of the QPM structure. We denote the mismatches of the
nonlinear material without modulation of the quadratic
nonlinearity ("bulk mismatches") as $\Delta b_1$ and $\Delta b_2$,
where $\Delta b_1=k_2-2k_1$ and $\Delta b_2=k_3-k_2-k_1$, and
choose the period of the QPM structure in order to satisfy the
phase-matching conditions $G_{m_1}=-\Delta b_1$ and
$G_{m_2}=-\Delta b_2$ . The two parametric processes, SHG and SFG,
are characterized by the wave-vector mismatches $\Delta
k_1=k_2-2k_1+G_{m_1}$ and $\Delta k_2=k_3-k_2-k_1+G_{m_2}$,
respectively, and they become simultaneously phase-matched for
this particular choice of the QPM period. A drawback of this
method is that it can satisfy simultaneously two phase-matching
conditions for discrete values of the optical wavelength only. In
particular, the values of the fundamental wavelength $\lambda$ for
the double phase-matching condition can be found from the relation
\begin{equation} \leqt{EqColl}
   \Delta b_2 / m_2 - \Delta b_1 / m_1 = 0,
\end{equation}
where both $\Delta b_1$ and $\Delta b_2$ are functions of the
wavelength. Equation~\reqt{EqColl} is valid for any pair of the
second-order parametric processes. For the third-harmonic
multistep-cascading process, Eq.~\reqt{EqColl} is transformed to
the following,
\begin{equation} \leqt{EqColl2}
   m_1 \left[3 n(3\omega) - 2 n(2\omega) - n(\omega) \right]
        - 2 m_2 \left[n(2\omega) - n(\omega) \right] = 0,
\end{equation}
where the arguments shows the wavelength dependence of the
refractive index. For a chosen pair of integer numbers
$(m_1,m_2)$, the required QPM period is found from the relation
$\Lambda=2\pi|m_1/\Delta b_1|$ or $\Lambda=2\pi|m_2/\Delta b_2|$.
Such a method was used for double phase-matching by many
researchers for the cascaded single-crystal third- and
fourth-harmonic generation (see details and references in
Table~\rtab{THG}). The highest efficiency achieved with the
uniform QPM structure so far is 19.3\%, and it is achieved by the
use of the combination of SHG (1st-order)/SFG(3rd-order) and is
reported in \cite{He:2002-944:CHIL}.

\subsubsection{Non-collinear case} \lsect{noncollinear}

Collinearity between the optical waves and the reciprocal vectors
of the QPM structure is an important requirement for achieving a
good overlapping of all the beams and a good conversion
efficiency. However, phase-matching is possible even in the case
when some of the waves propagate under a certain angle to the
direction of the reciprocal vectors of the QPM structure, i.e. for
the non-collinear case. With this method, the double
phase-matching processes can be realized in a broad spectral range
(\cite{Saltiel:2000-57:BJP}). Such a type of non-collinear
interaction will be efficient for the distances corresponding to
the overlap of the interacting beams.

\pict[1]{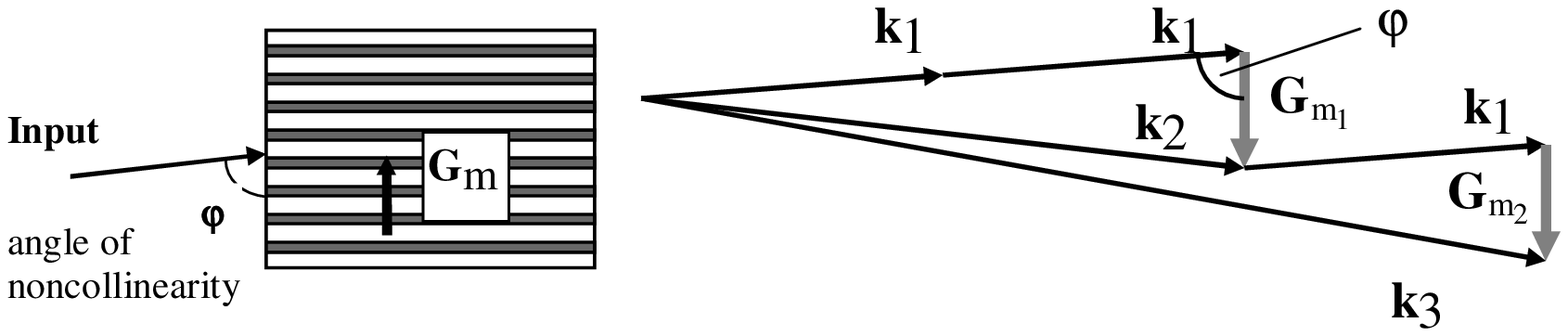}{THGangleQPM}{Geometry of the noncollinear
cascaded THG parametric process in an uniform QPM structure. }

As an example, we consider the THG multistep parametric process in
the noncollinear geometry as shown in  Fig.~\rpict{THGangleQPM}.
in this case, the simultaneous phase-matching of two parametric
processes $\omega+\omega=2\omega$ and $\omega+2\omega=3\omega$ is
required. We assume the fundamental wave at the input. As shown in
Fig.~\rpict{THGangleQPM}, for the first process the phase-
matching can be achieved by the use of the the reciprocal vector
$\mathbf{G}_{m_1}$, and the generated SH wave with the wavevector
$\mathbf {k_2}$ is not collinear to the fundamental wave:
$2\mathbf{k}_1-\mathbf{G}_{m_1}=\mathbf{k}_2$. The second process
can be phase-matched by using the vector $\mathbf{G}_{m_2}$:
$\mathbf{k}_1+\mathbf{k}_2-\mathbf{G}_{m_2}=\mathbf{k}_3$. From
Fig.~\rpict{THGangleQPM}, we can derive a result for the period of
the QPM structure that allows achieving the double phase-matching
of the processes of cascaded THG in a single QPM structure,
\begin{equation} \leqt{EqNoncol}
   \Lambda = 2 \pi \left[ \frac{2 m_1(k_3^2 - 9 k_1^2)
                                  - 3(m_1 + m_2) (k_2^2 - 4 k_1^2) }{
                               m_1(m_1 + m_2)(2 m_2 - m_1) }\right]^{1/2}.
\end{equation}
The conditions for the double phase-matching for the THG
multistep-cascading process in LiTaO$_3$ for the case when all
waves are polarized along the $z$ axis of the crystal were
considered by \cite{Saltiel:2000-57:BJP}.

\subsection{Non-uniform QPM structures} \lsect{QPMnonun}

Non-uniform QPM structures also allow the simultaneous
phase-matching of two parametric nonlinear processes. We consider
three types of such non-uniform QPM structures: (i) phase-reversed
QPM structures (\cite{Chou:1999-1157:OL}), (ii) periodically
chirped QPM structures (\cite{Bang:1999-1413:OL}), and (iii)
optical superlattices (\cite{Zhu:1999-1093:OQE,
Fradkin-Kashi:2002-23903:PRL}).

\subsubsection {Phase-reversed QPM structures} \lsect{QPMrevers}

\pict{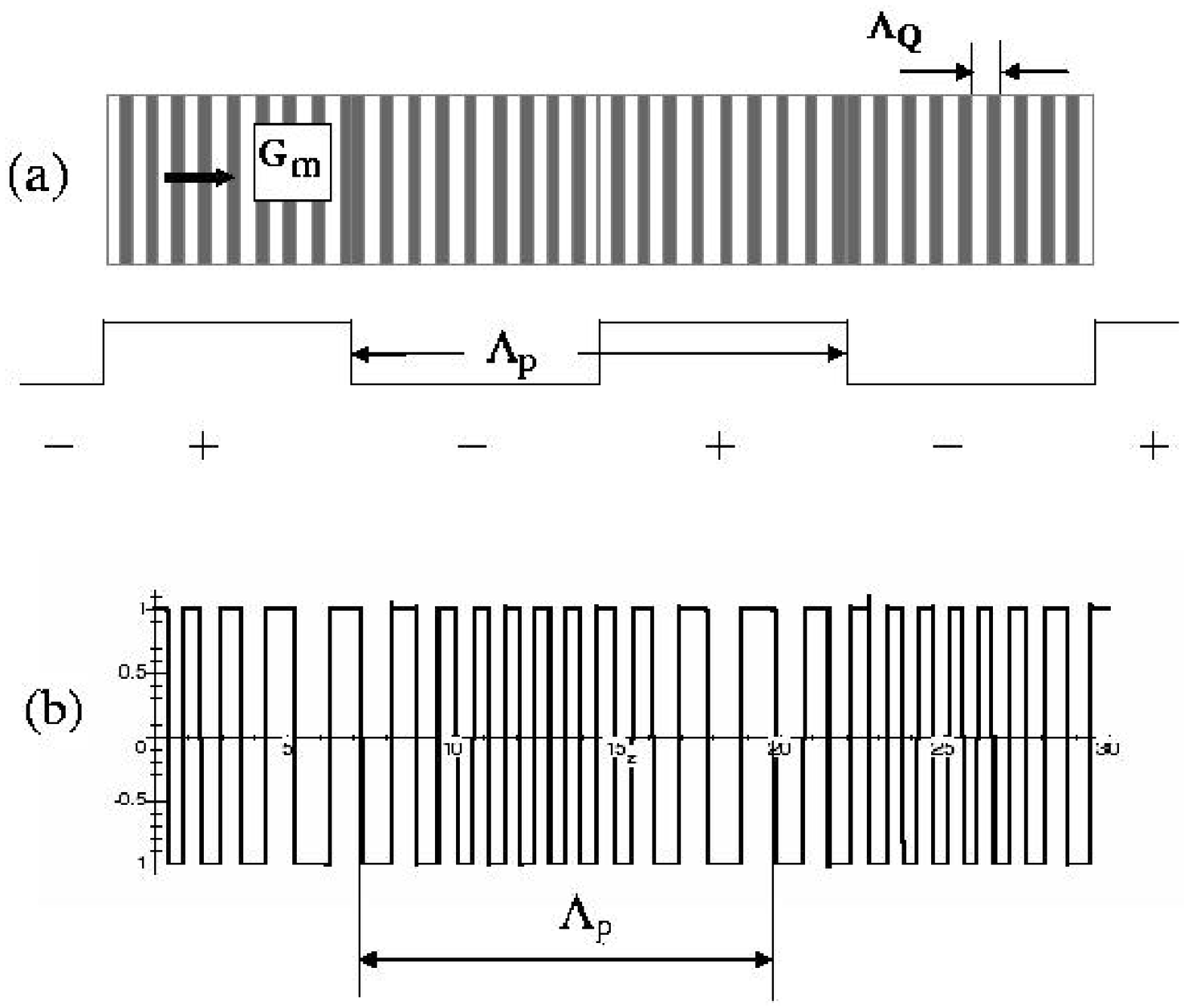}{QPMnonun}{Schematic of the non-uniform QPM
structures: (a)~phase-reversed QPM structure
(\cite{Chou:1999-1157:OL}), and (b)~periodically-chirped QPM
structure (\cite{Bang:1999-1413:OL}). }

The idea of {\em the phase-reversed QPM structures}
(\cite{Chou:1999-1157:OL}) is illustrated in
Fig.~\rpict{QPMnonun}(a). Such a structure can be explained as a
sequence of many equivalent uniform short QPM sub-structures with
the length $\Lambda_{\rm ph}/2$ connected in such a way that at
the place of the joint two end layers have the same sign of the
quadratic nonlinearity. Any two neighboring junctions have the
opposite signs of the $\chi^{(2)}$ nonlinearity. In other words,
the phase-reversed QPM structure can be thought of as a uniform QPM
structure with a change of the domain phase by $\pi$ characterized
by the second (larger) period $\Lambda_{\rm ph}$. Modulation of
the quadratic nonlinearity $d(z)$ in the phase-reversed QPM
structure with the filling factor $D=0.5$ can be described by the
response function
\begin{equation} \leqt{EqPhRev1}
   d(z) = d_0 (-1)^{{\rm int}(2z/\Lambda_Q)} (-1)^{{\rm int}(2z / \Lambda_{\rm
   ph})},
\end{equation}
and it can be expanded into the Fourier series,
\begin{equation} \leqt{EqPhRev2}
   d(z) = d_0 \sum_{l = -\infty}^{+\infty} g_l e^{iG_l z}
              \sum_{m = -\infty}^{+\infty} g_m e^{iF_m z}
        = d_0 \sum_{l = -\infty}^{+\infty} \sum_{m = -\infty}^{+\infty} g_{lm} e^{iG_{lm}z},
\end{equation}
where $g_0=0$, $g_{l \neq 0} = (2/\pi l)$, $g_{m \neq 0} = (2/\pi
m)$, $g_{lm}=g_l g_m$, $G_{l}=(2\pi/\Lambda_Q)l$, $F_{m}=
(2\pi/\Lambda_{\rm ph})m$, and $G_{lm}=(2\pi/\Lambda_Q)l +
(2\pi/\Lambda_{\rm ph})m$.

The phase-reversed QPM structure is characterized by a set of the
reciprocal vectors $\{G_{lm}\}$, which are collinear to the normal
of the periodic sequence with the magnitude depending on all
parameters: $l, m, \Lambda_Q, \Lambda_{\rm ph}$. Two of these
vectors can be chosen to phase-match two parametric processes
involved into the multistep-cascading, such that
$G_{{l_1}{m_1}}=-\Delta b_1$ and $G_{{l_2}{m_2}}=-\Delta b_2$ .
Then, the two QPM periods satisfying the double-phase-matching
condition can be found as follows,
\begin{equation} \leqt{EqPhRev3}
   \Lambda_Q = \left|\frac{2 \pi (l_2 m_1 - l_1 m_2)}{
                      m_2 \Delta b_1 - m_1 \Delta b_2}\right|,
      \qquad
   \Lambda_{\rm ph} = \left|\frac{2 \pi (l_2 m_1 - l_1 m_2)}{
                         l_2 \Delta b_1 - l_1 \Delta b_2}\right|,
\end{equation}

For the case of the THG multistep cascading process,
Eqs.~\reqt{EqPhRev3} are transformed into
\begin{equation} \leqt{EqPhRev4}
   \Lambda_Q = \left|\frac{\lambda (l_2 m_1 - l_1 m_2)}{
                      (m_1 - 2 m_2) n(\omega) + 2 (m_1 + m_2) n(2\omega)
                       - 3 m_1 n(3\omega)}\right|,
\end{equation}
\begin{equation} \leqt{EqPhRev5}
   \Lambda_{ph} = \left|\frac{\lambda (l_2 m_1 - l_1 m_2)}{
                    (l_1 - 2 l_2) n(\omega) + 2 (l_1 + l_2) n(2\omega)
                    - 3 l_1 n(3\omega)}\right|,
\end{equation}

In addition to Eqs.~\reqt{EqPhRev4} and~\reqt{EqPhRev5}, the
design shown in Fig.~\rpict{QPMnonun}(a) impose an additional
condition that the ratio $2\Lambda_{\rm ph}/\Lambda_{Q}$ is an
integer number. Nevertheless, the corresponding number of the
phase-matched wavelengths is larger than that achieved in the
uniform QPM structure for the collinear geometry. The
phase-reverse grating with arbitrary ratio $2\Lambda_{\rm
ph}/\Lambda_Q$ was considered theoretically by
\cite{Johansen:2002-393:OC} in the study of spatial parametric
solitons. The study of the effect of arbitrary $2\Lambda_{\rm
ph}/\Lambda_{Q}$ on the efficiency of different types of the
frequency conversion processes has not been investigated yet, to
the best of our knowledge. Experimentally, the phase-reversed QPM
structure was employed for the wavelength conversion
(\cite{Chou:1999-1157:OL}), and for the cascaded single-crystal
THG process (\cite{Liu:2001-6841:JJAP, Liu:2002-1676:JOSB}).

\subsubsection {Periodically chirped QPM structures} \leqt{QPMchirped}

{\em Periodically chirped QPM} structures, presented schematically
in Fig.~\rpict{QPMnonun}(b), have initially been suggested and
designed for increasing an effective (averaged) third-order
nonlinearity in quadratic media (\cite{Bang:1999-1413:OL}). We use
the terminology "periodically chirped QPM" in the analogy with the
chirped QPM structures that has a linear growth of the period
$\Lambda_{Q}$ along the structure. This type of structure is very
suitable for realizing the multiple phase-matching conditions. The
properties of the periodically chirped QPM structures have been
explored experimentally for achieving a larger bandwidth for the
wavelength converters (\cite{Asobe:2003-558:OL,
Gao:2004-557:IPTL}).

The periodically chirped QPM structure is characterized by the
modulated QPM period $\Lambda_Q$ that is itself a periodic
function of $z$ [see Fig.~\rpict{QPMnonun}(b)]:
\begin{equation} \leqt{EqPerChir1}
   \Lambda = \Lambda_Q + \varepsilon_0 \cos \left(\frac{2 \pi z}{ \Lambda_{\rm ch}}\right),
\end{equation}
Double phase-matching conditions written for this type of
structure allow to find two periods of the periodically chirped
QPM structure; the results are similar to the formulas that
describe the phase-reversed QPM structures,
\begin{equation} \leqt{EqPerChir2}
   \Lambda_Q = \left|\frac{2 \pi (l_2 m_1 - l_1 m_2)}{
                      m_2 \Delta b_1 - m_1 \Delta b_2}\right|,
\end{equation}
\begin{equation} \leqt{EqPerChir3}
   \Lambda_{\rm ch} = \left|\frac{2 \pi (l_2 m_1 - l_1 m_2)}{
                         l_2 \Delta b_1 - l_1 \Delta b_2}\right|,
\end{equation}
The main advantage of the use of the periodically-chirped QPM
structures is the possibility to satisfy the double phase-matching
conditions for any wavelength in a broad spectral range, i.e. for
any pair of $\Delta b_1$ and $\Delta b_2$.

\subsubsection{Quasi-periodic and aperiodic optical superlattices }
              \lsect{aperiodic}

\pict[0.4]{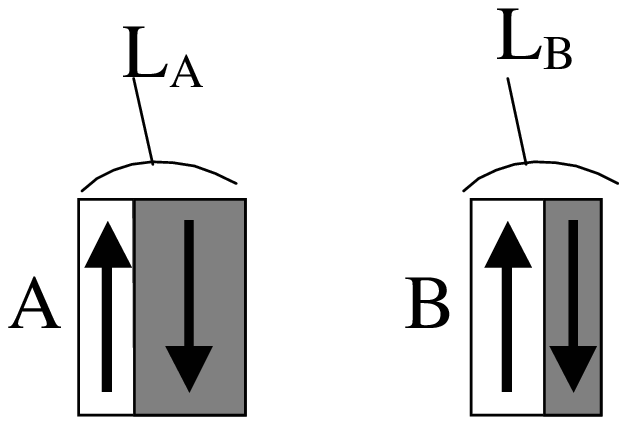}{QPMblocks}{Basic blocks of the
quasi-periodic and aperiodic QPM superlattices. }

Another method of the double phase-matching, studied extensively
both theoretically and experimentally, is based on the use of the
quasi-periodic optical superlattices (QPOS)
(\cite{Zhu:1999-1093:OQE, Fradkin-Kashi:1999-1649:IQE,
Fradkin-Kashi:2002-23903:PRL}) and aperiodic optical superlattices
(\cite{Gu:2000-7629:JAP}). In the most of the cases, QPOSs are
built with two two-component blocks A and B, as shown in
Fig.~\rpict{QPMblocks}, which are aligned in a Fibonacci-like or
more general quasi-periodic sequence. Importantly, each of the
blocks consists of two layers with the opposite sign of the
quadratic nonlinearity. In order to illustrate the possibility of
the double phase-matching in such structures we take, as an
example, the structure consisting of two blocks aligned in a
generalized sequence (\cite{Fradkin-Kashi:1999-1649:IQE}).
Modulation of the quadratic nonlinearity can be described by the
following Fourier expansion (\cite{Birch:1990-10398:PRB}):
\begin{equation} \leqt{QPOS1}
   d(z) = d_0 \sum_{m,n} f_{m,n} e^{iG_{m,n} z},
\end{equation}
where the reciprocal vectors are defined as $G_{m,n} =
2\pi(m+n\tau)/S$, where $S=\tau L_A+ L_B$.

For phase-matching of two parametric processes with the mismatch
parameters $\Delta b_1$ and $\Delta b_2$, we solve the system of
equations
\begin{equation} \leqt{QPOS2}
   \begin{array}{l} {\displaystyle
      G_{m_1,n_1} = 2 \pi (m_1 + n_1 \tau) / S = -\Delta b_1,
   } \\*[9pt] {\displaystyle
      G_{m_2,n_2}=2\pi(m_2+n_2\tau)/S=-\Delta b_2,
   } \end{array}
\end{equation}
and find the corresponding value of $S$ and $\tau$. The lengths of
the blocks $L_A$ and $L_B$ and a ratio of the sub-layers in each
block should be found by maximizing $f_{m_1,n_1}$ and
$f_{m_2,n_2}$. A resulting designed structure allows simultaneous
phase-matching in a broad spectral range without constrains on the
ratio $\Delta b_2/\Delta b_1$. Equations~\reqt{QPOS2} are also
valid for the Fibonacci-type QPOS lattices but, because the
parameter $\tau$ is fixed, the double phase-matching conditions
can be satisfied for a limited number of wavelengths
(\cite{Fradkin-Kashi:1999-1649:IQE}) defined from the equation
$\Delta b_1 / \Delta b_2 = (m_1 + n_1 \tau)/(m_2 + n_2 \tau)$.

In several papers \cite{Zhao:2003-1882:JAP, Gu:2000-7629:JAP}, and
\cite{Gu:1999-2175:APL} studied aperiodic one-dimensional optical
superlattices for the simultaneous phase-matching of several
parametric processes. In this case, the thickness of the layers
and their order can be found by solving an inverse problem
maximizing the efficiency of both the processes involved into the
multistep-cascading parametric interaction. Experimental results
that employed quasi-periodic and aperiodic optical lattices for
realizing the multistep parametric interactions are given in
Tables~\rtab{THG}, \rtab{OPO}, and \rtab{RGB}.

\subsection{Quadratic 2D nonlinear photonic crystals}
         \lsect{2D}

Nonlinear photonic crystals (NPCs) with a homogeneous change of
the linear refraction index and a two-dimensional (2D) periodic
variation of the nonlinear quadratic susceptibility have been
suggested by \cite{Berger:1998-4136:PRL, Berger:1999-366:ConfinedPhoton} as a 2D analog of the QPM structures. Both theoretical and experimental results published so far show that this kind of structures can  effectively be employed as a host media for realizing many different types of
multistep-cascading parametric processes (see, e.g., \cite{Broderick:2000-4345:PRL, Saltiel:2000-1204:OL,
deSterke:2001-539:OL, Chowdhury:2000-832:OL}). Below, we discuss
some of the possible application of these 2D structures for
multistep parametric interaction and multi-frequency generation.

\subsubsection{Phase-matching in two-dimensional QPM structures}
              \lsect{2Dmatching}

\pict[1]{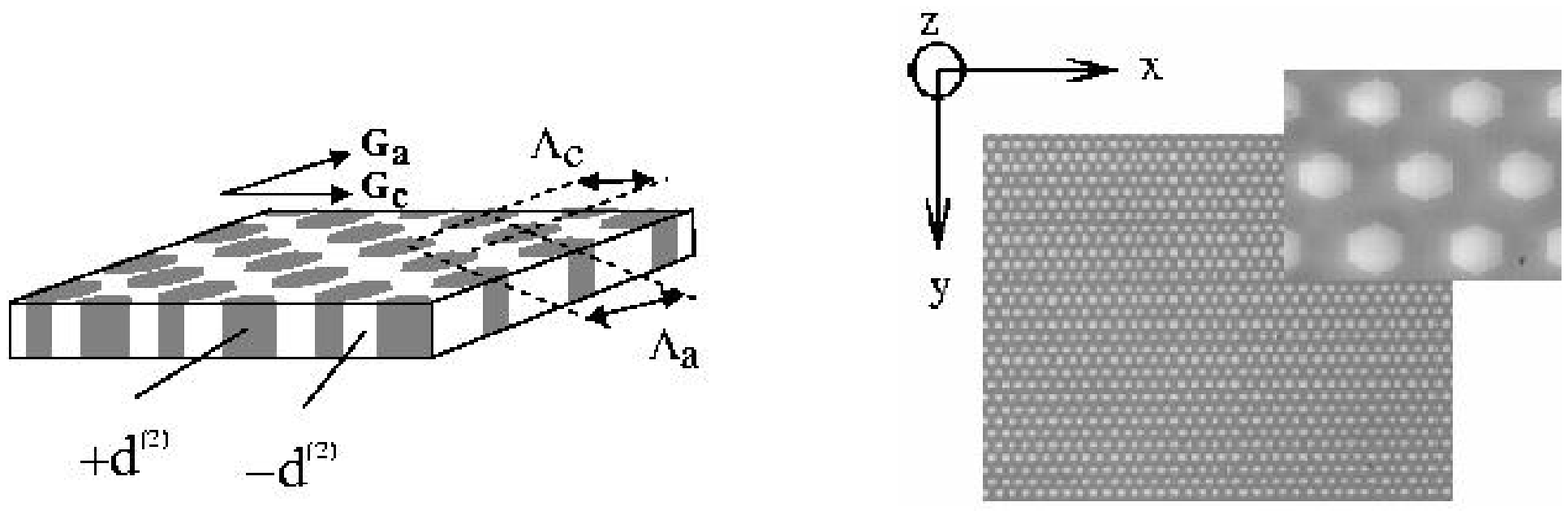}{NPC2d}{ Left: 2D nonlinear quadratic photonic
crystal composed of a periodic lattice of domains (gray) with the
reversed sign of $\chi^{(2)}$; $\mathbf{G_a}$ and $\mathbf{G_b}$
are the 2D reciprocal vectors. Right: Fabricated 2D
hexagonally-poled LiNbO$_3$ structure with the period 18.05 $\mu$m
(\cite{Broderick:2000-4345:PRL}). }

A schematic structure of 2D NPC is shown in Fig.~\rpict{NPC2d}. A
simple way to obtain the phase-matching conditions for 2D NPC is
to use a reciprocal lattice formed by the vectors $\mathbf{G}_a$
and $\mathbf{G}_c$, defined as $|G_a|=2\pi/\Lambda_a$ and
$|G_c|=2\pi/\Lambda_c$ . For a hexagonal lattice, $\Lambda_a=
\Lambda_c=a \sqrt{3}/2$, where $a$ is the distance between the
centers of two neighboring inverted volumes, the so-called lattice
spacing. All reciprocal vectors of the 2D NPC crystal are formed
by a simple rule, $\mathbf{G}_{m,n}=m\mathbf{G}_c+n\mathbf{G}_a$.
Any two vectors of this set can be used to compensate for the bulk
mismatch parameters $\Delta b_1$ and $\Delta b_2$, however, the
phase-matching conditions require, in most of the cases,
non-collinear parametric interactions.

The diagrams for calculating the phase-matching conditions for the
THG multistep-cascading process are presented in
Fig.~\rpict{THG2d} for (a) the process $\omega+ \omega=2\omega$
which is phase-matched by the reciprocal vector
$\mathbf{G}_{m_1,n_1}$, and (b) the process $\omega+
2\omega=3\omega$, which is phase-matched by the reciprocal vector
$\mathbf{G}_{m_2,n_2}$. Phase-matching is achieved by choosing the
lattice spacing $a$ and the angle of incidence $\beta$. 2D NPC can
be also used for simultaneous phase-matching of three nonlinear
processes, e.g. second-, third-, and fourth-harmonic generation
(\cite{Saltiel:2000-1204:OL}), or the generation of a pair of SH
waves and the third- (\cite{Karaulanov:2003-114:ProcSPIE}) or
fourth- (\cite{deSterke:2001-539:OL}) harmonic generation.
Experimentally, the simultaneous second-, third- and
fourth-harmonic generation was recently observed in 2D poled bulk
LiNbO$_3$ (\cite{Broderick:2000-4345:PRL} and
\cite{Broderick:2002-2263:JOSB}). The first experiment with
phase-matching in 2D NPC made in a poled LiNbO$_3$ waveguide slab
was reported in (\cite{Gallo:2003-75:ELL}). As shown in
\cite{He:2003-9943:JAP}, the method of a direct electron-beam
lithography can also be used for creating 2D nonlinear photonic
domain-reversed structures.

\pict{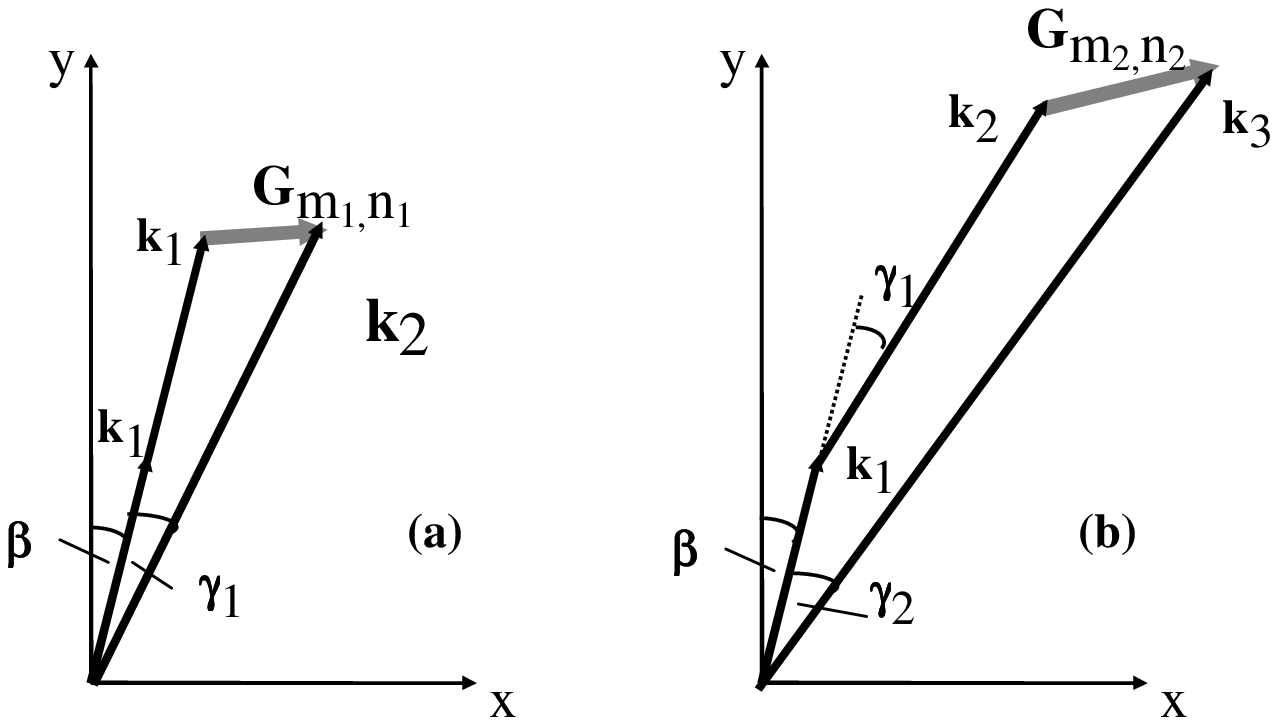}{THG2d}{(a,b) Diagrams of the double phase-matching
conditions for single-crystal THG in 2D NPC structures.}

The shapes of the inverted domain is an important property of the
2D NPC structures, and they can be employed for an effective
optimization of the parametric conversion processes. Optimization
of the domain shapes for the maximum efficiency of THG and FHG
processes was reported by \cite{Norton:2003-1008:OE}. For the case
of SHG in \cite{Lee:2003-194:ProcLEOS} the following three types
of the domain shapes were compared: hexagonal, circular, and
elliptical (aligned to the direction of SHG). The numerical
simulation made by the authors revealed that the
elliptically-poled domain pattern yields the highest frequency
conversion efficiency among these three types of the poling
structures. The role of the filling factor for the case of SHG was
studied both theoretically and experimentally in
\cite{Wang:2001-323:EPB,Ni:2003-4230:APL}.

\subsubsection{Multi-channel harmonic generation}
             \lsect{2Dharmonic}

\pict[1]{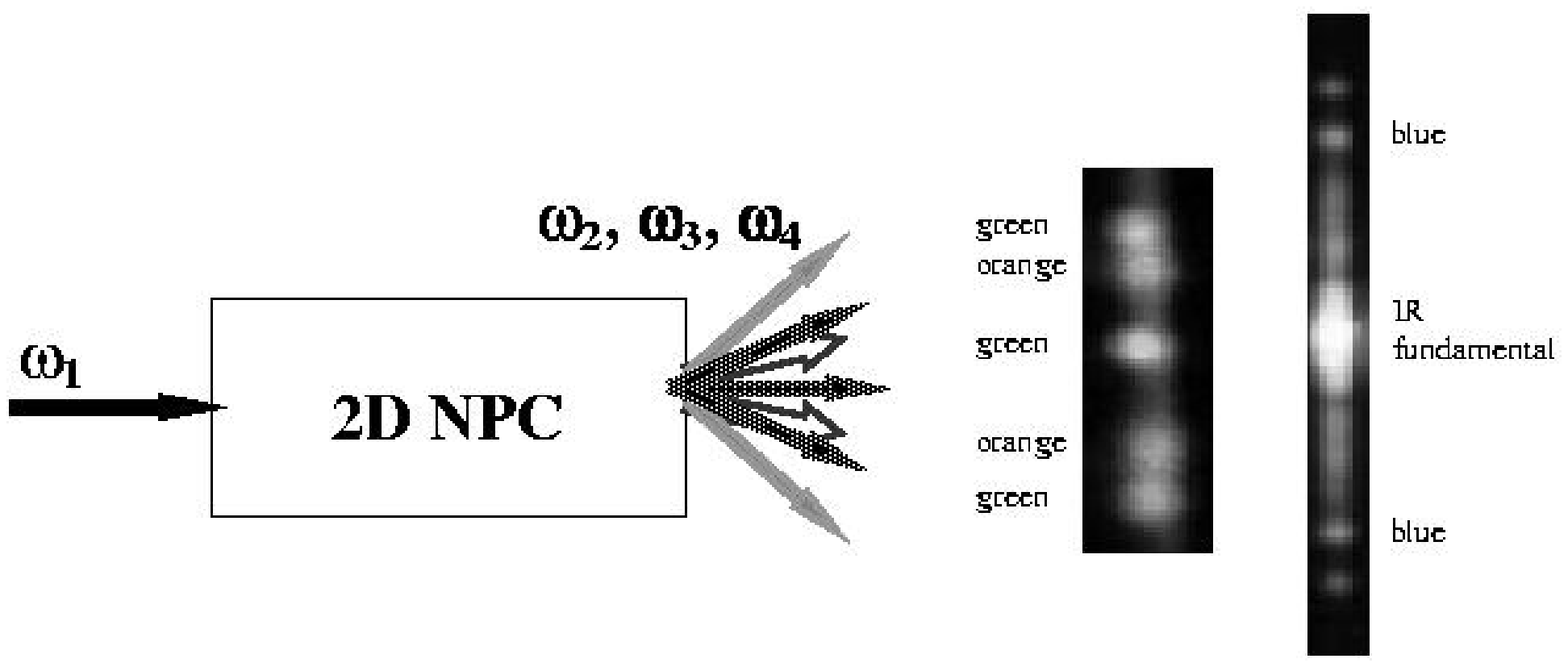}{harmonic2d}{Multiple second- ($\omega_2$),
third- ($\omega_3$), and fourth- ($\omega_4$) harmonic generation
schemes realized in a 2D NPC structure; orange~--- SHG (766 nm),
green~--- THG (511nm), blue~--- FHG (383 nm)
(\cite{Broderick:2000-4345:PRL}). }

In the first experimental work on the harmonic generation in the
2D NPC structures (\cite{Broderick:2000-4345:PRL} and
\cite{Broderick:2002-2263:JOSB}), it was noticed that each
generated harmonic has multiple outputs at different angles (see
Fig.~\rpict{harmonic2d}). The reason for that is that each
harmonic can be generated by using several different
phase-matching conditions. In several papers, an efficient method
was suggested to make use of this property and to combine the multiple
outputs in order to generate efficient harmonics with the
additional advantage of being collinear to the pump. A
phase-matching geometry for THG in 2DNPC, for which generated TH
wave is collinear to the input wave, was suggested
in~\cite{Karaulanov:2003-114:ProcSPIE}. This scheme uses the fact
that the TH wave is generated though two different channels
leading to the factor four improvement in comparison with the conventional collinear scheme.

A geometry for the multiple phase-matching of this THG
multichannel parametric interaction is presented in
Figs.~\rpict{channels2d}(a,b). The process starts with the
generation of a pair of the SH waves by employing the wave vectors
$\mathbf{k}_{2}'$ and $\mathbf{k}_{2}''$. The phase-matching
conditions are satisfied by two symmetric reciprocal vectors
$\mathbf{K}_{m,n}$ and $\mathbf{K}_{m,-n}$. Each SH wave interacts
again  with the fundamental wave [see Fig.~\rpict{channels2d}(b)]
via a pair of the phase-matched interaction with participation of
the reciprocal vectors $\mathbf{K}_{p,q}$ and $\mathbf{K}_{m,-q}$,
thus generating a pair of the collinear TH waves with one and the
same wave vector $\mathbf{k}_{3}$. The two TH waves interfere
constructively in the direction of the input fundamental wave,
resulting in  the overall higher TH efficiency (in the nondepleted
regime).

\pict[1]{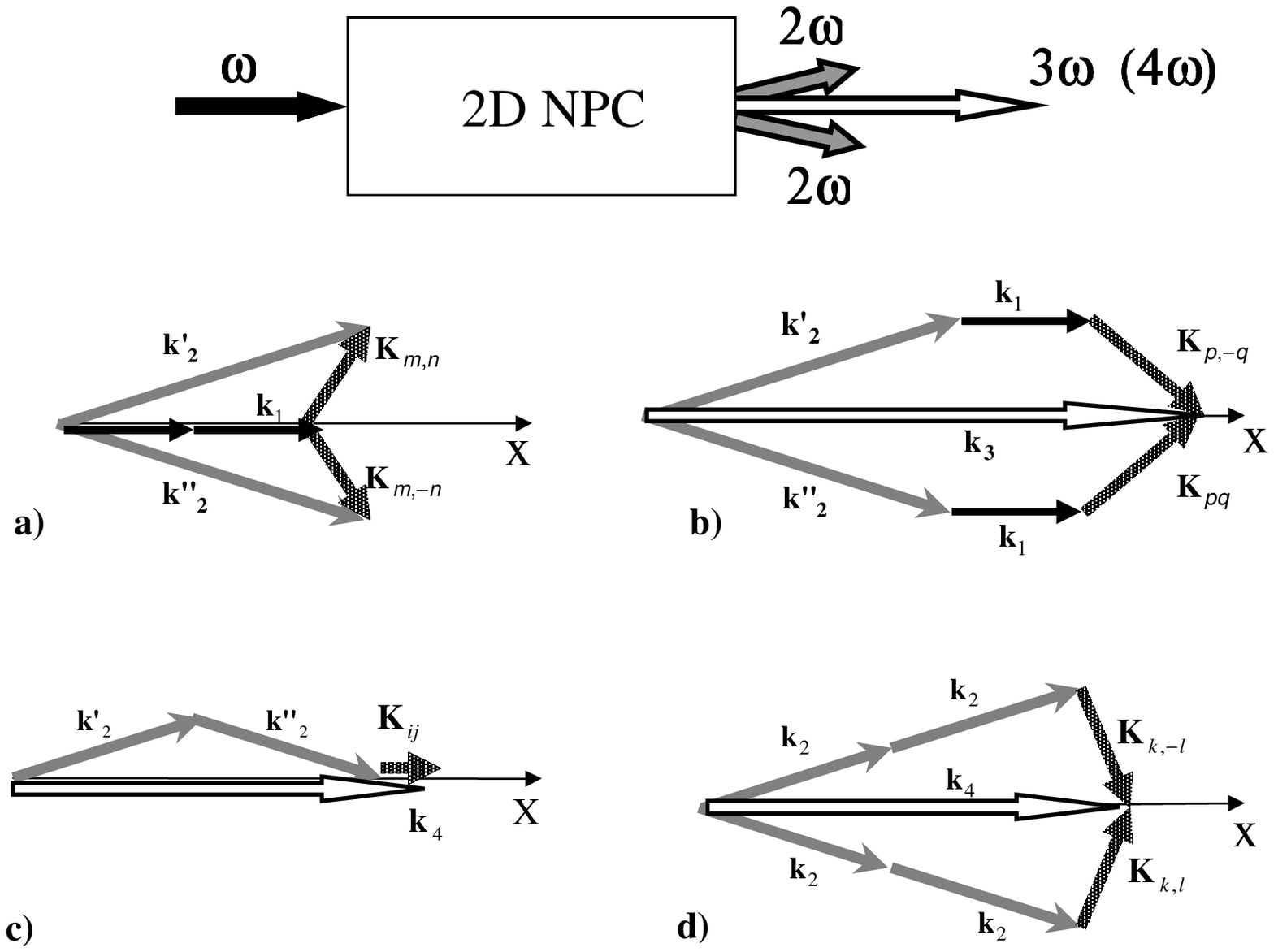}{channels2d}{ Multi-channel the third- and
forth-harmonic generation in 2D NPC structures: (a) the step
$\omega + \omega = 2\omega$; (b) the step $\omega + 2\omega =
3\omega$ for THG; (c,d) the step $2\omega + 2\omega = 4\omega$ for
FHG; $\mathbf{K}_{p,q}$, $\mathbf{K}_{i,j}$, $\mathbf{K}_{m,n}$,
and $\mathbf{K}_{k,l}$ are reciprocal wave-vectors of the lattice.
}

A scheme for a single-crystal FHG process in a 2D QPM structure
where the fundamental and FH waves are collinear was studied in
\cite{deSterke:2001-539:OL}. The parametric interaction includes
cascading of three phase-matched second-order processes that are
simultaneously phase-matched. \cite{deSterke:2001-539:OL}
demonstrated that the FHG efficiency achieved by this scheme is
larger than the comparable schemes by a factor that reaches four
at low input intensities. The phase-matching geometry for this
process is shown in Figs.~\rpict{channels2d}(c,d). Two SH waves
are generated at the first step [see Fig.~\rpict{channels2d}(a)],
and they interact non-collinearly generating a phase-matched FH
wave, which itself is collinear with the fundamental wave [see
Fig.~\rpict{channels2d}(c)]. As shown in
\cite{Norton:2003-188:OL}, there exist two more FHG channels that
are simultaneously phase-matched also leading to the generated FH
waves collinear to the input fundamental wave [see
Fig.~\rpict{channels2d}(d)]. Thus, in this case the generated FH
wave is produced in result of an constructive interference of
three FH waves generated at different phase-matching conditions,
but propagating in the same direction.

\subsubsection{Wave interchange and signal deflection}
          \lsect{2Dinterchange}

\pict{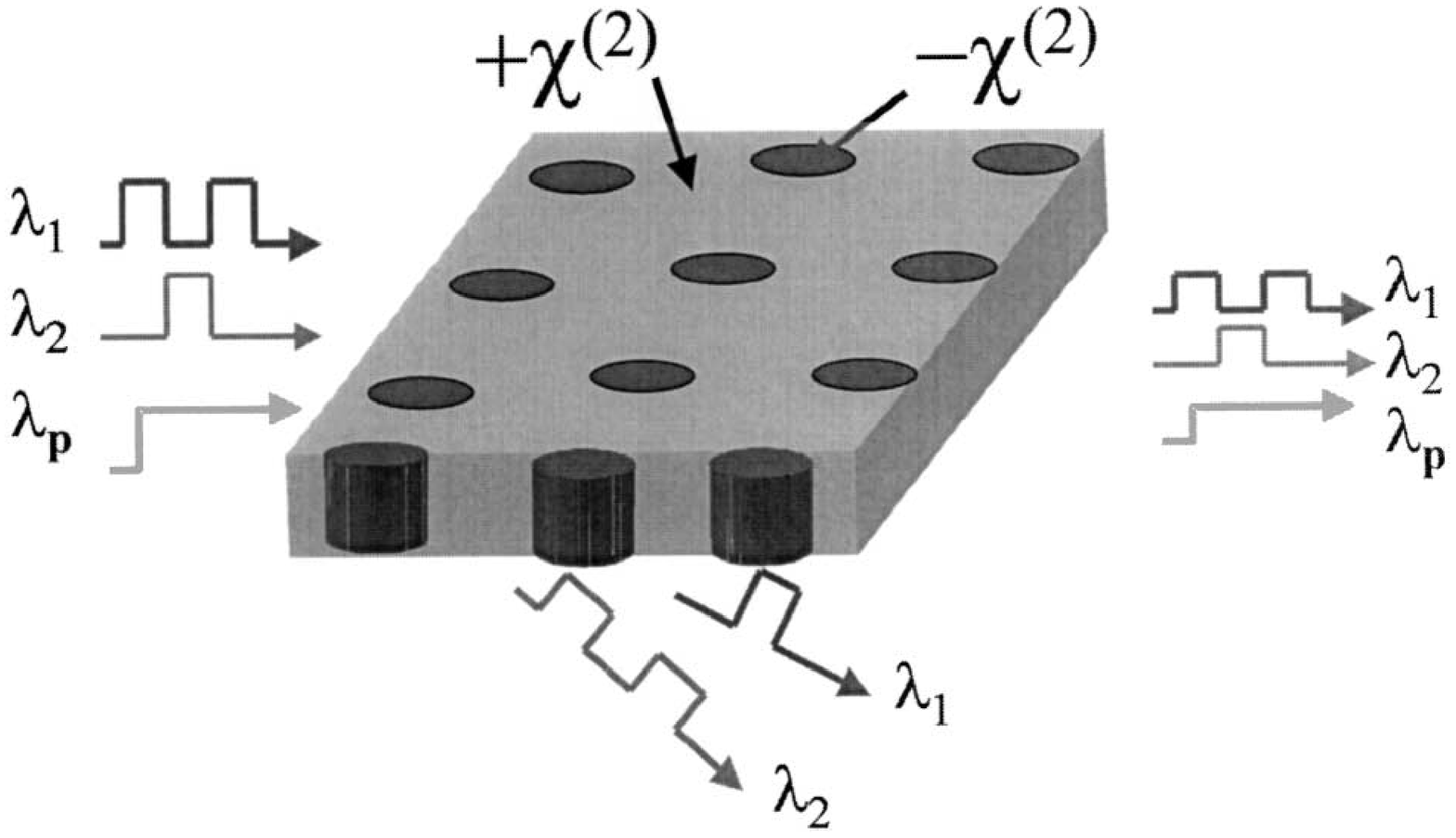}{interchange}{ Schematic of the simultaneous
optical wavelength interchange in a 2D NPC structure
(\cite{Chowdhury:2001-1353:OL}). }

The first experimental demonstration of the simultaneous optical
wavelength interchange by the use of a 2D NPC structure was
reported in \cite{Chowdhury:2001-1353:OL}. A nonlinear 2D lattice
fabricated in LiNbO$_3$ was designed to provide an interchange of
the waves with the wavelengths $\lambda_1 = 1535$nm and $\lambda_2
= 1555$nm. The pump was selected at $\lambda_p = 777.2$nm. The
wavelength interchange process takes place by means of the two
concurrent difference-frequency generation processes:
$\omega_p-\omega_1=\omega_2$ and $\omega_p-\omega_2=\omega_1$. The
two difference-frequency processes diffract the converted signals
from the unconverted ones, as depicted in
Fig.~\rpict{interchange}. The information carried by the input beam 1
will be carried by the output beam 2 and vice versa. In
this experiment, one of the main characteristics of the 2D NPC
phase-matching, i.e. non-collinearity of the parametric
interactions,  is used as a real advantage in separating two beams
at the output.

\pict{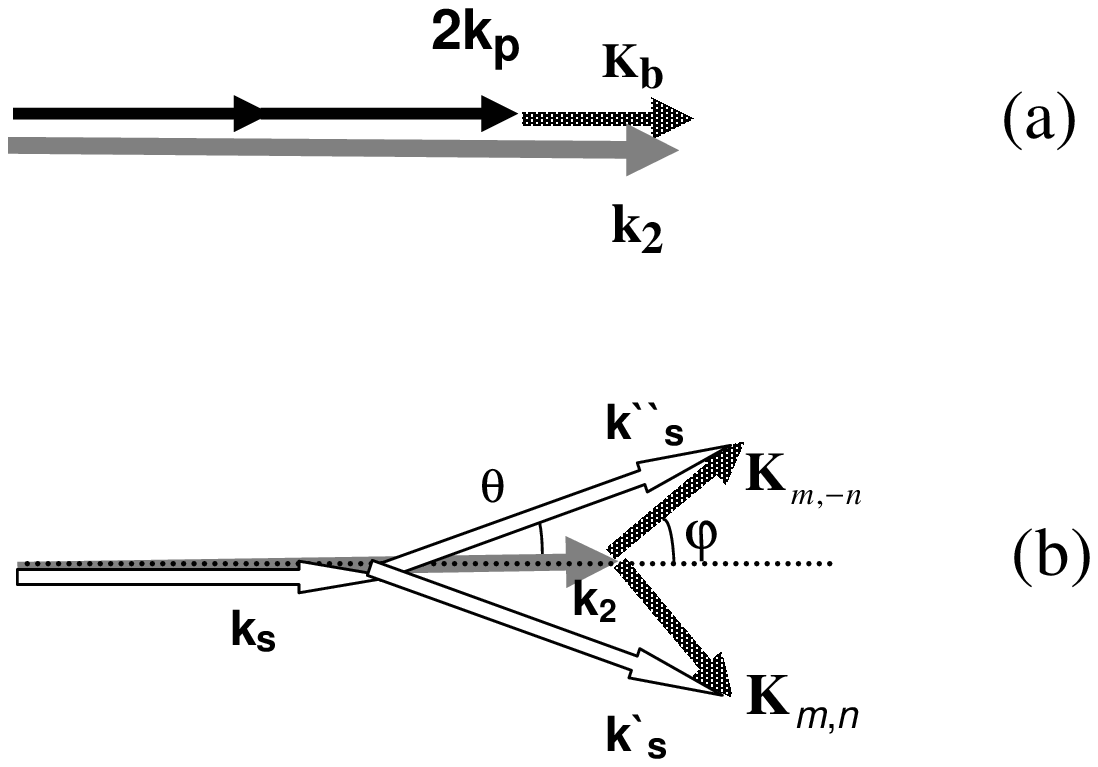}{matching2d}{Phase matching for the first (top)
and second (below) cascading steps, $y_{1p}y_{1p} - z_{2p}$,
$z_{2p}z_{1s} - z_{1s'}$. Drawings correspond to $n_{1z} < n_{2z}
< n_{1y}$ (\cite{Saltiel:2002-921:OL}). }

Another scheme proposed by \cite{Saltiel:2002-921:OL} is also
based on the advantage of non-collinearity of the parametric
interactions in the 2D geometry, and it demonstrates how the
signal wave can be deflected or split by the pump after the
interaction. The phase-matching conditions for this scheme are
shown in Fig.~\rpict{matching2d}. The collinear pump and signal
are at the same frequency, but they are polarized at the
orthogonal directions. If the pump carries some information, the
signal will be modulated according to this information after the
deflection. This interaction belongs to the two-color multistep
cascading processes discussed above. It can also be considered as
a spatial analog of the wavelength conversion process discussed
above.

In conclusion of this part of the review, we would like to mention
that several theoretical studies predicted that the double
phase-matched interactions such as the THG multistep parametric
process can be realized in photonic bandgap (PBG) structures with
the efficiency which should be by several orders higher than that
of the conventional nonlinear media of the similar length. In
particular, this includes an infinite PBG system as a host of the
second-order multistep cascading considered by
\cite{Konotop:1999-1370:JOSB}, cascaded THG in a finite PBG
crystal discussed in \cite{Centini:2001-46606:PRE}, and a
photonic-well type PBG system analyzed in
\cite{Shi:2002-3667:APL}. However, no experimental results on the
fabrication of PBG structures with strong nonlinear properties
were reported yet.

In conclusion, the part~\rsect{matching} presented a brief
overview of different techniques for achieving the simultaneous
phase-matching of several nonlinear parametric processes in
optical structures with a modulated second-order nonlinear
susceptibility. In all those cases, the double phase-matched
interaction becomes possible in a wide region of the optical
wavelengths provided the QPM structure used to achieve the
phase-matching conditions possesses \textit {one extra parameter}
(e.g., the modulation period in the chirped QPM structures, or the
second dimension, in the case of 2D nonlinear photonic crystals).
Some of the possible applications of the double phase-matching
processes include the simultaneous generation of several optical
frequencies in a single-crystal structure, multi-port frequency
conversion, etc. The results presented above look encouraging for
experimental feasibility of the predicted effects in the recently
engineered 1D and 2D periodic optical superlattices.

\section{Multi-color parametric solitons} \lsect{soliton}

In the previous sections, we have discussed the features of multiple parametric processes using the plane-wave and continuous-wave approximations. However, parametric interactions can strongly modify the dynamics of spatial beams or temporal pulses. The parametrically coupled waves in a medium with quadratic nonlinearity may experience mutual spatial focusing or temporal compression and lock together into a stationary state, {\em quadratic soliton} (see \cite{Sukhorukov:1988:NonlinearWave, Torner:1998-229:BeamShaping,  Kivshar:1997-451:AdvancedPhotonics, Etrich:2000-483:ProgressOptics, Torruellas:2001-127:SpatialOptical, Boardman:2001:SolitonDriven, Buryak:2002-63:PRP}, and references therein).

The first analysis of solitons supported by multistep parametric interactions was reported by \cite{Azimov:1987-229:IANF} who found that as many as seven waves can be trapped together, and such {\em multi-color solitons} may be remarkably stable, in particular in the presence of absorption. Since phase mismatch and strength of several parametric interactions can be engineered in nonlinear periodic structures (see Sec.~\rsect{matching}), there exists a flexibility in controlling the soliton properties, making them useful for potential applications including all-optical switching. In this section, we overview the properties of multistep parametric solitons that can form under the conditions of cascaded THG, two-color FWM, frequency conversion, and FHG.

\subsection{Third-harmonic parametric solitons}
       \lsect{solitonTHG}

\pict{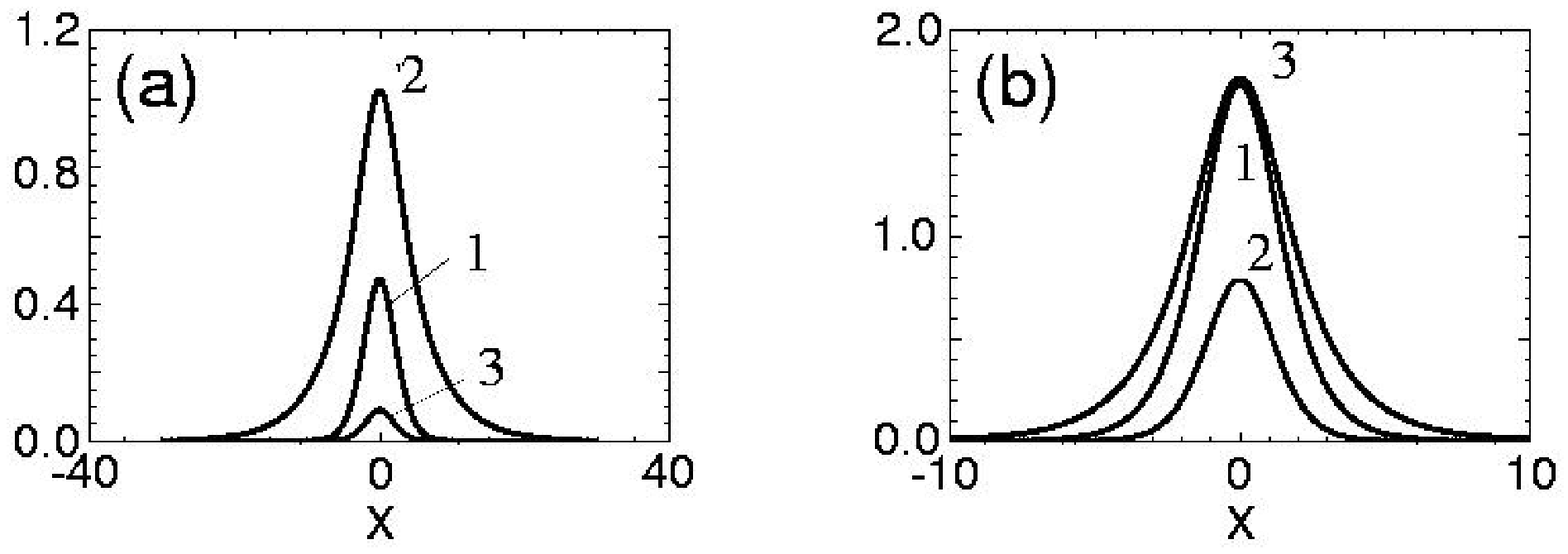}{THGsol}{
(a,b) Profiles of the THG parametric solitons for different values of phase mismatch (normalized units). Labels 1,2, and 3 indicate the number of harmonic.
(\cite{Kivshar:1999-759:OL})}

Generation of the TH wave through cascaded SHG and SFG processes was discussed in Sec.~\rsect{THG}. \cite{Komissarova:1996-1034:LP} demonstrated that these parametric interactions can result in simultaneous trapping of all the interacting waves and formation of a {\em three-colour parametric soliton}. \cite{Kivshar:1999-759:OL} performed a detailed investigation of these solitons and demonstrated their stability under various conditions.
Such solitons can be described by Eqs.~\reqt{THGeqns} with additional terms accounting for beam diffraction,
\begin{equation} \leqt{THGeqnsSol}
   \begin{array}{l} {\displaystyle
      \frac{d A_{1}}{d z}
      + \frac{i}{2 k_1} \nabla_{\perp}^2 A_1
      = - i \sigma_{1} A_{2} A_{1}^{\ast} e^{-i\Delta k_{\rm SHG}z}
      -   i \sigma_{3} A_{3} A_{2}^{\ast} e^{-i\Delta k_{\rm SFG}z} ,
   } \\*[9pt] {\displaystyle
      \frac{d A_{2}}{d z}
      + \frac{i}{2 k_2} \nabla_{\perp}^2 A_2
      = - i \sigma_{2} A_{1}^{2} e^{i\Delta k_{\rm SHG}z}
      -   i \sigma_{4} A_{3} A_{1}^{\ast} e^{-i\Delta k_{\rm SFG}z} ,
   } \\*[9pt] {\displaystyle
      \frac{d A_{3}}{d z}
      + \frac{i}{2 k_3} \nabla_{\perp}^2 A_3
      = - i \sigma_{5} A_{2} A_{1} e^{i\Delta k_{\rm SFG}z},
   } \end{array}
\end{equation}
where the contribution due to the direct THG process is neglected, assuming that the cascaded $\chi^{(2)}$ processes are dominant. The operator $\nabla_{\perp}^2$ acts on the spatial dimensions in the transverse plane (perpendicular to the beam propagation direction $z$).
 Similar equations can describe the formation of temporal solitons, where nonlinearity compensates for both the group-velocity mismatch and second-order dispersion effects (\cite{Huang:2001-418:CHI}).

Stationary propagation of solitons is possible only when there is no energy exchange between the constituent waves; this requires that the individual phase velocities are synchronized due to nonlinear coupling. Such solutions of Eqs.~\reqt{THGeqnsSol} have the form
$   A_m({\bf r},z) = B_m({\bf r}) e^{i m \beta z}$ ,
where $\beta$ is the nonlinear propagation constant and $B_m({\bf r})$ are the transverse soliton profiles. Spatial soliton properties were analyzed in the (1+1)-dimensional geometry, when the beam is confined by a planar waveguide and experiences diffraction only in one transverse direction so that $\nabla_{\perp}^2 = \partial^2/\partial x^2$. There exist two families of single-hump solitons, however solutions with higher power are always unstable. Soliton examples from the low-power branch are shown in Fig.~\rpict{THGsol}. It was also found that cascading interactions can lead to radiative decay of self-localized beams, however such quasi-solitons can demonstrate robust propagation for several diffraction length.

\cite{Lobanov:2002-1783:IANF, Lobanov:2003-407:IVR} developed a theoretical description of THG solitons in QPM structures, where all the frequency components oscillate along the propagation direction. The spectrum of these oscillations was determined analytically and numerically, and it was found that the solitons can be described by averaged equations with additional terms accounting for induced Kerr effects: self-phase modulation, cross-phase modulation, and third-harmonic generation.

\subsection{Two-color parametric solitons}
       \lsect{soliton2color}

\pict{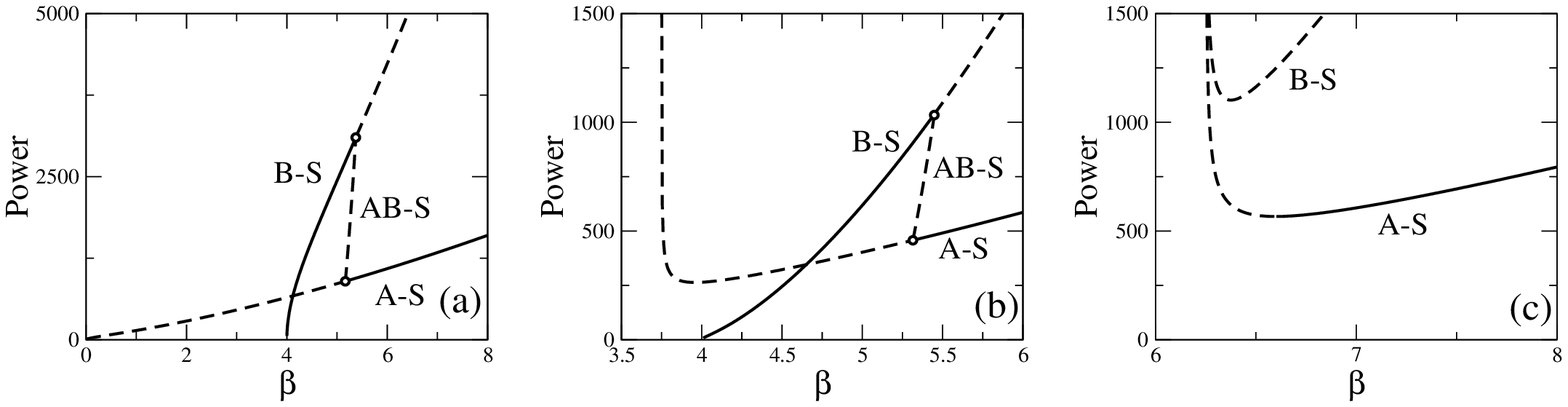}{2cPower}{
(a-c) Characteristic dependences of the total soliton power on the propagation constant corresponding to different phase mismatches for two-wave (A-S and B-S) and three-wave (AB-S) soliton families (normalized units).
Solid lines show stable solitons, and dashed~-- unstable; open circles mark the bifurcation points.
(the figures adapted from \cite{Sukhorukov:2000-4530:PRE})}

\pict{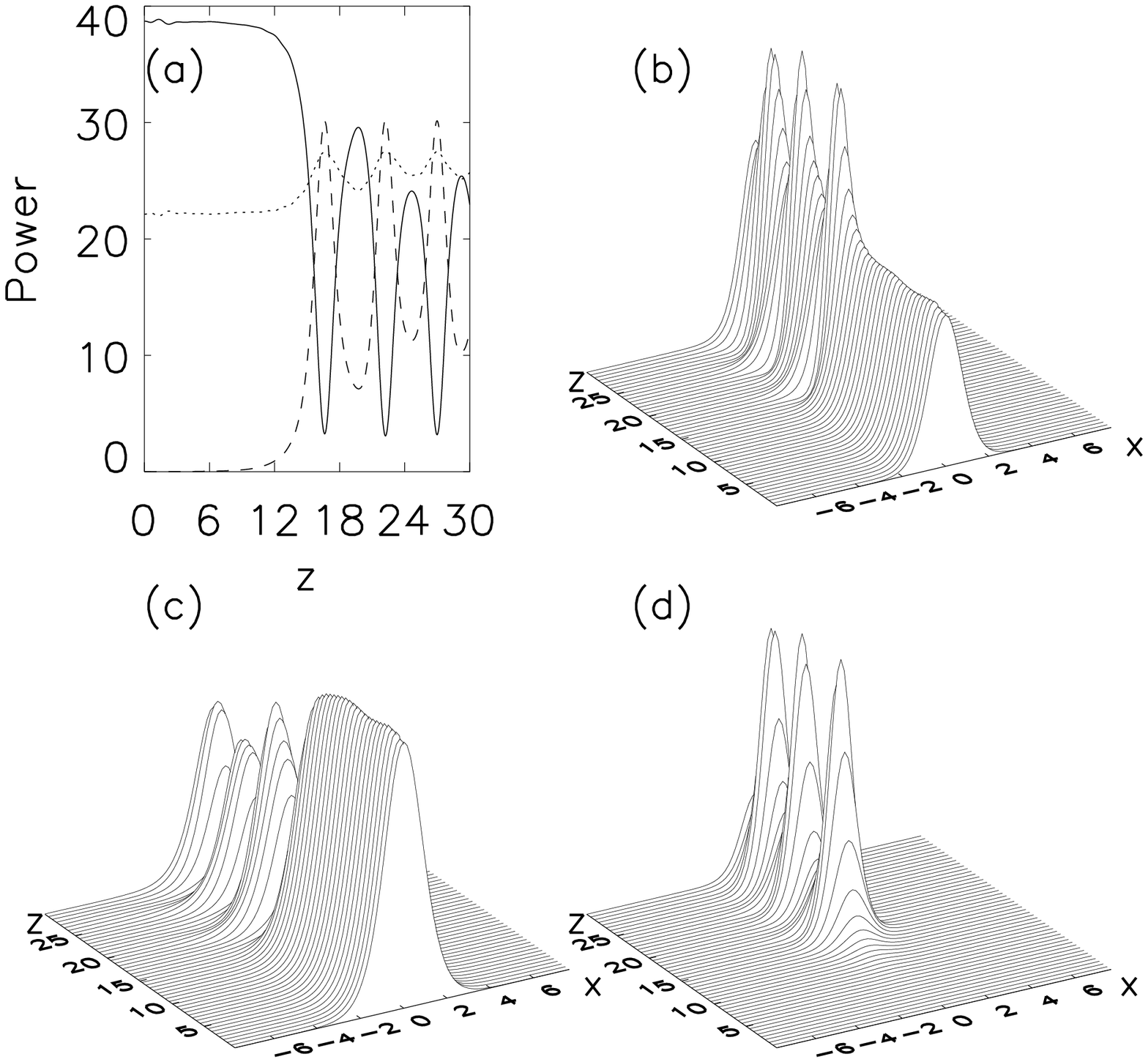}{2cDynamics}{
(a)~Change of the normalized power in FF (A, solid) and
SH (S, dotted) components, which initially constitute a two-wave
soliton, and in the guided mode (B, dashed), demonstrating amplification of a guided wave.
(b-d)~Evolution of the intensity profiles for (b)~the effective waveguide (SH), and (c,d)~A and B FF components, respectively.
(\cite{Kivshar:1999-5056:PRE})
}

As discussed above in Sec.~\rsect{2color}, two-colour multistep cascading can involve up to six steps corresponding to type I and type II interactions between two pairs of orthogonally polarized FF and SH waves. Existence of spatial quadratic solitons under these most general conditions was demonstrated by \cite{Boardman:1998-891:OQE} who found that, by changing the FF polarization at the input, the SH output can be precisely controlled. A possibility for achieving beam steering and switching based on collision of two solitons was demonstrated as well.

The general case involving four waves coupled by six parametric processes requires three phase-matching conditions to be fulfilled simultaneously, and the respective components of the $\chi^{(2)}$ susceptibility tensor should also be nonzero. Formation of three-wave solitons which require two phase-matching conditions may be easier to achieve, as indicated by the experimental results in the plane-wave regime (see references in Sec.~\rsect{2color}). \cite{Kivshar:1999-5056:PRE, Sukhorukov:2000-4530:PRE} considered formation of solitons involving two FF components (A and B) that are coupled together through a single SH component (S). The soliton formation can be described by the following set of coupled equations that are obtained in the slowly varying envelope approximation,
\begin{equation} \leqt{1}
 \begin{array}{l} {\displaystyle
    \frac{\partial A}{\partial z}
      + \frac{i}{2 k_A} \nabla_{\perp}^2 A
    = - i \sigma_{\rm AS} S A^{\ast} e^{-i\Delta k_{\rm AS} Z} ,
  } \\*[9pt] {\displaystyle
    \frac{\partial B}{\partial z}
    + \frac{i}{2 k_B} \nabla_{\perp}^2 B
    = - i \sigma_{\rm BS} S B^{\ast} e^{-i\Delta k_{\rm BS} z} ,
  } \\*[9pt] {\displaystyle
    \frac{\partial S}{\partial z}
    + \frac{i}{2 k_S} \nabla_{\perp}^2 S
    =- i \sigma_{\rm AS} A^2 e^{i\Delta k_{\rm AS} Z}
    - i \sigma_{\rm BS} B^2 e^{i\Delta k_{\rm BS} Z} ,
  } \end{array}
\end{equation}
where $\sigma_{\rm AS}$ and $\sigma_{\rm BS}$ are proportional to the elements of the second-order susceptibility tensor, as discussed in the previous sections, $\Delta k_{\rm AS} = k_{\rm S} - 2 k_{\rm A}$ and $\Delta k_{\rm BS} = k_{\rm S} - 2 k_{\rm B}$ are the wave-vector mismatch parameters for the A-S and B-S parametric interaction processes, respectively.

Similar to the case of the THG cascading, multi-component bright solitons are found in the form of stationary waves, which phases are locked together. When one of the FF waves is zero, then other two waves can form a type I quadratic soliton (A-S or B-S). However, these solitons may experience an instability associated with the amplification of the orthogonally polarized FF wave through the parametric decay instability of the SH component. Such changes of stability are associated with the bifurcation for two-wave to three-wave solitons, as shown in Fig.~\rpict{2cPower} for a (1+1)-dimensional case. We note that the Vakhitov-Kolokolov stability criterion (\cite{Vakhitov:1973-1020:IVR, Pelinovsky:1995-591:PRL}) can only be used as a necessary condition: branches with the negative power slope are unstable, but the positive slope does not guarantee stability.

Examples in Fig.~\rpict{2cPower} demonstrate the existence of stable solitons that have the same power but different polarizations of the FF waves. Such {\em multistability} is a sought-after property on nonlinear systems, since this may allow realization of controlled switching between different states. In Fig.~\rpict{2cDynamics}, we illustrate the development of the soliton instability which results in a power exchange between the two FF components. Such type of polarization switching was earlier predicted for plane waves by \cite{Assanto:1994-1720:OL}.

 The properties of two-color multistep cascading solitons were analyzed by \cite{Towers:1999-1738:OL, Towers:2000-2018:JOSB} for the case when two FF waves and the SH component are additionally coupled together through the type II parametric process. Due to this interaction, all solitons contain both FF waves, and transition between the states with different polarizations along the soliton family does not involve bifurcations. It was found that soliton multistability can be realized under appropriate conditions, both in planar (one-dimensional) and bulk (two-dimensional) configurations. On the other hand, \cite{Towers:2000-2018:JOSB} reported that no multistability occurs if the BB-S interaction is suppressed and the waves are coupled only through AA-S and AB-S processes.

\subsection{Solitons due to wavelength conversion}
       \lsect{solitonConversion}

\pict{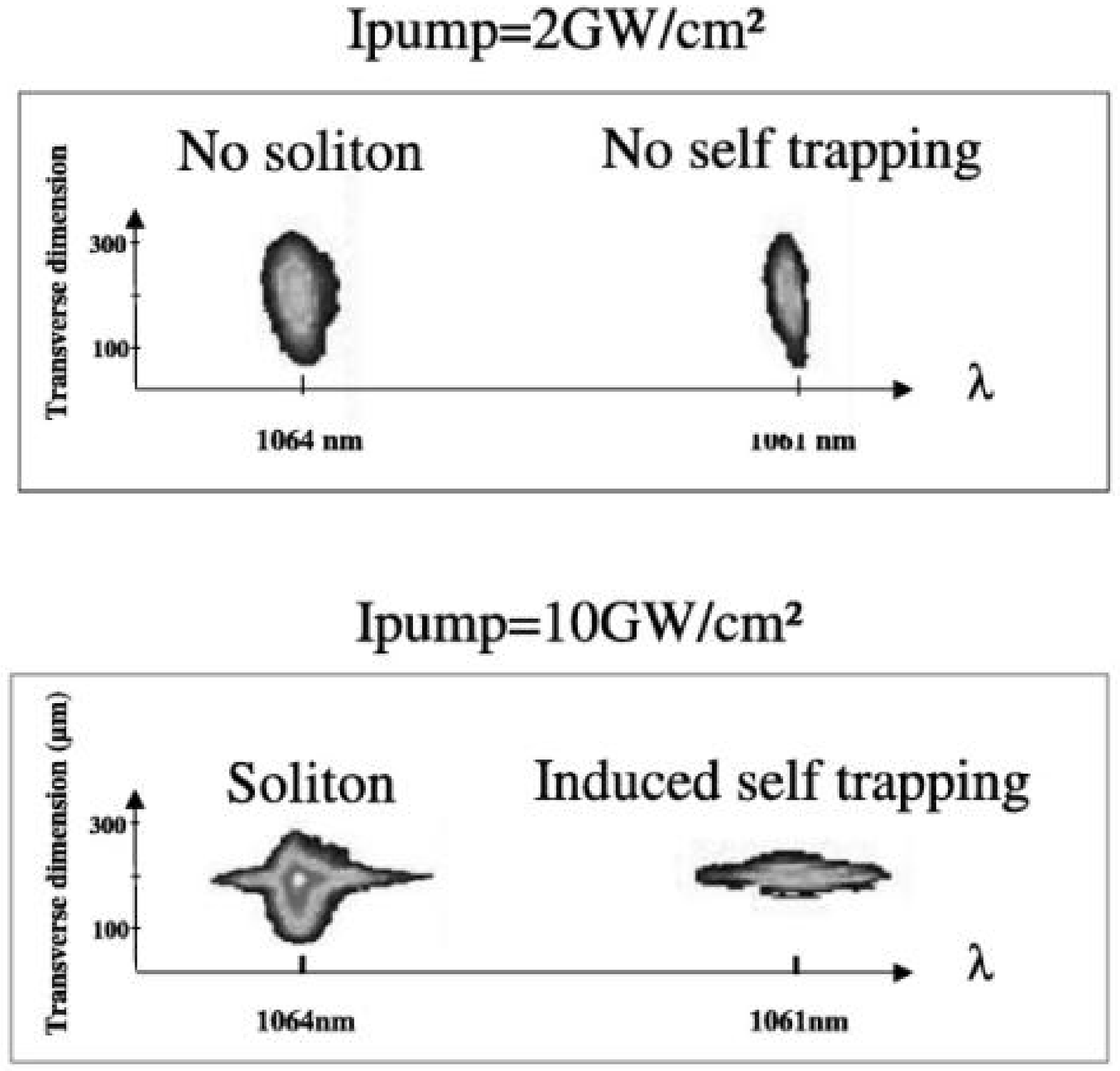}{solConversion}{
Beam profiles at the output of the 2cm long KTP crystal for low and high input intensities. Horizontal scale: wavelength; vertical scale: transverse dimension on the output pattern. Right part of the figure corresponds to the probe wave (1061 nm), the left part corresponds to the FF component of the soliton (pump at 1064 nm).
(\cite{Couderc:2002-421:OC})
}

Formation of spatial parametric solitons requires a strong coupling between the interacting waves, and this can be achieved when the parametric interactions are nearly phase-matched. In the cases of parametric THG and two-colour FWM, two (or even more) phase-matching conditions should be satisfied simultaneously. On the other hand, in the wavelength conversion scheme (see Sec.~\rsect{conversion}), only single phase-matching condition should be implemented, and this greatly simplifies the requirements for experimental observation of spatial beam localization and soliton formation under the conditions of multistep cascading. Indeed, the first experimental study was recently reported by \cite{Couderc:2002-421:OC} who demonstrated that a multi-color soliton composed of a pump and its second harmonic creates an effective waveguide that can trap a weak probe beam which frequency is slightly detuned from the pump wave (see Fig.~\rpict{solConversion}). This trapping is realized due to a parametric coupling of the probe with both the pump and harmonic waves, and additional side-band frequencies are generated in the process according to the principles of wavelength conversion. \cite{Couderc:2002-421:OC} demonstrated that evolution of a weak probe is governed by the equation,
\begin{equation} \leqt{solConversion}
   i \frac{\partial a}{\partial z}
    + \frac{1}{2 k_1} \nabla_{\perp}^2 a
    + \delta\omega^2 \frac{1}{2}
       \left. \frac{\partial^2 k}{\partial \omega^2}\right|_{\omega_1} a
    + \frac{\sigma^2}{\Delta k} \left( |A|^2 - |B|^2 \right) a
    = 0,
\end{equation}
where $A$ and $B$ are the profiles of mutually trapped pump wave and its second harmonic components, $\delta\omega$ is the frequency detuning of the probe, $\omega_1$ and $k_1$ are the frequency and the wavenumber of the fundamental frequency wave, and $\Delta k$ is the phase mismatch between the SH and FF components. The last term in Eq.~\reqt{solConversion} defines the profile of an effective waveguide which is experienced by the probe beam, and trapping can be realized when the overall sign is positive.
It has been demonstrated in earlier studies that the FF component ($|A|^2$) dominates in a quadratic soliton when $\Delta k >0$, whereas the SH component ($|B|^2$) becomes larger at negative mismatches, $\Delta k <0$. Therefore, the probe can be trapped in both cases, however anti-waveguiding effect may occur for particular values of the mismatches and pump intensities.

\subsection{Other types of multi-color parametric solitons}
       \lsect{solitonFHG}

\pict{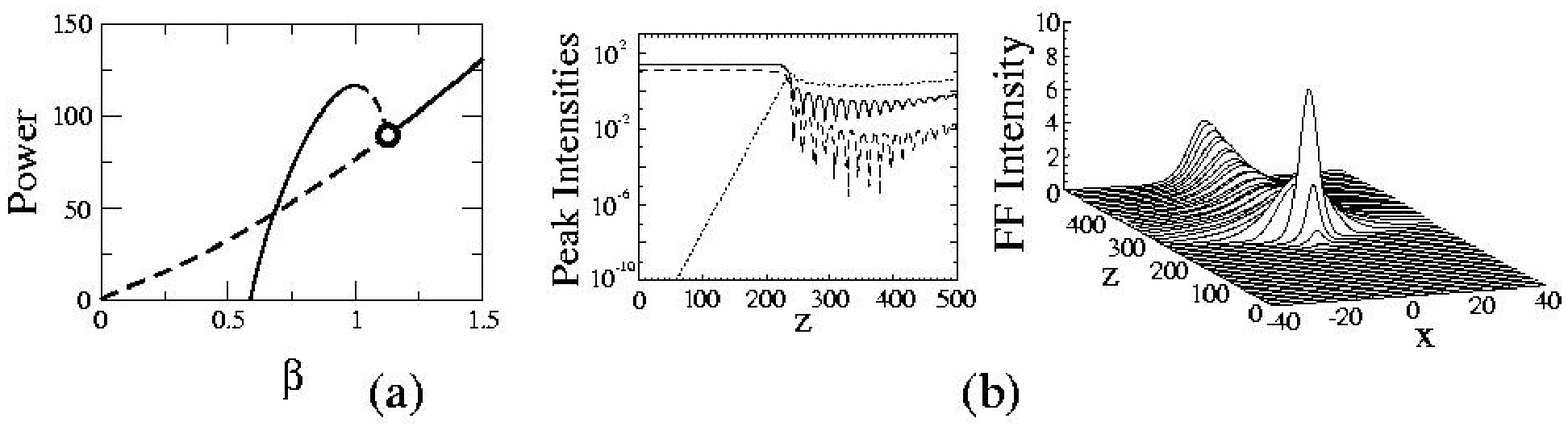}{solFHG}{
(a)~Thick~--- two-wave (SH + FH), and black~--- three-wave solitons; solid and dashed lines mark stable and unstable solutions, respectively. Open circle is the bifurcation point;
(b)~Development of a decay instability of a two-wave soliton corresponding to $\beta = 1$ in plot (a), and generation of a three-component soliton.
Dotted, solid, and dashed curves in the left plot show the FF, SH, and FH peak normalized intensities vs. distance, respectively.
(\cite{Sukhorukov:2001-34:PLA})
}

Formation of spatial parametric solitons due to FHG in a planar waveguide configuration was analyzed theoretically by \cite{Sukhorukov:2001-34:PLA}. It was found that, similar to the case of two-color parametric solitons (see Sec.~\rsect{soliton2color}), there can exist two-wave solitons (coupled second and fourth harmonics) and three-wave solitons containing all the three frequency components. Such coexistence of different solitons may give rise to multistability, as illustrated in Fig.~\rpict{solFHG}(a). A two-wave soliton can exhibit parametric decay instability due to the energy transfer into the fundamental-frequency wave, this process is illustrated in Fig.~\rpict{solFHG}(b).

\cite{Towers:2002-46620:PRE} performed a theoretical analysis of polychromatic solitons consisting of two low-frequency components $\omega_1$ and $\omega_2$ and three high-frequency waves, $2 \omega_1$, $2 \omega_2$, and $\omega_1 + \omega_2$. It was found that the power of a polychromatic soliton can be smaller compared to conventional two-wave parametric solitons, and this can be advantageous for applications. It was also demonstrated that polychromatic solitons can emerge after a collision of two-wave solitons, and soliton interactions may be used to implement a simple all-optical XOR logic gate.

\section{Conclusions} \lsect{conclusion}

Parametric interactions and phase matching are the key concepts in
nonlinear optics. The generation of new waves at the frequencies
which are not accessible by standard sources and the efficient
frequency conversion and manipulation are among the main goals of
the current research in nonlinear optics. The QPM technique is
becoming one of the leading technologies for the optical devices
based on the parametric wave interaction; it allows generating new
harmonics and can be made compatible with the operational
wavelengths of optical communication systems.

In this paper, we have presented, for the first time to our
knowledge, a systematic overview of the basic principles of the
simultaneous phase matching of two (or more) parametric processes
in different types of one- and two-dimensional nonlinear quadratic
optical lattices, the so-called multistep parametric interactions.
In particular, we have discussed different types of multiple
phase-matched processes in the engineered QPM structures and
two-dimensional nonlinear quadratic photonic crystals, as well as
the properties of multi-color optical solitons generated by the
multistep parametric processes. We have also summarized the most
important experimental demonstrations for the multi-frequency
generation due to multistep parametric processes. We believe that
such a comprehensive summary of the up-to-date achievements will
become a driving force for the future even more active research in
this exciting field of nonlinear optics.

\section{Acknowledgements} \lsect{acknowledgements}

This work was produced with the assistance of the Australian
Research Council under the ARC Centers of Excellence Program. The
Center for Ultra-high bandwidth Devices for Optical Systems
(CUDOS) is an ARC Center of Excellence. Solomon Saltiel thanks the
Nonlinear Physics Group for hospitality and the Research School of
Physical Sciences and Engineering at the Australian National
University for a grant of the senior visiting fellowship.

\begin{small}

\end{small}
\end{document}